\documentclass[a4paper,11pt]{article}
\pdfoutput=1 

\usepackage[usenames,dvipsnames,svgnames,table]{xcolor} 
\usepackage{jcapmod}
\usepackage{verbatim}

\usepackage[T1]{fontenc} 
\usepackage[latin9]{inputenc}
\definecolor{lightgray}{gray}{0.9}
\definecolor{Amber}{rgb}{1.0, 0.75, 0.0}
\definecolor{blizzardblue}{rgb}{0.67, 0.9, 0.93}

\usepackage{float}
\usepackage{graphicx}
\usepackage{hyperref}
\usepackage{amsmath}
\usepackage{amssymb}
\usepackage{microtype}
\usepackage[shortlabels]{enumitem}
\usepackage{cancel}
\usepackage{amsbsy}
\usepackage{color}
\usepackage[capitalise]{cleveref}
\usepackage{breakurl}
\usepackage{mathtools}
\usepackage{csquotes}
\usepackage{lmodern}
\usepackage{bm}
\usepackage{dsfont}
\usepackage{bbold}
\usepackage{empheq}

\usepackage{caption}
\usepackage{subcaption}

\pdfpageheight\paperheight
\pdfpagewidth\paperwidth

\usepackage[textwidth=0.71\paperwidth, textheight=0.8\paperheight]{geometry}

\renewcommand*{\vec}[1]{\bm{#1}}
\newcommand*{\unitvec}[1]{\vec{\hat{#1}}}
\newcommand*{\mat}[1]{\bm{\mathsf{#1}}}
\DeclareMathOperator{\diag}{diag}
\let\Re\relax
\let\Im\relax
\DeclareMathOperator{\Re}{Re}
\DeclareMathOperator{\Im}{Im}
\DeclareMathOperator{\Var}{Var}

\newcommand{\identity}{\ensuremath{\mathds{1}}}
\newcommand*{\transpose}[1]{\ensuremath{{#1}^{T}}}
\newcommand*{\inverse}[1]{\ensuremath{{#1}^{-1}}}
\newcommand{\integers}{\mathbb{Z}}

\newcommand*{\E}[1]{\texorpdfstring{\ensuremath{E_{#1}}}{E#1}}
\newcommand*{\Espace}{\texorpdfstring{\ensuremath{E^3}}{E(3)}}
\newcommand*{\A}[1]{\texorpdfstring{\ensuremath{A_{#1}}}{A#1}}
\newcommand{\slabh}{\texorpdfstring{\ensuremath{\E{16}^{(\mathrm{h})}}}{E16h}}
\newcommand{\slabi}{\texorpdfstring{\ensuremath{\E{16}^{(\mathrm{i})}}}{E16i}}
\newcommand{\Kdelta}{\delta^{(\textrm{K})}}
\newcommand{\Ddelta}{\delta^{(\textrm{D})}}
\newcommand{\setN}{\mathcal{N}}
\newcommand*{\Mod}[1]{\,(\mathrm{mod}\ #1)}
\newcommand{\DeltaYstarforell}{\DeltaYstar}
\newcommand{\DeltaYstar}{\Delta^{Y*}}
\newcommand{\vecnp}{\smash{\vec{n}'}}

\usepackage[table]{xcolor}

\allowdisplaybreaks

\makeatletter
\DeclareRobustCommand{\rcite}[1]{%
  \rcite@aux#1,\@nil{#1}%
}
\def\rcite@aux#1,#2\@nil#3{%
  \if\relax#2\relax
    Ref.~\cite{#3}%
  \else
    Refs.~\cite{#3}%
  \fi
}
\makeatother

\subheader{IFT-UAM/CSIC-23-75}

\title{Cosmic topology. Part IIa. Eigenmodes, correlation matrices, and detectability of orientable Euclidean manifolds}

\author[a,b]{Johannes R. Eskilt,}
\author[c,d,b]{Yashar Akrami,}
\author[e,f,g]{Stefano Anselmi,}
\author[d]{Craig J. Copi,}
\author[b]{Andrew H. Jaffe,}
\author[h]{Arthur Kosowsky,}
\author[d]{Deyan P. Mihaylov,}
\author[d,b]{Glenn D. Starkman,}
\author[d]{Andrius Tamosiunas,}
\author[d]{James B. Mertens,}
\author[d]{Pip Petersen,}
\author[d]{Samanta Saha,}
\author[d]{Quinn Taylor,}
\author[d]{and \"{O}zen\c{c} G\"{u}ng\"{o}r}

\collaboration{(COMPACT Collaboration)}
\collaborationImg{\includegraphics[width = 0.1\textwidth]{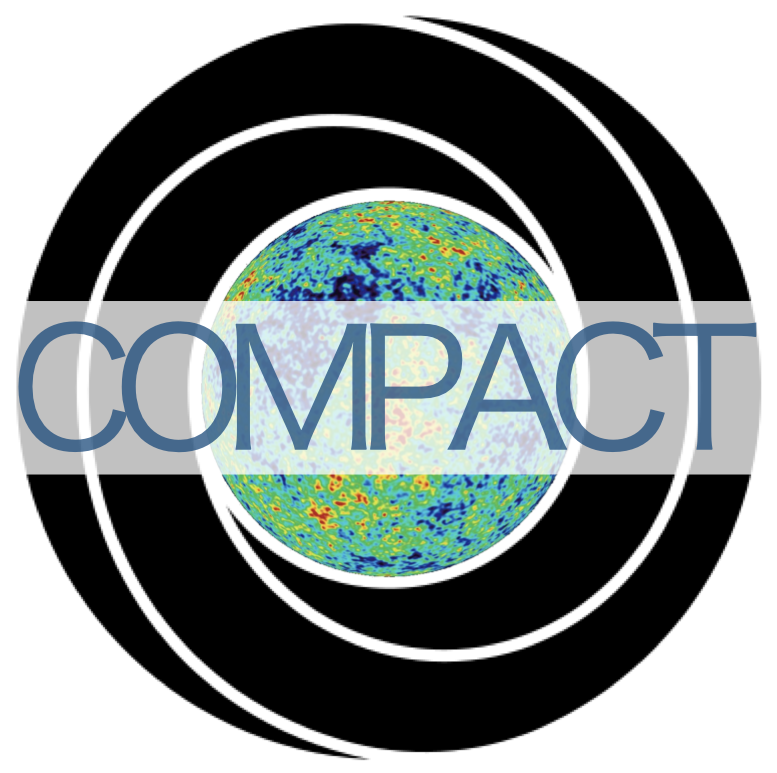}}

\affiliation[a]{Institute of Theoretical Astrophysics, University of Oslo, P.O. Box 1029 Blindern, N-0315 Oslo, Norway}
\affiliation[b]{Astrophysics Group \& Imperial Centre for Inference and Cosmology, Department of Physics, Imperial College London, Blackett Laboratory, Prince Consort Road, London SW7 2AZ, United Kingdom}
\affiliation[c]{Instituto de F\'isica Te\'orica (IFT) UAM-CSIC, C/ Nicol\'as Cabrera 13-15, Campus de Cantoblanco UAM, 28049 Madrid, Spain}
\affiliation[d]{CERCA/ISO, Department of Physics, Case Western Reserve University, 10900 Euclid Avenue, Cleveland, OH 44106, USA}
\affiliation[e]{Dipartimento di Fisica e Astronomia ``G. Galilei'', Universit\`a degli Studi di Padova, via Marzolo 8, I-35131 Padova, Italy}
\affiliation[f]{INFN, Sezione di Padova, via Marzolo 8, I-35131 Padova, Italy}
\affiliation[g]{LUTH, UMR 8102 CNRS, Observatoire de Paris, PSL Research University, Universit\'e Paris Diderot, 92190 Meudon, France}
\affiliation[h]{Department of Physics and Astronomy, University of Pittsburgh, Pittsburgh, PA 15260, USA}

\emailAdd{j.r.eskilt@astro.uio.no}
\emailAdd{yashar.akrami@csic.es}
\emailAdd{stefano.anselmi@pd.infn.it}
\emailAdd{craig.copi@case.edu}
\emailAdd{a.jaffe@imperial.ac.uk}
\emailAdd{kosowsky@pitt.edu}
\emailAdd{deyan.mihaylov@case.edu}
\emailAdd{glenn.starkman@case.edu}
\emailAdd{andrius.tamosiunas@case.edu}
\date{\today}

\abstract{
If the Universe has non-trivial spatial topology, 
observables depend on both the parameters of the spatial manifold and the position and orientation of the observer.
In infinite Euclidean space, most cosmological observables arise from the amplitudes of Fourier modes of primordial scalar curvature perturbations.
Topological boundary conditions  replace the full set of Fourier modes with specific linear combinations of selected Fourier modes as the  eigenmodes of the scalar Laplacian.
We present formulas for eigenmodes in orientable Euclidean manifolds with the topologies \E{1}--\E{6}, \E{11}, \E{12}, \E{16}, and \E{18} that encompass the full range of manifold parameters and observer positions, generalizing previous treatments.
Under the assumption that the amplitudes of primordial scalar curvature eigenmodes are independent random variables, for each topology we obtain the correlation matrices of Fourier-mode amplitudes (of scalar fields linearly related to the scalar curvature) and the correlation matrices of spherical-harmonic coefficients of such fields sampled on a sphere, such as the temperature of the cosmic microwave background (CMB).
We evaluate the detectability of these correlations given the cosmic variance of the observed CMB sky. We find that topologies where the distance to our nearest clone is less than about 1.2 times the diameter of the last scattering surface
of the CMB give a correlation signal that is larger than cosmic variance noise in the CMB. 
This implies that if cosmic topology is the explanation of large-angle anomalies in the CMB, then the distance to our nearest clone is not much larger than the diameter of the last scattering surface.
We argue that the topological information is likely to be better preserved in three-dimensional data, such as will eventually be available from large-scale structure  surveys.
}


\begin{document}
\maketitle
\flushbottom

\section{Introduction}
\label{secn:intro}

In the century since the proposal of general relativity (GR) \cite{Einstein:1916vd} as the dynamical theory of spacetime, and therefore of cosmology \cite{Einstein:1917ce}, we have widely come to view space as a three-dimensional (Riemannian) manifold with a geometry that is inhomogeneous on small scales but homogeneous and isotropic on large scales \cite{peebles:1993, Ostriker1995, mukhanov2005physical, Efstathiou2020}.
This geometry evolves according to the Einstein field equations, which are local second-order differential equations in which the evolution of the geometry is sourced by the stress-energy content of space.
Meanwhile, the evolution of that stress-energy content is governed by Euler-Lagrange equations that incorporate the influence of the geometry on the stress-energy.

It is useful to do a background/perturbation decomposition of the metric.
The largest-scale geometry is, per this view, given by the Friedman-Lema\^{i}tre-Robertson-Walker (FLRW) metric, 
\begin{equation}
    \label{eqn:FLRWmetric}
    \mathrm{d}s^2 = -\mathrm{d}t^2 + a^2(t) \left[\frac{\mathrm{d}r^2}{1-k(r/r_c)^2} + r^2(\mathrm{d}\theta^2 + \sin^2\theta \, \mathrm{d}\phi^2) \right]
\end{equation}
(in a convenient choice of coordinates, with $r_c$ the comoving curvature scale).
Deviations from this geometry on large scales are usually treated perturbatively, starting with a scalar-vector-tensor decomposition of the perturbations in which they are evolved analytically \cite{Mukhanov:1990me, peebles:1993} or numerically \cite{Giblin2016prl, Bernardeau2002pr}, or in a Newtonian approximation with N-body simulations \cite{Dehnen2011epjp, Angulo2022lrca}. On smaller scales, e.g., inside a galaxy, the deviations from \cref{eqn:FLRWmetric} may be highly non-linear, and \cref{eqn:FLRWmetric} may not even be a relevant approximation \cite{Benson2010pr}.

Equation \eqref{eqn:FLRWmetric} is the metric  (also known as the local geometry)  of a homogeneous isotropic space, whose sole dynamical variable is the scale factor $a(t)$.
There are only three such metrics, characterized by the curvature constant $k \in \{-1,0,+1\}$, representing, respectively,
homogeneous isotropic negatively curved (``hyperbolic'') space ($k=-1$, $H^3$),
homogeneous isotropic flat (``Euclidean'') space ($k=0$, \Espace), and
homogeneous isotropic positively curved (``spherical'') space ($k=+1$, $S^3$). 

The specific justification for this assumption of a homogeneous and isotropic background metric is a matter for some reconsideration, despite the near large-scale homogeneity and isotropy of our observed universe.
An inflationary perspective may prefer a nearly homogeneous and isotropic geometry, but the considerable evidence for violations of statistical isotropy in the cosmic microwave background (CMB) temperature data (see, e.g., \rcite{Planck:2013lks,Planck:2015igc,Schwarz:2015cma,Planck:2019evm,Abdalla:2022yfr} for reviews, and \rcite{Jones:2023ncn}) calls into question a strict enforcement of this received wisdom. 
Not all spatial three-manifolds admit a single homogeneous metric; those that do,  may admit one of the above three, or they may admit one of five others that are not isotropic \cite{Thurston1982ThreeDM}. Nevertheless, we leave this expansion of the suite of possible cosmological background metrics to future consideration.

There is also widespread misconception that the three FLRW metrics \eqref{eqn:FLRWmetric}  fully specify the three possible smooth three-spaces $H^3$, \Espace, and $S^3$.  This is not so.  Those are the covering spaces associated with these three metrics, i.e., the manifolds that admit those geometries and in which all closed loops can be deformed continuously to a point.
For each such geometry, there are many distinct manifolds that admit that geometry.
The richness of possible three-manifolds and their connections with possible spatial geometries has received much attention from mathematicians in recent decades (perhaps most notably the Thurston conjecture \cite{Thurston1982ThreeDM}, proven by Perelman \cite{Perelman2002arx, Perelman2003arx}), but exploring that richness is far beyond the scope of this paper (for an overview see \rcite{Lachieze-Rey1995pr, Luminet2007:arx}).
For our purposes, it will suffice to note that there are exactly 18 topologically distinct possibilities for flat space: \E{1}--\E{18} \cite{Lachieze-Rey1995pr, Luminet1999, Riazuelo2004:prd, Cornish:2003db, hitchman2009geometry}. For ten of these 18 topologies, \E{1}--\E{10}, the manifolds are compact --- they have finite three-volume when calculated with the FLRW metric ($k=0$);
\E{11}--\E{15} have finite area two-dimensional slices; 
\E{16} and \E{17} are compact in only one dimension;
\E{18} is the covering space and is infinite in every direction.
These are each described in detail below.
For $H^3$ and $S^3$ the topological possibilities are far richer --- there are a small number of classes of spherical three-manifolds each with either a finite number or a countable infinity of members; there is no known enumeration of the hyperbolic three-manifolds, even of the compact ones \cite{Thurston1982ThreeDM}.
This richness from the topologists' perspective should not be misunderstood by cosmologists as a statement that there are ``more'' curved three-manifolds than Euclidean ones --- the 18 flat topologies each require up to 6 real parameters to characterize a manifold, whereas the compact hyperbolic topologies have only one real parameter.

At any given time in the last century, a small group of cosmologists have been interested in the possibility that space is not the covering space of one of the three FLRW geometries, but rather is one of the many other possibilities.
This interest goes back at least to de Sitter \cite{deSitter1917mnras}, who remarked that Einstein's original $S^3$ cosmology would have been improved if situated in $S^3/\integers_2$ --- the three-sphere with opposite points identified. 
Interest has continued ever since, albeit tempered by the success of the inflationary paradigm for the early Universe (for an overview of recent developments see \rcite{COMPACT:2022gbl}).
Inflation addresses questions about the initial conditions of the Universe by invoking a period of accelerated expansion during which information about those initial conditions is stretched beyond the apparent horizon \cite{Tsujikawa2003, Linde2007, Baumann2008,Vasquez2018}.
If the Universe's non-trivial topology is an initial condition, an extended conventional inflationary period would make topology hard or impossible to detect, however, as discussed in \rcite{Linde:2004},   ``in many inflationary models based on string theory there is no exponential suppression of creation of topologically nontrivial compact flat or open inflationary universes'',  ``suggest[ing] ... that compact flat or open universes with nontrivial topology should be considered a rule rather than an exception''.

In the late 1990s and 2000s, attention turned to the possibility of gathering convincing evidence (or imposing strict  constraints) on cosmic topology from impending observations, especially the then-upcoming full-sky high-resolution survey of the CMB temperature fluctuations by the Wilkinson Microwave Anisotropy Probe (WMAP) \cite{WMAP:2003elm,WMAP:2012nax}.
Two principal observational approaches were proposed, studied, and eventually implemented, for both WMAP and {\it Planck} \cite{Cornish:1996kv,Cornish:1997ab,  Cornish:1997hz,Cornish:1997rp,deOliveira-Costa:2003utu,Cornish:2003db,Luminet2006brz,ShapiroKey:2006hm,Mota:2010jb,Bielewicz:2010bh,Bielewicz:2011jz,Vaudrevange:2012da,Aurich:2013fwa,Planck:2013okc, Planck:2015gmu,Luminet2016:uni,Starkman_Priv_Comm}.

The ``circles-in-the-sky'' method builds on the observation that in a topologically non-trivial universe with any closed spatial loop shorter than the diameter of the last scattering surface ($L_{\mathrm{LSS}}$) of the CMB, the LSS self-intersects \cite{Cornish:1996kv,Cornish:1997ab,  Cornish:1997hz,Cornish:1997rp}. Since the LSS is a thin spherical annulus centered on the observer,
that self-intersection is a circular locus of points visible to the observer in two distinct directions on the sky.  One can check all possible pairs of equal-radius circles and determine if the temperature patterns around any of them are more similar than would be likely in the covering space.

The ``Bayesian'' method relies on  comparing the pixel-pixel correlations in the observed CMB temperature map to those expected to be induced in manifolds with non-trivial topology (e.g., the paired circles), and to those expected in the covering space to determine which is the most likely underlying manifold \cite{Souradeep1998arx, Planck:2013okc, Planck:2015gmu}.

Both of these methods rely, for calculations of both the expected signal and its cosmic variance, on a statistical analysis of simulated realizations of the expected CMB sky in these topologically non-trivial manifolds.
Such simulations are produced by summing eigenmodes of the Laplacian with coefficients whose statistics are predicted by 
a theory for the generation of metric fluctuations in the early Universe and of their evolution.
This theory is generally taken to be inflation, 
and under certain conditions, which are usually taken to be applicable, results in the scale of any topology being inflated far beyond the current Hubble scale (so that functionally we can take our manifold to be the covering space of \Espace), effects of the initial conditions being ``inflated away'',  and the coefficients of the Laplacian eigenmodes of scalar fluctuations being statistically independent Gaussian random variables of zero mean with variance that depends only on the magnitude of the eigenvalue of the Laplacian and a power spectrum that is nearly scale-free.
We shall retain as an assumption this statistical characterization of the coefficients of the Laplacian eigenmodes, without specifically ascribing the fluctuations to inflation, and, clearly, without asserting that the topology scale is far beyond the Hubble scale.
The success of the usual Bayesian fits of covering-space inflationary $\Lambda$CDM model parameters to CMB temperature and $E$-mode polarization data, especially at $\ell\gtrsim 50$, 
and the failure so far to detect any ``primordial non-Gaussianity'' \cite{Planck:2013wtn,Planck2015pnongaus,Planck:2019kim,Meerburg2019baas, DAmico:2022gki}
suggest that this Gaussian hypothesis is at least a reasonable approximation for those $\ell$.
Nevertheless, we should certainly be aware of this limitation, and the associated assumption of an inflation-inspired nearly-scale-free power spectrum.

The program of searching for topology therefore relies explicitly on developing analytic expressions for Laplacian eigenmodes in the manifolds of interest --- or at least on identifying rapid algorithms for numerical calculation of the eigenmodes.
So far, this is not known to be possible in most $H^3$ or $S^3$ manifolds, but is straightforward, at least in principle, in all \Espace\ manifolds and select $S^3$ manifolds \cite{Inoue1999:cqg,Lehoucq2002:cqg, Weeks2006:cqg}.
Not surprisingly, the program to search for topology therefore included the derivation of such eigenmodes for the \Espace\ manifolds \cite{Riazuelo2004:prd}.
In this case, the eigenmodes are calculable finite linear combinations of certain covering-space eigenmodes --- Fourier modes.
In addition to calculating the eigenmodes themselves, one has need for many purposes of their expansion in a spherical coordinate basis, giving their contributions to individual spherical harmonics, and to the mode-mode or pixel-pixel correlation matrices on the sky for hypothetical observers.

Unfortunately, this program of calculating Laplacian eigenmodes on Euclidean three-manifolds was incompletely implemented.
While eigenmodes were calculated for representative examples of all Euclidean manifolds, what is needed is a comprehensive exhaustive description of all eigenmodes for all possible manifolds.
For each of the 17 non-trivial \Espace\  topologies, we must specify at least one real parameter to fully fix the manifold, 
specify the Laplacian eigenmodes, and
enable statistical predictions for cosmological observables.
Moreover, because the topological boundary conditions are not translation invariant,
observers in different locations in the same manifold can have different statistical expectations for observables in most topologies.
Because the boundary conditions are not invariant under rotations, the expectations for observables are not statistically isotropic in any manifold of non-trivial topology. Statistical anisotropy, and potentially statistical inhomogeneity, are key observational features of a topologically non-trivial universe.

The circles-in-the-sky searches for topology performed on WMAP \cite{deOliveira-Costa:2003utu,Cornish:2003db,ShapiroKey:2006hm,Mota:2010jb,Bielewicz:2010bh,Bielewicz:2011jz,Vaudrevange:2012da,Aurich:2013fwa} and {\it Planck} data \cite{Planck:2013kqc,Planck:2015gmu} (as well as a  general unpublished search analogous to \rcite{Vaudrevange:2012da}) constrain the shortest non-trivial closed loop through the Earth to have a length greater than $98.5\%$ of the diameter of the LSS\@.
The translation of this limit to limits on the parameters characterizing the manifolds \E{i} is underway; the results for the orientable manifolds \E{1}--\E{6}, \E{11}, \E{12}, and \E{16} are in \rcite{COMPACT:2022nsu}, and the non-orientable manifolds will be presented in an upcoming paper \cite{COMPACT:2023tbd}.

The Bayesian limits presented in \rcite{Planck:2013okc,Planck:2015gmu} are, typically, mildly more constraining than the circles-in-the-sky limit where both apply.
However, the Bayesian limits rely on the mode-mode correlation matrix, which depends on the geometric parameters of the manifold, as well as (for \E{2}--\E{17}) on the location of the observer.
The limits therefore only apply to the special values of the topological parameters and specific origin choices that were considered.
How they extend even to the neighborhoods of those special values is unclear without further study.

In this paper, we remedy the previously incomplete characterization of the orientable Euclidean manifolds (\E{1}-\E{6}, \E{11}, \E{12}, \E{16}, and \E{18}), providing general expressions for their eigenmodes, allowing for general parametrizations of the geometry of the space and for arbitrary observer location.
We also present formulae for the mode-mode correlation matrices on the sphere and other useful quantities.
In so doing, we will find that we are implementing known cases that were however omitted from previous attempts to model all possible manifolds.

These mathematical results will be central to future work  modeling the consequences of non-trivial topology, and searching exhaustively for the topology of the Universe.
In the meantime, they enable us to forecast the detectability of topology in the CMB and eventually in other cosmological observables.
Previous works \cite{Fabre:2013wia,Planck:2013okc,Planck:2015gmu} have employed the Kullback-Leibler (KL) divergence \cite{kullback1951, kullback1959information} to quantify the information available in the CMB to distinguish between a topologically non-trivial universe and a topologically trivial one.
We show here that the inclusion of all necessary topological parameters does not qualitatively change the conclusion that non-trivial topology should be detectable in the CMB if the length $L_{\mathrm{topology}}$ of the shortest closed geodesic through us is less than the diameter of the LSS, $d_\mathrm{LSS}$, but probably becomes undetectable if $L_{\mathrm{topology}}\gtrsim 1.2 d_\mathrm{LSS}$.  We argue that this should be expected to change once high-quality three-dimensional data (i.e., tomographic intensity mapping or deep-enough galaxy surveys) become available.
On the other hand, if the large-angle anomalies in the CMB are caused by cosmic topology (one of the very few promising sources of large-angle violation of statistical isotropy), then this result reassures us that $L_{\mathrm{topology}}\lesssim1.2 d_\mathrm{LSS}$ and there is a reasonable prospect of discovering cosmic topology, and possibly identifying the specific spatial manifold, from the CMB alone.

In \cref{secn:topologiesmanifolds-general} we discuss the notation and review some general properties of the topologies and manifolds of \Espace. \cref{secn:topologiesmanifolds} discusses the key properties of the orientable Euclidean manifolds.
In \cref{secn:eigenmodes}, we present the eigenmodes of the scalar Laplacian along with the Fourier and spherical-harmonic correlation matrices. \cref{secn:numerical_results} contains our numerical results and a number of representative examples of spherical-harmonic covariance matrices for the compact orientable manifolds with selected values of their parameters, as well as the corresponding KL divergence and a useful new off-diagonal signal-to-noise statistic.

The GitHub repository associated with this study is publicly available at \url{https://github.com/CompactCollaboration}. Codes will be deposited there as publicly usable versions become available.

\section{Topologies and manifolds of \Espace: general considerations}
\label{secn:topologiesmanifolds-general}
 
The isometry group of Euclidean three-space \Espace, denoted by $E(3)$, includes arbitrary rotations and reflections (i.e., elements of the orthogonal group in three dimensions $O(3)$), arbitrary translations, and all products of these.
A group element of $E(3)$ is freely acting on \Espace\ if it takes no point of \Espace\ to itself.
These are comprised of translations, rotations about arbitrary axes followed by translations with a component parallel to that axis (``corkscrew motions''), reflections across planes followed by translations with components parallel to the plane (``glide reflections''), and certain products of these.
The non-trivial \Espace\ topologies \E{i} are formed by modding out $E(3)$ by a discrete subgroup $\Gamma^{\E{i}}$ of freely acting elements, i.e., $E(3)\to E(3)/\Gamma^{\E{i}}$ (see, e.g., \rcite{hempel20043, hirsch2012differential}).

There are 18 distinct possibilities for $\Gamma^{\E{i}}$ (including the trivial group), leading to the 18 distinct topologies \E{i} for $i\in\{1,\ldots,18\}$ \cite{Lachieze-Rey1995pr, Luminet1999, hitchman2009geometry, Riazuelo2004:prd}.
These $\Gamma^{\E{i}}$ are  characterized in whole or in part by the specific selection of $O(3)$ elements that each involves, but in all cases, when constructing actions of $\Gamma^{\E{i}}$, there are translations that must be characterized by up to 6 real parameters.
If space is a Euclidean manifold, then these degrees of freedom are physical, and a general description of the manifold must include them.\footnote{
    In addition, to fully describe observational properties, one must typically specify the position and orientation of the observer, which introduces up to 6 more real parameters.
}

The situation can be illustrated in the most familiar case --- a simple three-torus, which is referred to as \E{1}.
Its symmetry group $\Gamma^{\E{1}}$ is generated by three pure translations (i.e., the only element of $O(3)$ involved is the identity), which we will represent as $g^{\E{1}}_i$, a translation by $\vec{T}^{\E{1}}_i$, for $i\in\{1,2,3\}$:
\begin{equation}
    g^{\E{1}}_i: \vec{x} \to \vec{x} + \vec{T}^{\E{1}}_i.
\end{equation}

\begin{figure}[t]
     \centering
     \begin{subfigure}[b]{0.5\textwidth}
         \centering
         \includegraphics[scale=1.0]{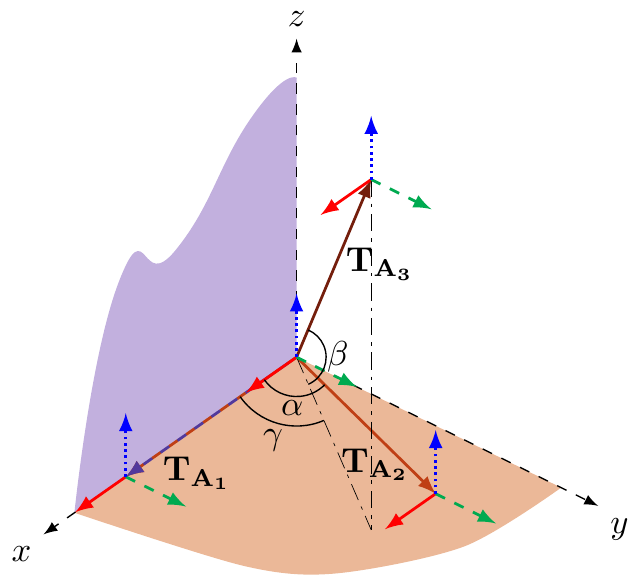}
         \caption{\(\E{1}\)}
         \label{fig:E1}
     \end{subfigure}%
     \begin{subfigure}[b]{0.5\textwidth}
         \centering
         \includegraphics[scale=1.0]{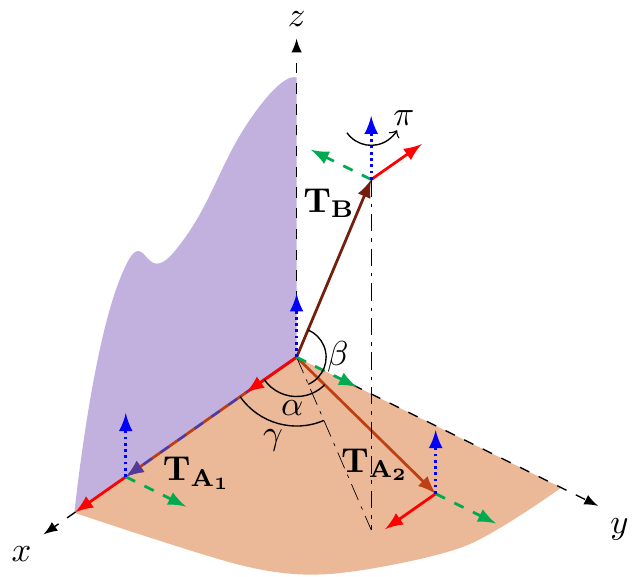}
         \caption{\(\E{2}\)}
         \label{fig:E2}
     \end{subfigure}
     \newline
     \begin{subfigure}[b]{0.5\textwidth}
         \centering
         \includegraphics[scale=1.0]{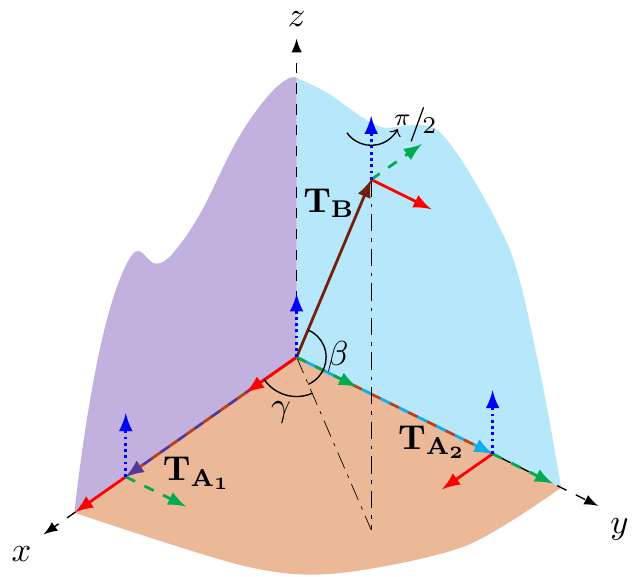}
         \caption{\(\E{3}\)}
         \label{fig:E3}
     \end{subfigure}%
     \begin{subfigure}[b]{0.5\textwidth}
         \centering
         \includegraphics[scale=1.0]{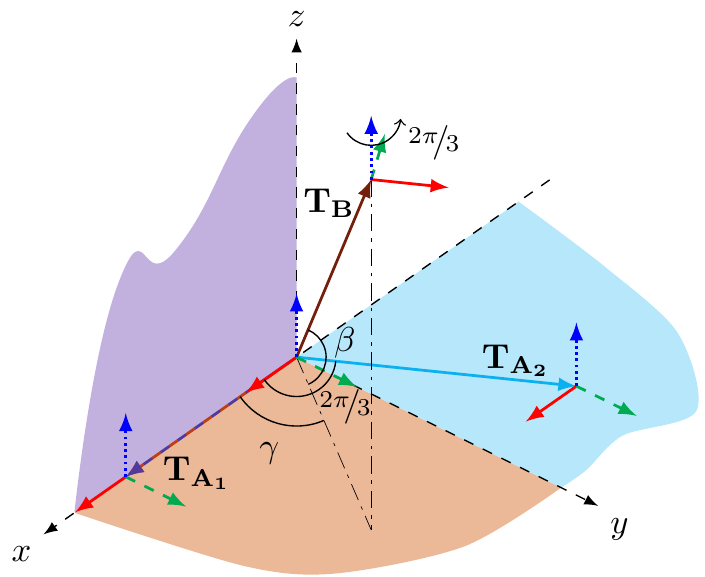}
         \caption{\(\E{4}\)}
         \label{fig:E4}
     \end{subfigure}
     \newline
     \begin{subfigure}[b]{0.5\textwidth}
         \centering
         \includegraphics[scale=1.0]{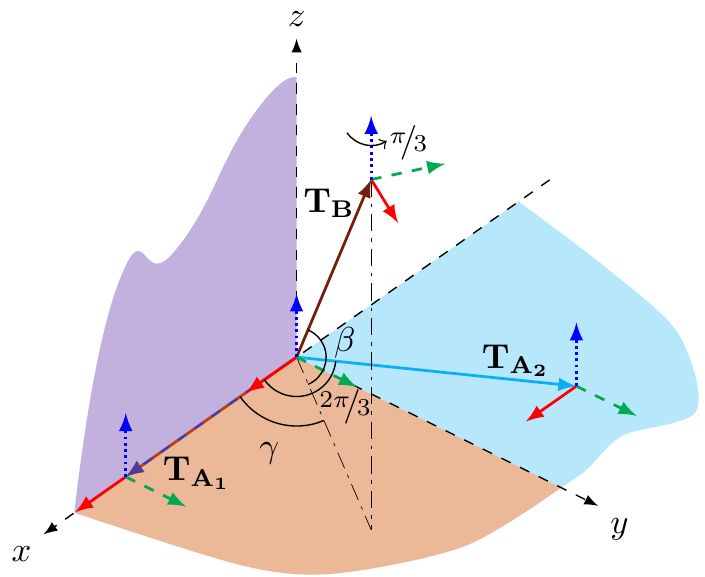}
         \caption{\(\E{5}\)}
         \label{fig:E5}
     \end{subfigure}%
     \begin{subfigure}[b]{0.5\textwidth}
         \centering
         \includegraphics[scale=1.0]{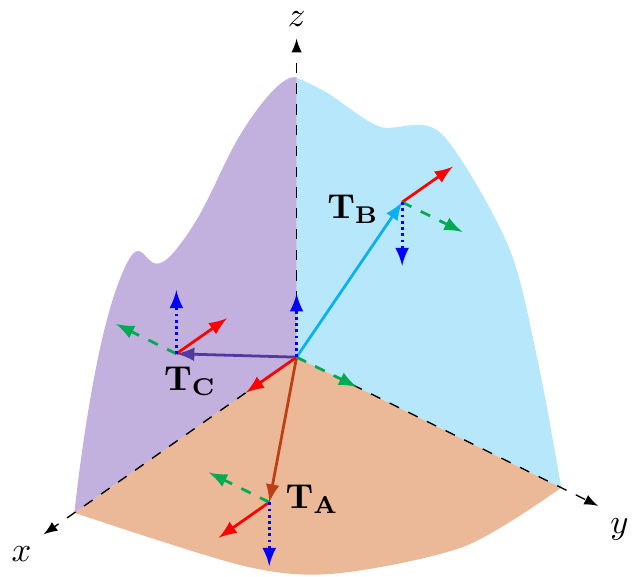}
         \caption{\(\E{6}\)}
         \label{fig:E6}
     \end{subfigure}
    \caption{Diagrams showing the actions of the generators for the topologies \E{1}-\E{6}. In each subdiagram, an observer at the origin is represented by an orthogonal triad $\hat{x}$, $\hat{y}$, $\hat{z}$ shown as  short red, green, and blue arrows, respectively. Translation vectors  for each of the three generators are rooted at the origin, and may be labelled $T_{A_{i}}$, $T_{A}$, $T_{B}$, and $T_{C}$, depending on the details of the topology. The angles $\alpha$, $\beta$, and $\gamma$ mark their orientation.  At the head of each of those translation vectors is another red-green-blue triad representing one of the observer's topological clones, with the amount indicated by which they have been rotated compared to the observer at the origin.
    The colored coordinate planes are provided only as visual aids.
    }
    \label{ologyplots}
\end{figure}

\begin{figure}[t]
     \centering
     \begin{subfigure}[b]{0.5\textwidth}
         \centering
         \includegraphics[scale=1.0]{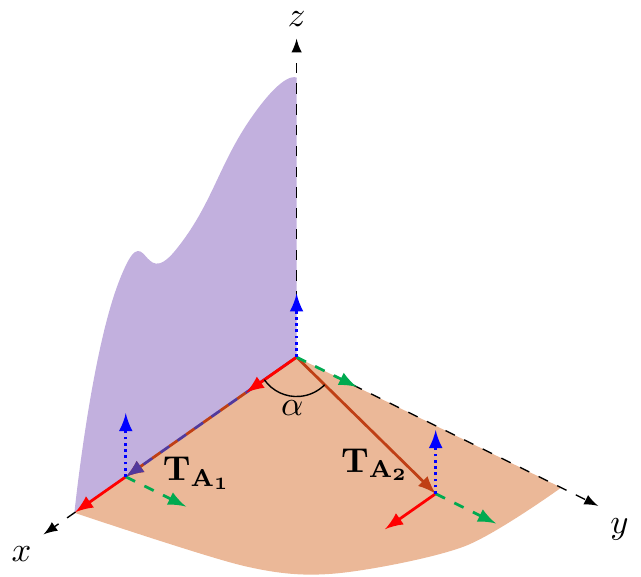}
         \caption{\(\E{11}\)}
         \label{fig:E11}
     \end{subfigure}%
     \begin{subfigure}[b]{0.5\textwidth}
         \centering
         \includegraphics[scale=1.0]{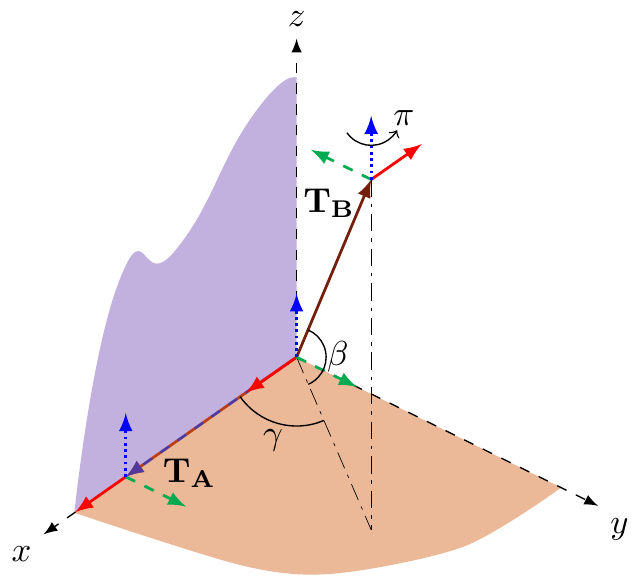}
         \caption{\(\E{12}\)}
         \label{fig:E12}
     \end{subfigure}
     \newline
     \begin{subfigure}[b]{0.5\textwidth}
         \centering
         \includegraphics[scale=1.0]{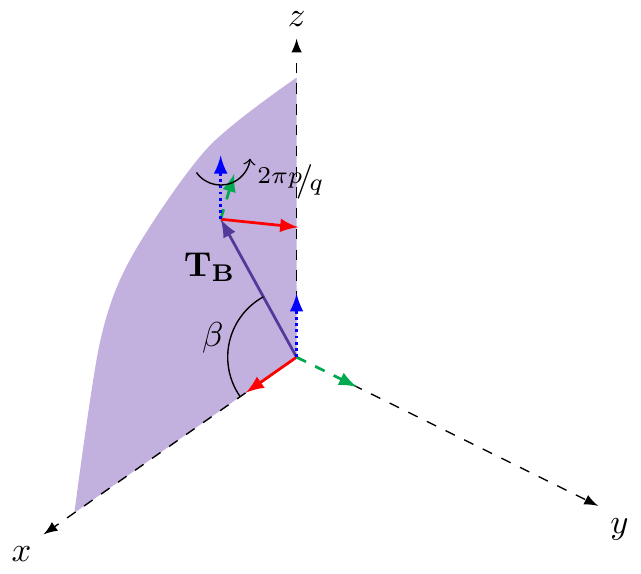}
         \caption{\(\E{16}\)}
         \label{fig:E16}
     \end{subfigure}%
     \begin{subfigure}[b]{0.5\textwidth}
         \centering
         \includegraphics[scale=1.0]{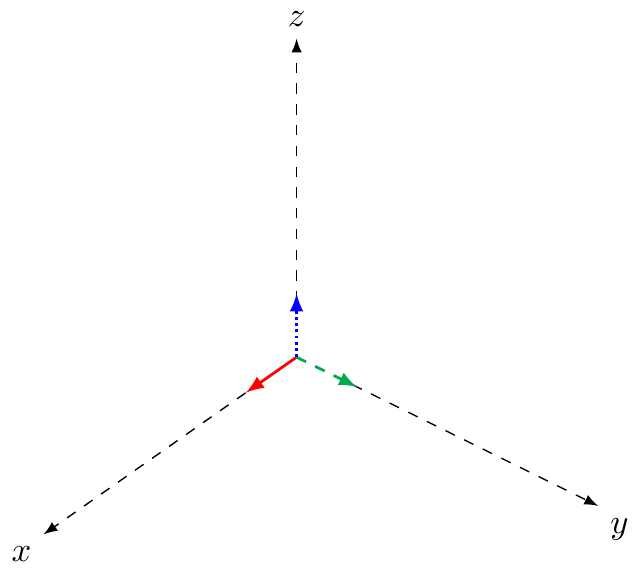}
         \caption{\(\E{18}\)}
         \label{fig:E18}
     \end{subfigure}
    \caption{Diagrams as in \cref{ologyplots}, but showing the actions of the generators for the topologies \E{11}, \E{12}, \E{16}, and \E{18}. 
    }
    \label{ologyplotsnoncompact}
\end{figure}

\cref{ologyplots,ologyplotsnoncompact} illustrate the actions of the generators for \E{1} and some of the other Euclidean topologies. The simplest, special case is the cubic three-torus where the translations are orthogonal and of equal length, e.g., $\vec{T}^{\E{1}}_1 = L \unitvec{e}_1$, 
    $\vec{T}^{\E{1}}_2 = L \unitvec{e}_2$, 
    $\vec{T}^{\E{1}}_3 = L \unitvec{e}_3$,
with $\{\unitvec{e}_1,\unitvec{e}_2,\unitvec{e}_3\}$ a set of three orthogonal unit vectors.
A general element of $\Gamma^{\E{1}}$ is a product of integer powers of these $g^{\E{1}}_i$, i.e., it is a translation by an integer linear combination of these three translations,
\begin{equation}
  T^{\E{1}}_{\vec{n}}=n_1\vec{T}^{\E{1}}_1 + n_2\vec{T}^{\E{1}}_2 + n_3\vec{T}^{\E{1}}_3.
\end{equation}

An observer in \E{1} will perceive themselves to have a lattice of ``clones'' displaced from themselves by these vectors $T^{\E{1}}_{\vec{n}}$, for all sets of integers $\{n_1,n_2,n_3\}$.
They would also perceive any object they see around them to also have clones displaced from its closest instance by these same vectors.
This lattice of clones is a real physical phenomenon and is observable if $L$ is small enough (see, e.g., \rcite{Sokolov:1974,Fang:1983,Fagundes:1987,Lehoucq:1996qe,Roukema:1996cu,Weatherley:2003,Fujii:2011ga,Fujii:2013xsa}).
The observer might choose to interpret such a situation as living in a finite volume, for example, the cube with corners at
\begin{align}
    \label{eqn:cubecorners}
    &\{\vec{0}, 
    \vec{T}^{\E{1}}_1, 
    \vec{T}^{\E{1}}_2,
    \vec{T}^{\E{1}}_1+\vec{T}^{\E{1}}_2, \\ \nonumber
    &\quad \vec{T}^{\E{1}}_3,
    \vec{T}^{\E{1}}_1+\vec{T}^{\E{1}}_3,
    \vec{T}^{\E{1}}_2+\vec{T}^{\E{1}}_3, 
    \vec{T}^{\E{1}}_1+\vec{T}^{\E{1}}_2+\vec{T}^{\E{1}}_3 \}
\end{align}
and with all opposite faces identified.
Alternately they might consider themselves to be living in the covering space of \Espace\ (i.e., \E{18}) with any pair of points differing by a $g_{\vec{n}}$ identified with one another, i.e., space is simply tiled by this cube.
Further, they might call this cube their ``fundamental domain'' (FD) or unit cell \cite{hitchman2009geometry}.
These are equivalent descriptions of the same reality, and we often use them interchangeably in describing topologically non-trivial universes.

\newgeometry{textwidth=0.8\paperwidth, textheight=0.85\paperheight}
\restoregeometry

Although in this simple case, we associate the cube with this \E{1}, the shape of the FD is not itself an observable,
and is certainly not a physical property of the manifold.
In two dimensions this was perhaps most famously made evident by the many interesting FD shapes represented by the Dutch artist M.C.~Escher --- birds, fish, \textit{etc} \cite{arthive}.
What is physical is the set of group elements in $\Gamma^{\E{1}}$, or, equivalently, the relative locations (and orientations) of the ``clones'' of a given point in space --- i.e., its images under elements of $\Gamma^{\E{1}}$.
Each location in a manifold will therefore have a Dirichlet domain: the set of all points in the manifold that are closer to them than to any of their clones.

For example, in the specific case given above with $\vec{T}^{\E{1}}_i=L\unitvec{e}_i$, 
an observer located at  $\frac{1}{2}(\vec{T}^{\E{1}}_1+\vec{T}^{\E{1}}_2+\vec{T}^{\E{1}}_3)$ will have the cube with corners given by \eqref{eqn:cubecorners} as their Dirichlet domain.
Different observers will have different Dirichlet domains centered on themselves.
Generically, the shape of the Dirichlet domain will also change with the observer location.\footnote{
    \E{1} and the other homogeneous topologies, \E{11}, \slabh, and \E{18}, are special cases since an observer located at any point in the manifold has the same lattice of clones, and therefore all observers have the same shape Dirichlet domain.}

Connected with this ambiguity of the shape of the FD, we
note that there are many distinct choices of $\vec{T}^{\E{1}}_i$ that lead to the exact same lattice of clones.
For example, we could replace $\vec{T}^{\E{1}}_3$ by $\vec{T}^{\E{1}\prime}_3\equiv\vec{T}^{\E{1}}_3+\vec{T}^{\E{1}}_1$.
This results in the precisely identical set of clones, even though $\vec{T}^{\E{1}}_1$, $\vec{T}^{\E{1}}_2$, and $\vec{T}^{\E{1}\prime}_3$
are not the same length and not orthogonal. The FD defined as in \eqref{eqn:cubecorners} is not a cube, nor even a rectangular prism, but a parallelepiped;
it is not the Dirichlet domain of any observer.

Indeed, $\vec{T}^{\E{1}}_i$ (for $i\in\{1,2,3\}$) need not be orthogonal, nor of equal length ---
they can be any three linearly independent vectors, with \eqref{eqn:cubecorners} forming a general parallelepiped,
\begin{equation} 
    \vec{T}^{\E{1}}_{1} = \begin{pmatrix} L_{1x} \\ L_{1y} \\ L_{1z} \end{pmatrix} , \quad
    \vec{T}^{\E{1}}_{2} = \begin{pmatrix} L_{2x} \\ L_{2y} \\ L_{2z} \end{pmatrix} , \quad
    \vec{T}^{\E{1}}_{3} = \begin{pmatrix} L_{3x} \\ L_{3y} \\ L_{3z} \end{pmatrix} .
\end{equation}
However, we would prefer to choose $\vec{T}^{\E{1}}_i$ so that \eqref{eqn:cubecorners} is the Dirichlet domain of some observer.
This will constrain the relative values of $L_{iw}$ with $w\in\{x,y,z\}$ in ways that we discuss in detail below for each of the orientable topologies in \cref{secn:topologiesmanifolds}.

Three of these nine degrees of freedom $L_{iw}$ are degenerate with the three Euler angles describing the orientation of the coordinate system.
While a search for topology will need to account for the orientation of the observer's coordinates, it is convenient, when cataloging manifolds, and when simulating the possible signals, to remove these degrees of freedom.
Generically, we can order the three vectors by their length (from shortest to longest), and then choose the shortest of the three $\vec{T}^{\E{1}}_{1}$ to point in the $x$-direction, and $\vec{T}^{\E{1}}_{2}$ to lie in the $xy$-plane, so that $L_{1y}=L_{1z}=L_{2z}=0$, i.e.,
\begin{equation} 
    \label{eqn:E1translations}
    \vec{T}^{\E{1}}_{1} = \begin{pmatrix} L_{1x} \\ 0 \\ 0 \end{pmatrix}, \quad
    \vec{T}^{\E{1}}_{2} = \begin{pmatrix} L_{2x} \\ L_{2y} \\ 0 \end{pmatrix}, \quad
    \vec{T}^{\E{1}}_{3} = \begin{pmatrix} L_{3x} \\ L_{3y} \\ L_{3z} \end{pmatrix}.
\end{equation}
We may alternatively choose to write them as
\begin{equation} 
    \label{eqn:E1translations_intro}
    \vec{T}^{\E{1}}_{1} = L_1 \begin{pmatrix} 1 \\ 0 \\ 0 \end{pmatrix}, \quad
    \vec{T}^{\E{1}}_{2} = L_2 \begin{pmatrix} \cos\alpha \\ \sin\alpha \\ 0 \end{pmatrix}, \quad
    \vec{T}^{\E{1}}_{3} = L_3 \begin{pmatrix} \cos\beta\cos\gamma \\ \cos\beta\sin\gamma \\ \sin\beta \end{pmatrix}.
\end{equation}
Both of these parametrizations can be useful.

Thus the group $\Gamma^{\E{1}}$ associated with \E{1} has 6 real parameters and all allowed choices of these parameters result in the same \E{1} topology, but generically they result in different lattices of clones and so are physically distinct (and distinguishable) manifolds.
However, there are equivalence classes, each with a countably infinite number of members, as we can replace these three vectors by any three linearly independent integer linear combinations of these three vectors without changing the lattice of clones.
Thus if we are trying to characterize the allowed possibilities for $\Gamma$ without double counting we must take care in choosing the ranges of the parameters.
In this case of \E{1} with the alternative parametrization \eqref{eqn:E1translations_intro}, we could require that $0<L_1\leq L_2\leq L_3$.
However, that is not sufficient to prevent double-counting the physical parameters of the manifold.
We need to insist that $\vec{T}^{\E{1}}_{i}$ cannot be shortened by adding integer linear combinations of 
$\vec{T}^{\E{1}}_{j}$ and $\vec{T}^{\E{1}}_{k}$
(with $i,j,k$ distinct elements of $\{1,2,3\}$).
Details are provided below in \cref{secn:topologyE1}.

The lattice of clones of any given point is not rotationally invariant --- rotational invariance, and thus statistical isotropy, is not an expected property of the Universe for an observer in \E{1} \cite{thurston2014three}.
The orientation of the observer relative to the lattice of clones is an observable.
This is important to keep in mind when making use of the results of this paper --- we will choose a particular orientation of the Cartesian coordinate axes (e.g., as reflected in \eqref{eqn:E1translations_intro}), but there is no reason for that to coincide with the orientation of the observer's coordinate system.
The Euler angles of the rotation between the manifold's coordinate system and the observer's coordinate system must be varied.

In the case of \E{1}, the lattice of an observer's clones does not depend on the location of the observer.
This is not generally true.
Generically for all \E{i} and with respect to an arbitrary location $\vec{x}^{\E{i}}_0$, each generator $g_{a_j}$ of $\Gamma^{\E{i}}$ acts on a point $\vec{x}$ in the manifold as
\begin{equation}
    \label{eqn:actionofgenerator}
    g_{a_j}^{\E{i}}: \vec{x} \to \mat{M}^{\E{i}}_a (\vec{x}-\vec{x}_0^{\E{i}}) + \vec{T}^{\E{i}}_{a_j} + \vec{x}^{\E{i}}_0 ,
\end{equation}
where $\mat{M}^{\E{i}}_a \in O(3)$ and $\vec{T}^{\E{i}}_{a_j}$ is a translation vector appropriate for the given topology \E{i}.
We use the index $a$ to distinguish among the up to three distinct $\mat{M}^{\E{i}}_a \in O(3)$ and the index $j$ to label the distinct vectors $\vec{T}^{\E{i}}_{a_j}$ for a given $\mat{M}^{\E{i}}_a$.
The origin $\vec{x}^{\E{i}}_0$ is the position (i.e., relative to some arbitrary coordinate origin) of a point on the axis about which $\mat{M}^{\E{i}}_a$ rotates or on the plane across which it reflects.
Since the $\mat{M}^{\E{i}}_a$ are such that the axes about which they rotate or the normals to the planes across which they reflect are orthogonal \cite{Thurston1982ThreeDM}, we can choose a single $\vec{x}_0^{\E{i}}$ for all the generators. (This is why $\vec{x}_0^{\E{i}}$ needs neither $a$ nor $j$ labels.)

All elements of $\Gamma^{\E{i}}$ can be obtained by successive actions of these generators and their inverses.
For \E{1}--\E{10} three generators are required (one can choose to include extra redundant generators, though we mostly refrain from doing so in this work).
These are the compact Euclidean topologies.
For \E{11}--\E{15} two generators are required; for \E{16} and \E{17} one generator is required; \E{18} is the covering space of the Euclidean geometry  \Espace, for which no generators are required.

In the case of \E{1} described above, $\mat{M}^{\E{i}}_a$ was the identity for all three generators.
More generally the generators can be chosen so that each $\mat{M}^{\E{i}}_a$ is one of:  
  the identity,  a rotation about a coordinate axis, or the reflection of a single coordinate.
$\vec{T}^{\E{i}}_{a_j}$ can never be $\vec{0}$, if it were then $g_{a_j}^{\E{i}}$ would not be freely acting.
As described above,
the three types of generators are referred to as translations ($\mat{M}^{\E{i}}_a=\identity$), corkscrew motions ($\mat{M}^{\E{i}}_a$ a rotation), and glide reflections ($\mat{M}^{\E{i}}_a$ a reflection).

In the case where one or more of the $\mat{M}^{\E{i}}_a$ is not the identity, the manifold is not homogeneous, i.e., the lattice of clones of an observer depends on the location of the invariant axis/plane of $\mat{M}^{\E{i}}_a$ relative to the observer, as encoded in $\vec{x}_0^{\E{i}}$.
One way to understand this inhomogeneity is that a change of choice of origin changes $\vec{x}_0^{\E{i}}$, which can change $\vec{T}^{\E{i}}_{a_j}$.

For example, consider a shift of origin by $-\vec{v}$, which takes  $\vec{x}^{\E{i}}_0\to\vec{x}^{\prime\E{i}}_0 \equiv\vec{x}^{\E{i}}_0 + \vec{v}$.
In this case, we would rewrite the generator \eqref{eqn:actionofgenerator} as

\begin{equation}
    \label{eqn:actionofgeneratoraxisshift}
    g_{a_j}^{\E{i}}: \vec{x} \to 
        \mat{M}^{\E{i}}_a (\vec{x}-\vec{x}_0^{\prime\E{i}}) + \left(\vec{T}^{\E{i}}_{a_j} - (\identity - \mat{M}^{\E{i}}_a)\vec{v}\right) + \vec{x}_0^{\prime\E{i}}, 
\end{equation}
so that
\begin{equation}
    \vec{T}^{\E{i}}_{a_j} \to \vec{T}^{\E{i}}_{a_j} - (\identity - \mat{M}^{\E{i}}_a)\vec{v}.
\end{equation}
This is precisely the statement that observers at different locations have different clone lattices, 
or equivalently that the shape of the Dirichlet domain depends on the position of the observer (see, e.g., \rcite{COMPACT:2022nsu} for a wider discussion).

For certain purposes, it might be useful to use the shift in origin  to ``simplify'' the set of $\vec{T}^{\E{i}}_{a_j}$, for example, to set certain components to zero, or to equate certain components to one another.
However, $\identity - \mat{M}^{\E{i}}_a$ always has some eigenvectors with eigenvalue $0$, and these components of $\vec{T}^{\E{i}}_{a_j}$ are unaffected by such shifts:
\begin{itemize}
    \item If $\mat{M}^{\E{i}}_a$ is a proper rotation about an axis, 
    then the component of $\vec{T}^{\E{i}}_{a_j}$ normal to the plane of rotation cannot be altered by shifting $\vec{x}^{\E{i}}_0$, but the components in the plane of rotation can be adjusted.
    For example, if $\mat{M}^{\E{i}}_a=\mat{R}_{\unitvec{z}}(\theta)$, then we can shift the origin such that $\vec{T}^{\E{i}}_{a_j}\|\unitvec{z}$, absorbing the other components of $\vec{T}^{\E{i}}_{a_j}$.
    In other words, we can shift the origin of the coordinate system so that it lies on the axis of rotation.
    \item If $\mat{M}^{\E{i}}_a$ is a reflection across a plane, then the components of $\vec{T}^{\E{i}}_{a_j}$ in the plane of reflection cannot be altered by shifting $\vec{x}^{\E{i}}_0$, but the components normal to the plane of reflection can be adjusted.
    For example, if $\mat{M}^{\E{i}}_a=\diag(1,-1,1)$, then $\identity-\mat{M}^{\E{i}}_a=\diag(0,2,0)$ and we can absorb only the $y$-component of $\vec{T}^{\E{i}}_{a_j}$, leaving $\vec{T}^{\E{i}}_{a_j}$ to be a general vector in the $xz$-plane.
    In other words, we can shift  the origin of the coordinate system to lie in the plane of reflection.
    \item If $\mat{M}^{\E{i}}_a=\identity$, then $\identity - \mat{M}^{\E{i}}_a=0$ and none of the components of $\vec{T}^{\E{i}}_{a_j}$ can be absorbed: $\vec{T}^{\E{i}}_{a_j}$ remains an arbitrary vector.
\end{itemize}
If more than one of the $\mat{M}^{\E{i}}_a$ is not the identity, then their axes/planes must be orthogonal to one another (see, for example, \rcite{Thurston1982ThreeDM}); since there are never more than three distinct $\mat{M}^{\E{i}}_a$, the associated axes/planes can always be taken to be parallel to coordinate axes/planes.

This orthogonality will also simplify the choices of which $\vec{T}^{\E{i}}_{a_j}$ to modify using the freedom to choose the coordinate origin.
A shift in origin, with the resulting shift in the positions of the axes of rotation and planes of reflection associated with the generators $g^{\E{i}}_{a_j}$, changes the $\vec{T}^{\E{i}}_{a_j}$, and results in a different lattice of clones for an observer located at the new origin versus the old one.
It also results in a different Dirichlet domain.
We might have been tempted to interpret the unit cell of the clone lattice or the observer's Dirichlet domain as ``the shape of the Universe''.
This would then lead to the conclusion that the shape of the Universe depends on the choice of origin.
However ``the shape of the Universe'' is ambiguous.
What is physical is the lattice of clones (and their orientations) seen by an observer for themselves (and for any other objects in the manifold); except for \E{1}, \E{11}, $\slabh$, and \E{18}, this lattice depends on where the observer is located relative to the invariant axes/planes of the $\mat{M}^{\E{i}}_a$.

From a mathematical point of view, we could use our freedom to choose the origin to eliminate or relate as many as three of the components of $\vec{T}^{\E{i}}_{a_j}$.
While this ability to simplify the $\vec{T}^{\E{i}}_{a_j}$ may prove useful for enumerating manifolds or for simulating cosmological observables,
for an observer, the most sensible choice of origin is  likely to be their own position, which may be very far from the point one would choose to yield a simplified set of generators.
We therefore preserve both $\vec{x}^{\E{i}}_0$ and $\vec{T}^{\E{i}}_{a_j}$ in our expressions for eigenmodes, and comment appropriately.

One important property of manifolds is their orientability or non-orientability.
Loosely, a manifold is orientable if a right-handed triad remains right-handed when carried around all possible closed loops; it is non-orientable if there are closed loops for which it becomes left-handed when carried around them.
The properties of Euclidean manifolds of various topologies are discussed below and summarized in \cref{tab:properties}.

\section{Properties of orientable Euclidean topologies}
\label{secn:topologiesmanifolds}

\begin{table}
    \newcommand{\highlightcolor}{yellow}
    \begin{tabular}{clcccc} \hline
        \textbf{Symbol} & \textbf{Name} &  \textbf{Compact} & \textbf{Orientable} & \textbf{Homogeneous} & \textbf{Isotropic} \\ 
        & & \textbf{Dimensions} & & & \\ \hline
         \rowcolor{\highlightcolor} \E{1} & 3-torus & 3 & Yes & Yes & No \\
         \rowcolor{\highlightcolor} \E{2} & Half-turn& 3 & Yes & No & No \\
         \rowcolor{\highlightcolor} \E{3} & Quarter-turn & 3 & Yes & No & No \\
         \rowcolor{\highlightcolor} \E{4} & Third-turn & 3 & Yes & No & No \\
         \rowcolor{\highlightcolor} \E{5} & Sixth-turn & 3 & Yes & No & No \\
         \rowcolor{\highlightcolor} \E{6} & Hantzsche-Wendt & 3 & Yes & No & No \\ \hline
         \E{7} & Klein space & 3 & No & No & No \\
         \E{8} & --- (horizontal flip) & 3 & No & No & No \\
         \E{9} & --- (vertical flip) & 3 & No & No & No \\
         \E{10} & --- (half-turn) & 3 & No & No & No \\ \hline
         \rowcolor{\highlightcolor} \E{11} & Chimney space & 2 & Yes & Yes & No \\
         \rowcolor{\highlightcolor} \E{12} & --- (half-turn) & 2 & Yes & No & No \\ \hline
         \E{13} & --- (vertical flip) & 2 & No & No & No \\
         \E{14} & --- (horizontal flip) & 2 & No & No & No \\
         \E{15} & --- (half-turn + flip) & 2 & No & No & No \\ \hline
         \rowcolor{\highlightcolor} \slabh & Slab (unrotated) & 1 & Yes & Yes & No \\
         \rowcolor{\highlightcolor} \slabi & Slab (rotated) & 1 & Yes & No & No \\ \hline
         \E{17} & Slab (flip) & 1 & No & No & No \\ \hline
         \rowcolor{\highlightcolor} \E{18} & Covering space & 0 & Yes & Yes & Yes \\ \hline
    \end{tabular}
    \caption{Properties of the 18 three-dimensional Euclidean topologies.
    The orientable topologies, the focus of this work, are highlighted. Manifolds are labelled as homogeneous if the statistics of observables are independent of observer location, and isotropic if they are independent of observer orientation.
    }
    \label{tab:properties}
\end{table}

The 18 Euclidean topologies can be classified by their number of compact dimensions and whether they are orientable, homogeneous, and/or isotropic.
The topologies, with their names, symbols, and properties, are listed in \cref{tab:properties}. The balance of this paper concerns only topologies with orientable manifolds: the fully compact (\E{1}--\E{6}, illustrated in \cref{ologyplots}), those with compact-cross-sectional area (\E{11}--\E{12}, see \cref{ologyplotsnoncompact}), those that are compact in one dimension and orientable (\E{16}, see \cref{ologyplotsnoncompact}),  and the covering space \E{18}, as highlighted in Table~\ref{tab:properties}.
The other topologies will be addressed in upcoming papers.

In this section, we summarize the important features of each of these topologies organized as follows.
First, we list its important properties as summarized in Table~\ref{tab:properties}.
Next we provide an action of the generators $g_{a_j}^{\E{i}}$ of its associated discrete subgroup  $\Gamma^{\E{i}}$ of $E(3)$.
In other words, we specify the matrices $\mat{M}^{\E{i}}_a$ and the associated non-zero translation vectors $\vec{T}^{\E{i}}_{a_j}$ that characterize a manifold of each topology.
For orientable manifolds, we can always choose the generators such that all the matrices $\mat{M}^{\E{i}}_a$ are either the identity, so that $g^{\E{i}}_{a_j}$ is a pure translation, or a rotation about an axis parallel to a coordinate axis through some origin $\vec{x}^{\E{i}}_0$, so that $g^{\E{i}}_{a_j}$ is a ``corkscrew motion'' (e.g., see \rcite{Thurston1982ThreeDM}).
With this restriction to orientable manifolds, none of the generators is a glide reflection, so $\mat{M}^{\E{i}}_a\in SO(3)$.

Within each set of topologies with the same number of compact dimensions, there is exactly one for which all of its generators, and thus all the elements of $\Gamma^{\E{i}}$, are pure translations.
These are the 3-torus, \E{1}, with three compact dimensions; the chimney space, \E{11}, with two compact dimensions; the unrotated slab space, \slabh, with one compact dimension; and, trivially, the covering space (i.e., the full Euclidean space), \E{18}, with no compact dimensions.
The other topologies in each set can be viewed as ``roots'' of \E{1}, \E{11}, or \slabh.

As noted above, for each group $\Gamma^{\E{i}}$, 
either each $SO(3)$ matrix $\mat{M}^{\E{i}}_a$ of its group elements, and in particular of its generators, 
is itself the identity, or 
there is a positive integer $N$ such that $(\mat{M}^{\E{i}}_a)^N=\identity$.
For \slabi\ this is only true if the associated rotation angle is a rational multiple of $2\pi$.
Thus the generator applied $N$ times is a pure translation, and
we are always able to construct a subgroup of $\Gamma^{\E{i}}$ of the same rank composed of such pure translations.
For \E{1}--\E{6} this subgroup is rank 3, for \E{11} and \E{12} it is rank 2, and for \slabi\ it is rank 1.
In other words, those integer-powers of generators generate an associated homogeneous manifold:
    for \E{2}--\E{6} we call this \E{1} the ``associated \E{1}'' of this manifold; 
    for \E{12} it is called the associated \E{11};
    \slabi\ has an associated \slabh\ only for rotation angles that are rational multiples of $2\pi$.

We provide the associated \E{1}, \E{11}, or \slabh\ of each manifold.
It has the important property that it is homogeneous, i.e., every observer agrees on it.
This can prove useful.
For example, as detailed in \rcite{COMPACT:2022nsu} for the compact manifolds, we can construct a unit cell from these translations.
In this case, when we center that cell on the origin and construct the $3\times3\times3$ block of neighboring unit cells, then the nearest clone to any point will always be located within that neighborhood.
The associated \E{1} is also used to calculate the volume of the compact manifolds, which are provided next.

As remarked above, the action of the generators is affected by the choice of orientation and origin of the coordinate system in conjunction with the orientation and origin of an observer's coordinate system.
The choices made are contained in the description of each manifold and fall into two broad categories:
\begin{itemize}
    \item The orientation of the coordinate system used in the action of the generators allows for the simplification of the translation vectors $\vec{T}^{\E{i}}_{a_j}$ and/or to fix the ratios of some of their parameters. In particular, we will first use the rotational freedom to fix the axis associated with any corkscrew motions to be along a coordinate axis.
    Next, when additional rotational freedom remains, we use it to fix one of the components of a translation vector.
    \item Shifting the origin of the coordinate system, $\vec{x}_0$, allows us to freely adjust the two components of $\vec{T}^{\E{i}}_{a_j}$ perpendicular to it.
\end{itemize}
We describe how these are implemented and components could be adjusted by the freedom to shift the origin.

Care must be taken when varying the parameters in generators to ensure that choices are not redundant, i.e., that choices of parameters that appear different actually generate the same lattice of clones.
A list of conditions is provided to allow one to  vary the parameters over all allowed values without ``double-counting''.

As noted above, the fundamental domain is commonly used as a tool to describe homogeneous spaces, but it is observer-dependent in inhomogeneous ones.
Due to this, convenient representations of the fundamental domain are given for the homogeneous manifolds, \E{1}, \E{11}, and \slabh, but not for any of the others.

\subsection{\E{1}: 3-torus}
\label{secn:topologyE1}

\noindent \textit{Properties}: As listed in \cref{tab:properties}, manifolds of this topology are compact, orientable, homogeneous, and anisotropic.
Further, all the compact topologies are roots of \E{1}. \\

\noindent
\textit{Generators}: The generators are given by
\begin{align} 
    \label{eqn:E1generalT}
    & \mat{M}^{\E{1}}_A = \identity, \quad \mbox{with} \nonumber \\
    & \vec{T}^{\E{1}}_{\A{1}} = \begin{pmatrix} L_{\A{1}x} \\ 0 \\ 0 \end{pmatrix} \equiv \vec{T}^{\E{1}}_1, \quad    
    \vec{T}^{\E{1}}_{\A{2}} = \begin{pmatrix} L_{\A{2}x} \\ L_{\A{2}y} \\ 0 \end{pmatrix} \equiv \vec{T}^{\E{1}}_2, \quad    
    \vec{T}^{\E{1}}_{\A{3}} = \begin{pmatrix} L_{\A{3}x} \\ L_{\A{3}y} \\ L_{\A{3}z}\end{pmatrix} \equiv \vec{T}^{\E{1}}_3 .
\end{align}
We may also write the translation vectors as
\begin{equation} 
    \label{eqn:E1generalTalt}
    \vec{T}^{\E{1}}_{\A{1}} = L_{\A{1}} \begin{pmatrix} 1 \\ 0 \\ 0 \end{pmatrix}, \quad
    \vec{T}^{\E{1}}_{\A{2}} = L_{\A{2}} \begin{pmatrix} \cos\alpha \\ \sin\alpha, \\ 0 \end{pmatrix}, \quad
    \vec{T}^{\E{1}}_{\A{3}} = L_{\A{3}} \begin{pmatrix} \cos\beta\cos\gamma \\ \cos\beta\sin\gamma \\ \sin\beta \end{pmatrix} .
\end{equation}
Here and throughout, when written in the alternative form we will always choose the lengths to be positive (here meaning $0 < L_{\A{i}}$) with the orientation of the vector determined by the angles (here $\alpha$, $\beta$, and $\gamma$).
The alternative names $\vec{T}^{\E{1}}_i$ introduce a notation that will be useful for the rest of the non-trivial topologies, \E{2}--\E{17}. \\

Since \E{1} is homogeneous, all the generators are pure translations associated with the same matrix $\mat{M}^{\E{1}}_A=\identity$; the ``$A$'' label is extraneous and can become cumbersome.
For compactness of expressions we will often drop the $A$ in the subscript of $L_{\A{i}w}$ and $L_{\A{i}}$ so that
\begin{align}
    L_{iw} &\equiv L_{\A{i}w}, \quad \mbox{for } i\in \{1,2,3\}, \, w\in\{x,y,z\}; \nonumber \\
    L_{i} &\equiv L_{\A{i}}, \quad \mbox{for } i\in \{1,2,3\} .
\end{align}
\\
\noindent \textit{Volume}:  
\begin{align}
    \label{eqn:E1volume}
    V_{\E{1}} 
        &= \vert(\vec{T}^{\E{1}}_{{1}}\times\vec{T}^{\E{1}}_{{2}})\cdot\vec{T}^{\E{1}}_{{3}}\vert \\
        &= \vert L_{{1}x} L_{{2}y} L_{{3}z}\vert =
        L_{{1}} L_{{2}} L_{{3}} \vert \sin{\alpha}\sin{\beta} \vert . \nonumber
\end{align}

\noindent \textit{Origin}: Since $\mat{M}^{\E{1}}_A=\identity$ the manifold is homogeneous and the clone lattice of an observer is independent of the location of the observer. \\

\noindent \textit{Real parameters (6 independent)}: There are 6 independent parameters required to fully define \E{1}.
Since \E{1} is homogeneous they are all required, none can be traded for shifts of the origin. \\

\noindent \textit{Parameter ranges}: We want to ensure that we do not double-count parameter choices that appear different but actually generate the same lattice of clones.
To this end, we choose our coordinate system such that the shortest translation vector is $\vec{T}^{\E{1}}_{\A{1}}$ and is oriented along the $+\unitvec{x}$ direction, while the two shortest translation vectors,
$\vec{T}^{\E{1}}_{\A{1}}$ and $\vec{T}^{\E{1}}_{\A{2}}$, define the $xy$-plane, with the $y$-component of $\vec{T}^{\E{1}}_{\A{2}}$ positive.
$\vec{T}^{\E{1}}_{\A{3}}$ is then an arbitrary vector subject to the following conditions.
This serves as the base set of conditions that will be applied to all the compact orientable spaces, \E{1}--\E{6}.
\begin{enumerate}
    \item $0< L_{\A{1}x}=L_{\A{1}}$, $0 < L_{\A{2}y} \equiv L_{2y}$, and $0 \neq L_{\A{3}z} \equiv L_{3z}$, i.e., choice of orientation;
    \item $0<\vert\vec{T}^{\E{1}}_{\A{1}}\vert/\vert\vec{T}^{\E{1}}_{\A{3}}\vert \leq \vert\vec{T}^{\E{1}}_{\A{1}}\vert/\vert\vec{T}^{\E{1}}_{\A{2}}\vert \leq 1$, i.e., choice of ordering;
    \item $\vert\vec{T}^{\E{1}}_{\A{2}}\cdot\vec{T}^{\E{1}}_{\A{1}}\vert \leq \frac{1}{2}\vert\vec{T}^{\E{1}}_{\A{1}}\vert^2$, i.e., $\vec{T}^{\E{1}}_{\A{2}}$ cannot be shortened by adding or subtracting $\vec{T}^{\E{1}}_{\A{1}}$;\footnote{
        Strictly speaking, the constraint is $-\frac{1}{2}\vert\vec{T}^{\E{1}}_{\A{1}}\vert^2 < \vec{T}^{\E{1}}_{\A{2}}\cdot\vec{T}^{\E{1}}_{\A{1}} \leq \frac{1}{2}\vert\vec{T}^{\E{1}}_{\A{1}}\vert^2$.
        To keep these conditions terse we will continue to use the stated shorter form.
        A similar simplification holds for the conditions when written in terms of the parameters.}
    \item $\vert\vec{T}^{\E{1}}_{\A{3}}\cdot\vec{T}^{\E{1}}_{\A{1}}\vert \leq \frac{1}{2}\vert\vec{T}^{\E{1}}_{\A{1}}\vert^2$, i.e., $\vec{T}^{\E{1}}_{\A{3}}$ cannot be shortened by adding or subtracting $\vec{T}^{\E{1}}_{\A{1}}$;
    \item $\vert\vec{T}^{\E{1}}_{\A{3}}\cdot\vec{T}^{\E{1}}_{\A{2}}\vert \leq \frac{1}{2}\vert\vec{T}^{\E{1}}_{\A{2}}\vert^2$, i.e., $\vec{T}^{\E{1}}_{\A{3}}$ cannot be shortened by adding or subtracting $\vec{T}^{\E{1}}_{\A{2}}$;
    \item  $\vert\vec{T}^{\E{1}}_{\A{3}}\cdot(\vec{T}^{\E{1}}_{\A{1}}\pm\vec{T}^{\E{1}}_{\A{2}})\vert \leq \frac{1}{2}\vert\vec{T}^{\E{1}}_{\A{1}}\pm\vec{T}^{\E{1}}_{\A{2}}\vert^2$, i.e., $\vec{T}^{\E{1}}_{\A{3}}$ cannot be shortened by adding or subtracting $\vec{T}^{\E{1}}_{\A{1}}\pm\vec{T}^{\E{1}}_{\A{2}}$;
    \item  $\vert\vec{T}^{\E{1}}_{\A{1}}\cdot(\vec{T}^{\E{1}}_{\A{2}}\pm\vec{T}^{\E{1}}_{\A{3}})\vert \leq \frac{1}{2}\vert\vec{T}^{\E{1}}_{\A{2}}\pm\vec{T}^{\E{1}}_{\A{3}}\vert^2$, i.e., $\vec{T}^{\E{1}}_{\A{1}}$ cannot be shortened by adding or subtracting $\vec{T}^{\E{1}}_{\A{2}}\pm\vec{T}^{\E{1}}_{\A{3}}$;
    \item  $\vert\vec{T}^{\E{1}}_{\A{2}}\cdot(\vec{T}^{\E{1}}_{\A{3}}\pm\vec{T}^{\E{1}}_{\A{1}})\vert \leq \frac{1}{2}\vert\vec{T}^{\E{1}}_{\A{3}}\pm\vec{T}^{\E{1}}_{\A{1}}\vert^2$, i.e., $\vec{T}^{\E{1}}_{\A{2}}$ cannot be shortened by adding or subtracting $\vec{T}^{\E{1}}_{\A{3}}\pm\vec{T}^{\E{1}}_{\A{1}}$.
\end{enumerate}
In terms of the parameters, these conditions become:
\begin{enumerate}
    \item $0 < \alpha < \pi$ so that $0 < \sin\alpha \leq 1$ (recall that by definition $0<L_i$ for $i\in\{1, 2, 3\}$); 
    \item $0<L_{1}/L_{3}\leq L_{1}/L_{2}\leq 1$;
    \item $\vert \cos\alpha \vert \leq \frac{1}{2} L_{1} / L_{2}$;
    \item $\vert\cos\beta\cos\gamma\vert \leq \frac{1}{2}L_{1} / L_{3}$;
    \item $\vert\cos\beta \cos(\alpha-\gamma)\vert \leq \frac{1}{2} L_{2} / L_{3}$;
    \item $L_{{3}}\vert\cos\beta\vert\vert L_{2}\cos(\alpha-\gamma) \pm L_{1} \cos\gamma\vert \leq
    \frac{1}{2}\left[L_{1}^2 + L_{2}^2 \pm 2L_{1}L_{2}\cos\alpha\right]$;
    \item  $L_{1}\vert L_{2}\cos\alpha\pm L_{{3}}\cos\beta\cos\gamma\vert \leq \frac{1}{2} \left[ L_{2}^2 + L_{3}^2 \pm 2 L_{2} L_{3} \cos\beta\cos(\alpha-\gamma)\right]$;
    \item  $L_{2} \vert L_{3} \cos\beta\cos(\alpha-\gamma)\pm L_{1}\cos\alpha\vert \leq \frac{1}{2}\left[L_{1}^2 + L_{3}^2 \pm 2 L_{1} L_{3} \cos\beta\cos\gamma \right].$
\end{enumerate}

\noindent \textit{Convenient fundamental domain}: A convenient choice of FD is a parallelepiped, the vertices of which are any base point and seven clones.

There are two convenient choices of base point:
\begin{enumerate}[(A)]
    \item Origin-centered FD.
    \begin{itemize}
        \item Base point:
        $\vec{x}_A\equiv\frac{1}{2}(-\vec{T}^{\E{1}}_{\A{1}}-\vec{T}^{\E{1}}_{\A{2}}-\vec{T}^{\E{1}}_{\A{3}})$;
        \item 3 other corners of the bottom face:
        \begin{itemize}
            \item $\vec{x}_B=g^{\E{1}}_{\A{1}}\vec{x}_A=\frac{1}{2}(\vec{T}^{\E{1}}_{\A{1}}-\vec{T}^{\E{1}}_{\A{2}}-\vec{T}^{\E{1}}_{\A{3}})$;
           \item $\vec{x}_C=g^{\E{1}}_{\A{2}}g^{\E{1}}_{\A{1}}\vec{x}_A=\frac{1}{2}(\vec{T}^{\E{1}}_{\A{1}}+\vec{T}^{\E{1}}_{\A{2}}-\vec{T}^{\E{1}}_{\A{3}})$;
            \item $\vec{x}_D=g^{\E{1}}_{\A{2}}\vec{x}_A=\frac{1}{2}(-\vec{T}^{\E{1}}_{\A{1}}+\vec{T}^{\E{1}}_{\A{2}}-\vec{T}^{\E{1}}_{\A{3}})
            $;
        \end{itemize}
        \item Top face:
        \begin{itemize}
            \item 
            $\vec{x}_E=g^{\E{1}}_{\A{3}}\vec{x}_A=\vec{x}_A+\vec{T}^{\E{1}}_{\A{3}}$;
            \item
            $\vec{x}_F=g^{\E{1}}_{\A{3}}\vec{x}_B=\vec{x}_B+\vec{T}^{\E{1}}_{\A{3}}$;
            \item
            $\vec{x}_G=g^{\E{1}}_{\A{3}}\vec{x}_C=\vec{x}_C+\vec{T}^{\E{1}}_{\A{3}}$;
            \item
            $\vec{x}_H=g^{\E{1}}_{\A{3}}\vec{x}_D=\vec{x}_D+\vec{T}^{\E{1}}_{\A{3}}$.
        \end{itemize}
    \end{itemize}
\item Origin-rooted FD.
    \begin{itemize}
        \item Base point:
        $\vec{x}_A\equiv\vec{0}$;
        \item 3 other corners of $z=0$ face:
        \begin{itemize}
            \item $\vec{x}_B=g^{\E{1}}_{\A{1}}\vec{x}_A=\vec{T}^{\E{1}}_{\A{1}}$;
           \item $\vec{x}_C=g^{\E{1}}_{\A{2}}g^{\E{1}}_{\A{1}}\vec{x}_A=\vec{T}^{\E{1}}_{\A{1}}+\vec{T}^{\E{1}}_{\A{2}}$;
            \item $\vec{x}_D=g^{\E{1}}_{\A{2}}\vec{x}_A=\vec{T}^{\E{1}}_{\A{2}};
            $
        \end{itemize}
        \item $z\neq0$ face: same as ``top face'' from case (A).
    \end{itemize}
\end{enumerate}

\subsection{\E{2}: Half-turn space}
\label{secn:topologyE2}

\noindent \textit{Properties}: As listed in \cref{tab:properties}, this manifold is compact, orientable, inhomogeneous, and anisotropic. \\

\noindent \textit{Generators}: In general (see \cref{app:E2}) the generators of \E{2} can be written as\footnote{Here and throughout all rotations will be treated as active.}
\begin{align}
    \label{eqn:E2generalT}
    & \mat{M}^{\E{2}}_A = \identity, \quad \mat{M}^{\E{2}}_B = \mat{R}_{\unitvec{z}}(\pi) = \diag(-1,-1,1), \quad \mbox{with} \nonumber \\
    &\vec{T}^{\E{2}}_{\A{1}} = \begin{pmatrix} L_{\A{1}x} \\ 0 \\ 0 \end{pmatrix} 
        \equiv \vec{T}^{\E{2}}_1, \quad
    \vec{T}^{\E{2}}_{\A{2}} = \begin{pmatrix} L_{\A{2}x} \\ L_{\A{2}y} \\ 0 \end{pmatrix}
        \equiv \vec{T}^{\E{2}}_2, \quad
    \vec{T}^{\E{2}}_{B} = \begin{pmatrix} L_{Bx} \\ L_{By} \\ L_{Bz} \end{pmatrix}.
\end{align}
Alternatively, we can write the translation vectors as
\begin{equation}
    \label{eqn:E2generalTalt}
    \vec{T}^{\E{2}}_{\A{1}} = L_{\A{1}} \begin{pmatrix} 1 \\ 0 \\ 0 \end{pmatrix}, \quad 
    \vec{T}^{\E{2}}_{\A{2}} = L_{\A{2}} \begin{pmatrix} \cos\alpha \\ \sin\alpha \\ 0 \end{pmatrix}, \quad
    \vec{T}^{\E{2}}_{B} = L_B \begin{pmatrix} \cos\beta\cos\gamma \\ \cos\beta\sin\gamma \\ \sin\beta \end{pmatrix}.
\end{equation}
Similar to \E{1}, we often simplify the notation when working with parameters of \E{2} by dropping the $A$ label and instead using
\begin{align}
    L_{iw} &\equiv L_{\A{i}w}, \quad \mbox{for } i\in \{1,2\}, \, w\in\{x,y,z\}; \nonumber \\
    L_{i} &\equiv L_{\A{i}}, \quad \mbox{for } i\in \{1,2\} .
\end{align}

\noindent\textit{Associated \E{1}:}
In addition to $\vec{T}^{\E{2}}_1$ and $\vec{T}^{\E{2}}_2$ defined above, a third independent translation is
\begin{equation}
    \label{eqn:E2assocE1}
    g^{\E{2}}_3 \equiv (g_B^{\E{2}})^2: \vec{x} \to \vec{x} + \vec{T}^{\E{2}}_3 ,
\end{equation}
for
\begin{equation}
    \vec{T}^{\E{2}}_3 \equiv \begin{pmatrix} 0 \\ 0 \\ 2L_{Bz} \end{pmatrix} = 2 L_B \begin{pmatrix} 0 \\ 0 \\ \sin\beta \end{pmatrix}.
\end{equation}
The three vectors $\vec{T}^{\E{2}}_1$, $\vec{T}^{\E{2}}_2$, and $\vec{T}^{\E{2}}_3$ define the associated \E{1}. \\

\noindent \textit{Volume}:
\begin{equation}
    \label{eqn:VE2}
    V_{\E{2}} = \frac{1}{2}\vert(\vec{T}^{\E{2}}_{1}\times\vec{T}^{\E{2}}_{2})\cdot\vec{T}^{\E{2}}_{3}\vert 
      = \vert L_{1x} L_{2y} L_{Bz} \vert
      = L_{1} L_{2} L_B \vert \sin\alpha\sin\beta\vert.
\end{equation}
\\
\noindent \textit{Tilts versus origin position}:
When shifting the origin, ${x}^{\E{2}}_{0x}$ and ${x}^{\E{2}}_{0y}$ change $L_{Bx}$ and $L_{By}$, but $L_{Bz}$ is unaffected.
Equivalently, when shifting $L_B$, $\beta$, and $\gamma$ are changed while holding $L_B\sin\beta$ fixed.
This shows that shifting the origin is equivalent to tilting the translation vector associated with the rotation, $\vec{T}^{\E{2}}_B$.

Explicitly, since $\mat{M}^{\E{2}}_A=\identity$, neither $\vec{T}^{\E{2}}_{\A{1}}$ nor  $\vec{T}^{\E{2}}_{\A{2}}$ is affected by shifts in $\vec{x}_0^{\E{2}}$.
In contrast, tilting $\vec{T}^{\E{2}}_B$ out of the $\unitvec{z}$ direction is equivalent to a shift of origin; i.e., $\beta$ and $\gamma$ can be adjusted (for example to $\pi/2$) by the choice of $\vec{x}^{\E{2}}_{0x}$ and $\vec{x}^{\E{2}}_{0y}$.
In particular, one can set $\beta=\pi/2$, so that $\vec{T}^{\E{3}}_B=L_B\unitvec{e}_z$.
Of course, the observer will then not sit at the origin of coordinates and may be up to $\vert\vec{T}^{\E{2}}_{\A{1}}+\vec{T}^{\E{2}}_{\A{2}}\vert$ away in the $xy$-plane.
(Their $z$ position is immaterial.) \\

\noindent\textit{Real parameters (6 independent):}
There are 6 independent parameters required to fully define \E{2}.
As noted above, some are redundant with shifting the origin, or equivalently, a tilt.
Thus we have:
\begin{itemize}
    \item $L_{1x}$, $L_{2x}$, $L_{2y}$, and $L_{Bz}$ are intrinsic parameters of the manifold;
    \item $L_{Bx}$ and $L_{By}$ can be traded for ${x}^{\E{2}}_{0x}$ and ${x}^{\E{2}}_{0y}$;
    \item the standard (special origin, i.e., ``untilted'') form is $0=L_{2x}=L_{Bx}=L_{By}$.
\end{itemize}
In terms of the alternative parameter form we have:
\begin{itemize}
    \item $L_{{1}}$, $L_{{1}}/L_{{2}}$, $L_{{1}}/L_B$, and $\alpha$ are intrinsic parameters of the manifold;
    \item $\beta$ and $\gamma$ can be traded for ${x}^{\E{2}}_{0x}$ and ${x}^{\E{2}}_{0y}$;
    \item the standard (special origin, i.e., ``untilted'') form is $\alpha=\pi/2$, $\beta=\pi/2$;
\end{itemize}

\noindent \textit{Parameter ranges}: We want to ensure that we do not double-count parameter choices that appear different but actually generate the same lattice of clones.  Similar to \E{1}, we therefore require:
\begin{enumerate}
    \item $0< L_{\A{1}x}=L_{\A{1}}$, $0 < L_{\A{2}y} \equiv L_{2y}$, and $0 \neq L_{Bz}$, i.e., choice of orientation;
    \item $0< \vert\vec{T}^{\E{2}}_{\A{1}}\vert/\vert\vec{T}^{\E{2}}_{\A{2}}\vert \leq 1$, i.e., choice of ordering (note, in contrast to \E{1}, we cannot constrain $\vert\vec{T}^{\E{2}}_B\vert$ since it is identified not as the longest vector but as the one associated with $\mat{M}^{\E{2}}_B\neq\identity$);
    \item $\vert\vec{T}^{\E{2}}_{\A{2}}\cdot\vec{T}^{\E{2}}_{\A{1}}\vert \leq \frac{1}{2}\vert\vec{T}^{\E{2}}_{\A{1}}\vert^2$, i.e., $\vec{T}^{\E{2}}_{\A{2}}$ cannot be shortened by adding or subtracting $\vec{T}^{\E{2}}_{\A{1}}$;
    \item $\vert\vec{T}^{\E{2}}_B\cdot\vec{T}^{\E{2}}_{\A{1}}\vert \leq \frac{1}{2}\vert\vec{T}^{\E{2}}_{\A{1}}\vert^2$, i.e., $\vec{T}^{\E{2}}_B$ cannot be shortened by adding or subtracting $\vec{T}^{\E{2}}_{\A{1}}$;
    \item $\vert\vec{T}^{\E{2}}_B\cdot\vec{T}^{\E{2}}_{\A{2}}\vert \leq \frac{1}{2}\vert\vec{T}^{\E{2}}_{\A{2}}\vert^2$, i.e., $\vec{T}^{\E{2}}_B$ cannot be shortened by adding or subtracting $\vec{T}^{\E{2}}_{\A{2}}$;
    \item  $\vert\vec{T}^{\E{2}}_B\cdot(\vec{T}^{\E{2}}_{\A{1}}\pm\vec{T}^{\E{2}}_{\A{2}})\vert \leq \frac{1}{2}\vert\vec{T}^{\E{2}}_{\A{1}}\pm\vec{T}^{\E{2}}_{\A{2}}\vert^2$, i.e., $\vec{T}^{\E{2}}_B$ cannot be shortened by adding or subtracting $\vec{T}^{\E{2}}_{\A{1}}\pm\vec{T}^{\E{2}}_{\A{2}}$;
    \item $\vert\vec{T}^{\E{2}}_{\A{1}}\cdot(\vec{T}^{\E{2}}_{\A{2}}\pm\vec{T}^{\E{1}}_3)\vert \leq \frac{1}{2}\vert\vec{T}^{\E{2}}_{\A{2}}\pm\vec{T}^{\E{2}}_3\vert^2$, automatically satisfied by condition (3) and since $\vec{T}^{\E{2}}_{\A{1}}\cdot\vec{T}^{\E{2}}_3=0$ (note that it is $\vec{T}^{\E{2}}_3$ that appears here, not $\vec{T}^{\E{2}}_B$);
    \item  $\vert\vec{T}^{\E{2}}_{\A{2}}\cdot(\vec{T}^{\E{2}}_3\pm\vec{T}^{\E{2}}_{\A{1}})\vert \leq \frac{1}{2}\vert\vec{T}^{\E{2}}_3\pm\vec{T}^{\E{2}}_{\A{1}}\vert^2$, automatically satisfied by condition (3) and since $\vec{T}^{\E{2}}_{\A{2}}\cdot\vec{T}^{\E{2}}_3=0$ (note that it is $\vec{T}^{\E{2}}_3$ that appears here, not $\vec{T}^{\E{2}}_B$).
\end{enumerate}

In terms of the parameters, the necessary conditions become:
\begin{enumerate}
    \item $0 < \alpha < \pi$ so that $0 < \sin\alpha \leq 1$, $0 < \vert \beta \vert \leq\pi/2$ so that $0 \neq \sin\beta$, and $0\leq\gamma<2\pi$;
    \item $0< L_{1}/L_{2} \leq 1$,  $0 < L_{1}/L_B < \infty $;
    \item $\vert \cos\alpha \vert \leq \frac{1}{2}L_{1} / L_{2}$;
    \item $\vert\cos\beta\cos\gamma\vert \leq \frac{1}{2}L_{1} / L_B$;
    \item $\vert\cos\beta \cos(\alpha-\gamma)\vert \leq \frac{1}{2}L_{2} / L_B$;
    \item $L_B\vert\cos\beta\vert\vert L_{2}\cos(\alpha-\gamma) \pm L_{1} \cos\gamma\vert \leq
    \frac{1}{2}\left[L_{1}^2 + L_{2}^2 \pm 2L_{1}L_{2}\cos\alpha\right]$.
\end{enumerate}

\subsection{\E{3}: Quarter-turn space}
\label{secn:topologyE3}

\textit{Properties}: As listed in \cref{tab:properties}, this manifold is compact, orientable, inhomogeneous, and anisotropic. \\

\noindent \textit{Generators}:  In general (see \cref{app:E3}) the generators of \E{3} can be written as
\begin{align}
    \label{eqn:E3generalT}
    & \mat{M}^{\E{3}}_A = \identity, \quad \mat{M}^{\E{3}}_B = \mat{R}_{\unitvec{z}}(\pi/2) =
        \begin{pmatrix}
        0 & -1 & 0\\
        1 &  \hphantom{-}0 & 0\\
        0 &  \hphantom{-}0 & 1
        \end{pmatrix} ,
        \quad \mbox{with}
        \nonumber \\
    & \vec{T}^{\E{3}}_{\A{1}} = \begin{pmatrix} L_{A} \\ 0 \\ 0 \end{pmatrix} \equiv \vec{T}^{\E{3}}_1, \quad
    \vec{T}^{\E{3}}_{\A{2}} = \begin{pmatrix} 0 \\ L_{A} \\ 0 \end{pmatrix} \equiv \vec{T}^{\E{3}}_2, \quad
    \vec{T}^{\E{3}}_{B} = \begin{pmatrix} L_{Bx} \\ L_{By} \\ L_{Bz} \end{pmatrix} .
\end{align}
Alternatively, we can write the translation vectors as
\begin{equation}
    \vec{T}^{\E{3}}_{\A{1}} = L_{A} \begin{pmatrix} 1 \\ 0 \\ 0 \end{pmatrix}, \quad
    \vec{T}^{\E{3}}_{\A{2}} = L_{A} \begin{pmatrix} 0 \\ 1 \\ 0 \end{pmatrix}, \quad
    \vec{T}^{\E{3}}_{B} = L_B \begin{pmatrix} \cos\beta\cos\gamma \\ \cos\beta\sin\gamma \\ \sin\beta \end{pmatrix} .
\end{equation}
Note that the $\vec{T}^{\E{3}}_{\A{1}}$ and $\vec{T}^{\E{3}}_{\A{2}}$ are the same length and orthogonal to one another.
\\
\noindent \textit{Associated \E{1}}: In addition to $\vec{T}^{\E{3}}_1$ and $\vec{T}^{\E{3}}_2$ defined above, a third independent translation follows from $(\mat{M}^{\E{3}}_B)^4=\identity$:
\begin{equation}
    g^{\E{3}}_3 \equiv (g^{\E{3}}_B )^4:\vec{x}\to \vec{x} + \vec{T}^{\E{3}}_3,
\end{equation}
for
\begin{equation}
    \label{eqn:E3assocE1}
    \vec{T}^{\E{3}}_3 \equiv \begin{pmatrix} 0 \\ 0 \\ 4 L_{Bz} \end{pmatrix}
      = 4 L_B \begin{pmatrix} 0 \\ 0 \\ \sin\beta \end{pmatrix}.
\end{equation}
\\
\noindent \textit{Volume}:
\begin{equation}
    \label{eqn:VE3}
    V_{\E{3}} = \frac{1}{4}
        \vert(\vec{T}^{\E{3}}_{{1}}\times\vec{T}^{\E{3}}_{{2}})\cdot\vec{T}^{\E{3}}_{3}\vert 
      = L_A^2 L_B |\sin\beta|.
\end{equation}
\\
\noindent \textit{Tilts versus origin position}: As in \E{2}, a shift of origin will change $L_{Bx}$ and $L_{By}$ but will not affect $L_{Bz}$.
This again leads to a tilt being equivalent to a shift of origin, which allows us to choose $\vec{T}^{\E{3}}_B = L_B \unitvec{e}_z$ at the expense of the observer no longer being at the origin. \\

\noindent\textit{Real parameters (4 independent)}: There are 4 independent parameters required to fully define \E{3}.
As in \E{2} some are redundant with shifting the origin.
Thus we have:
\begin{itemize}
    \item $L_A$ and $L_A/L_B$ are intrinsic parameters of the manifold;
    \item $L_{Bx}$ and $L_{By}$, or equivalently $\beta$ and $\gamma$, can be traded for ${x}^{\E{3}}_{0x}$ and ${x}^{\E{3}}_{0y}$;
    \item the standard (special origin, i.e., ``untilted'') form is $0 = L_{Bx} = L_{By}$, or equivalently $\beta=\pi/2$ (with $\gamma$ irrelevant).
\end{itemize}

\noindent \textit{Parameter ranges}: We want to ensure that we do not double-count parameter choices that appear different but actually generate the same lattice of clones.
Similar to \E{1}, we therefore require:
\begin{enumerate}
    \item $0<L_{A}$ by definition as in \E{1} and $0 \neq L_{Bz}$, i.e., choice of orientation;
    \item $\vert\vec{T}^{\E{3}}_{\A{1}}\vert/\vert\vec{T}^{\E{3}}_{\A{2}}\vert =1$, automatically enforced by parametrization (note that we cannot constrain $\vert\vec{T}^{\E{3}}_B\vert$ since it is identified not as the longest vector but as the one associated with $\mat{M}^{\E{3}}_B\neq\identity$);
    \item $\vert\vec{T}^{\E{3}}_{\A{2}}\cdot\vec{T}^{\E{3}}_{\A{1}}\vert =0$; automatically enforced by parametrization;
    \item $\vert\vec{T}^{\E{3}}_B\cdot\vec{T}^{\E{3}}_{\A{1}}\vert \leq \frac{1}{2}\vert\vec{T}^{\E{3}}_{\A{1}}\vert^2$, i.e., $\vec{T}^{\E{3}}_B$ cannot be shortened by adding or subtracting $\vec{T}^{\E{3}}_{\A{1}}$;
    \item $\vert\vec{T}^{\E{3}}_B\cdot\vec{T}^{\E{3}}_{\A{2}}\vert \leq \frac{1}{2}\vert\vec{T}^{\E{3}}_{\A{2}}\vert^2$, i.e., $\vec{T}^{\E{3}}_B$ cannot be shortened by adding or subtracting $\vec{T}^{\E{3}}_{\A{2}}$;
    \item $\vert\vec{T}^{\E{3}}_B\cdot(\vec{T}^{\E{3}}_{\A{1}}\pm\vec{T}^{\E{3}}_{\A{2}})\vert \leq \frac{1}{2}\vert\vec{T}^{\E{3}}_{\A{1}}\pm\vec{T}^{\E{3}}_{\A{2}}\vert^2$, automatically enforced given conditions (4) and (5);
    \item $\vert\vec{T}^{\E{3}}_{\A{1}}\cdot(\vec{T}^{\E{3}}_{\A{2}}\pm\vec{T}^{\E{3}}_3)\vert = 0$, automatically enforced (note that it is $\vec{T}^{\E{3}}_3$ that appears here, not $\vec{T}^{\E{3}}_B$);
    \item $\vert\vec{T}^{\E{3}}_{\A{2}}\cdot(\vec{T}^{\E{3}}_3\pm\vec{T}^{\E{3}}_{\A{1}})\vert = 0$, automatically enforced (note that it is $\vec{T}^{\E{3}}_3$ that appears here, not $\vec{T}^{\E{3}}_B$).
\end{enumerate}

In terms of the parameters, the necessary conditions become:
\begin{enumerate}
    \item $0 < L_A$, $0 < \vert \beta \vert \leq\pi/2$ so that $0 \neq \sin\beta$, and $0\leq\gamma<2\pi$;
    \item $0< L_A/L_B < \infty $;
    \stepcounter{enumi} 
    \item $\vert L_{Bx} \vert \leq \frac{1}{2} L_A$, or equivalently $\cos\beta \vert \cos\gamma \vert \leq \frac{1}{2} L_A/ L_B$;
    \item $\vert L_{By} \vert \leq \frac{1}{2} L_A$, or equivalently $\cos\beta \vert \sin\gamma \vert \leq \frac{1}{2} L_A/ L_B$.
\end{enumerate}

\subsection{\E{4}: Third-turn space}
\label{secn:topologyE4}

\textit{Properties}: As listed in \cref{tab:properties}, this manifold is compact, orientable, inhomogeneous, and anisotropic. \\

\noindent \textit{Generators}: In general (see \cref{app:E4}) the generators of \E{4} can be written as
\begin{align}
    \label{eqn:E4generalT}
    & \mat{M}^{\E{4}}_A = \identity, \quad \mat{M}^{\E{4}}_B = \mat{R}_{\unitvec{z}}(2\pi/3) =
        \begin{pmatrix}
        -1/2 & -\sqrt{3}/2 & 0\\
        \sqrt{3}/2 & -1/2 & 0\\
        0 & \hphantom{-}0 & 1
        \end{pmatrix} ,
        \quad \mbox{with}
        \nonumber \\
    & \vec{T}^{\E{4}}_{\A{1}} = \begin{pmatrix} L_{A} \\ 0 \\ 0 \end{pmatrix} \equiv \vec{T}^{\E{4}}_1, \quad
    \vec{T}^{\E{4}}_{\A{2}} = \begin{pmatrix} -L_{A}/2 \\ \sqrt{3} L_{A} / 2 \\ 0 \end{pmatrix} \equiv \vec{T}^{\E{4}}_2, \quad
    \vec{T}^{\E{4}}_{B} = \begin{pmatrix} L_{Bx} \\ L_{By} \\ L_{Bz} \end{pmatrix} .
\end{align}
Notice that $\vec{T}^{\E{4}}_{\A{2}} = \mat{R}_{\unitvec{z}}(2\pi/3) \vec{T}^{\E{4}}_{\A{1}}$.
It can also be useful to define and use
\begin{equation}
    \vec{T}^{\E{4}}_{\A{3}} \equiv \mat{R}_{\unitvec{z}}(-2\pi/3) \vec{T}^{\E{4}}_{\A{1}} = \begin{pmatrix} -L_{A}/2 \\ -\sqrt{3} L_{A} / 2 \\ 0 \end{pmatrix} = - \vec{T}^{\E{4}}_{\A{1}} - \vec{T}^{\E{4}}_{\A{2}}.
\end{equation}
Alternatively, we can write the translation vectors as
\begin{equation}
    \vec{T}^{\E{4}}_{\A{1}} = L_{A} \begin{pmatrix} 1 \\ 0 \\ 0 \end{pmatrix}, \quad
    \vec{T}^{\E{4}}_{\A{2}} = L_{A} \begin{pmatrix} -1/2 \\ \sqrt{3}/2 \\ 0 \end{pmatrix}, \quad
    \vec{T}^{\E{4}}_{B} = L_B \begin{pmatrix} \cos\beta\cos\gamma \\ \cos\beta\sin\gamma \\ \sin\beta \end{pmatrix} .
\end{equation}
\\
\noindent \textit{Associated \E{1}}: In addition to $\vec{T}^{\E{4}}_1$ and $\vec{T}^{\E{4}}_2$ defined above, a third independent translation follows from $(\mat{M}^{\E{4}}_B)^3=\identity$:
\begin{equation}
    g^{\E{4}}_3 \equiv (g^{\E{4}}_B )^3:\vec{x}\to \vec{x} + \vec{T}^{\E{4}}_3,
\end{equation}
for
\begin{equation}
    \label{eqn:E4assocE1}
    \vec{T}^{\E{4}}_3 \equiv \begin{pmatrix} 0 \\ 0 \\ 3 L_{Bz} \end{pmatrix}
      = 3 L_B \begin{pmatrix} 0 \\ 0 \\ \sin\beta \end{pmatrix}.
\end{equation}
\\
\noindent \textit{Volume}:
\begin{equation}
    \label{eqn:VE4}
    V_{\E{4}}= \frac{1}{3}
    \vert(\vec{T}^{\E{4}}_{{1}}\times\vec{T}^{\E{4}}_{{2}})\cdot\vec{T}^{\E{4}}_{3}\vert 
    = \frac{\sqrt{3}}{2} L_A^2 L_B |\sin\beta|.
\end{equation}
\\
\noindent \textit{Tilts versus origin position}: As in \E{2}, a shift of origin will change $L_{Bx}$ and $L_{By}$ but will not affect $L_{Bz}$.
This again leads to a tilt being equivalent to a shift of origin, which allows us to choose $\vec{T}^{\E{4}}_B = L_B \unitvec{e}_z$ at the expense of the observer no longer being at the origin. \\

\noindent \textit{Real parameters (4 independent)}: There are 4 independent parameters required to fully define \E{4}.
As in \E{2}, some are redundant with shifting the origin.
Thus we have:
\begin{itemize}
    \item $L_A$ and $L_A/L_B$ are intrinsic parameters of the manifold;
    \item $L_{Bx}$ and $L_{By}$, or equivalently $\beta$ and $\gamma$, can be traded for ${x}^{\E{4}}_{0x}$ and ${x}^{\E{4}}_{0y}$;
    \item the standard (special origin, i.e., ``untilted'') form is $0 = L_{Bx} = L_{By}$, or equivalently $\beta=\pi/2$ (with $\gamma$ irrelevant).
\end{itemize}

\noindent \textit{Parameter ranges}: We want to ensure that we do not double-count parameter choices that appear different but actually generate the same lattice of clones.
Similar to \E{1}, we therefore require:
\begin{enumerate}
    \item $0 < L_A$ and $0 \neq L_{Bz}$, i.e., choice of orientation;
    \item $\vert\vec{T}^{\E{4}}_{\A{1}}\vert/\vert\vec{T}^{\E{4}}_{\A{2}}\vert = 1$, automatically enforced by parametrization (note that we cannot constrain $\vert\vec{T}^{\E{4}}_B\vert$ since it is identified not as the longest vector but as the one associated with $\mat{M}^{\E{4}}_B\neq\identity$);
    \item $\vert\vec{T}^{\E{4}}_{\A{2}}\cdot\vec{T}^{\E{4}}_{\A{1}}\vert = \frac{1}{2} \vert\vec{T}^{\E{4}}_{\A{1}}\vert^2$, automatically enforced by parametrization;
    \item $\vert\vec{T}^{\E{4}}_B\cdot\vec{T}^{\E{4}}_{\A{1}}\vert \leq \frac{1}{2}\vert\vec{T}^{\E{4}}_{\A{1}}\vert^2$, i.e., $\vec{T}^{\E{4}}_B$ cannot be shortened by adding or subtracting $\vec{T}^{\E{4}}_{\A{1}}$;
    \item $\vert\vec{T}^{\E{4}}_B\cdot\vec{T}^{\E{4}}_{\A{2}}\vert \leq \frac{1}{2}\vert\vec{T}^{\E{4}}_{\A{2}}\vert^2$, i.e., $\vec{T}^{\E{4}}_B$ cannot be shortened by adding or subtracting $\vec{T}^{\E{4}}_{\A{2}}$;
    \item $\vert\vec{T}^{\E{4}}_B\cdot(\vec{T}^{\E{4}}_{\A{1}}\pm\vec{T}^{\E{4}}_2)\vert \leq \frac{1}{2}\vert\vec{T}^{\E{4}}_{\A{1}}\pm\vec{T}^{\E{4}}_{\A{2}}\vert^2$, i.e., $\vec{T}^{\E{4}}_B$ cannot be shortened by adding or subtracting $\vec{T}^{\E{4}}_{\A{1}}\pm\vec{T}^{\E{4}}_{\A{2}}$;
    \item $\vert\vec{T}^{\E{4}}_{\A{1}}\cdot(\vec{T}^{\E{4}}_{\A{2}}\pm\vec{T}^{\E{4}}_3)\vert \leq \frac{1}{2}\vert\vec{T}^{\E{4}}_{\A{2}}\pm\vec{T}^{\E{4}}_3\vert^2$, automatically enforced given that $\vec{T}^{\E{4}}_{\A{1}}\cdot\vec{T}^{\E{4}}_3=0$ and $\vec{T}^{\E{4}}_{\A{2}}\cdot\vec{T}^{\E{4}}_3=0$ (note that it is $\vec{T}^{\E{4}}_3$ that appears here, not $\vec{T}^{\E{4}}_B$);
    \item $\vert\vec{T}^{\E{4}}_{\A{2}}\cdot(\vec{T}^{\E{4}}_3\pm\vec{T}^{\E{4}}_{\A{1}})\vert \leq \frac{1}{2}\vert\vec{T}^{\E{4}}_3\pm\vec{T}^{\E{4}}_{\A{1}}\vert^2$, automatically enforced given that $\vec{T}^{\E{4}}_{\A{1}}\cdot\vec{T}^{\E{4}}_3=0$ and $\vec{T}^{\E{4}}_{\A{2}}\cdot\vec{T}^{\E{4}}_3=0$ (note that it is $\vec{T}^{\E{4}}_3$ that appears here, not $\vec{T}^{\E{4}}_B$).
\end{enumerate}

In terms of the parameters the necessary conditions become:
\begin{enumerate}
    \item $0 < L_A$, $0 < \vert \beta \vert \leq\pi/2$ so that $0 \neq \sin\beta$, and $0\leq\gamma<2\pi$;
    \item $0< L_A/L_B < \infty $;
    \stepcounter{enumi}
    \item $\cos\beta \vert \cos\gamma\vert \leq \frac{1}{2} L_A / L_B$;
    \item $\cos\beta \vert \cos(\gamma+\pi/3)\vert \leq \frac{1}{2} L_A / L_B$;
    \item $\cos\beta \vert \cos(\gamma-\pi/3)\vert \leq \frac{1}{2} L_A / L_B$
    and
    $\cos\beta \vert \cos(\gamma+\pi/6) \vert \leq \frac{\sqrt{3}}{2} L_A / L_B$.
\end{enumerate}

These conditions can be written compactly as $0 < L_A$, $0 \neq L_{Bz}$, and
\begin{enumerate}
    \item $-\frac{1}{2} L_A < L_{Bx} \leq 0$ and $-L_A + L_{Bx} < \sqrt{3} L_{By} \leq L_A + 3 L_{Bx}$, or
    \item $0 < L_{Bx} \leq \frac{1}{2} L_A$ and $-L_A + 3 L_{Bx} < \sqrt{3} L_{By} \leq L_A + L_{Bx}$.
\end{enumerate}

\subsection{\E{5}: Sixth-turn space}
\label{secn:topologyE5}

\textit{Properties}: As listed in \cref{tab:properties}, this manifold is compact, orientable, inhomogeneous, and anisotropic. \\

\noindent \textit{Generators}: In general (see \cref{app:E5}) the generators of \E{5} can be written as
\begin{align}
    \label{eqn:E5generalT}
    & \mat{M}^{\E{5}}_A = \identity, \quad \mat{M}^{\E{5}}_B = \mat{R}_{\unitvec{z}}(\pi/3) =
        \begin{pmatrix}
        1/2 & -\sqrt{3}/2 & 0\\
        \sqrt{3}/2 & 1/2 & 0\\
        0 & \hphantom{-}0 & 1
        \end{pmatrix} ,
        \quad \mbox{with}
        \nonumber \\
    & \vec{T}^{\E{5}}_{\A{1}} = \begin{pmatrix} L_{A} \\ 0 \\ 0 \end{pmatrix} \equiv \vec{T}^{\E{5}}_1, \quad
    \vec{T}^{\E{5}}_{\A{2}} = \begin{pmatrix} -L_{A}/2 \\ \sqrt{3} L_{A} / 2 \\ 0 \end{pmatrix} \equiv \vec{T}^{\E{5}}_2, \quad
    \vec{T}^{\E{5}}_{B} = \begin{pmatrix} L_{Bx} \\ L_{By} \\ L_{Bz} \end{pmatrix} .
\end{align}
Notice that $\vec{T}^{\E{5}}_{\A{2}} = -\mat{R}_{\unitvec{z}}(-\pi/3) \vec{T}^{\E{5}}_{\A{1}}$.
It can also be useful to define and use
\begin{equation}
    \vec{T}^{\E{5}}_{\A{3}} \equiv -\mat{R}_{\unitvec{z}}(\pi/3) \vec{T}^{\E{5}}_{\A{1}} = \begin{pmatrix} -L_{A}/2 \\ -\sqrt{3} L_{A} / 2 \\ 0 \end{pmatrix} = - \vec{T}^{\E{5}}_{\A{1}} - \vec{T}^{\E{5}}_{\A{2}}.
\end{equation}
Alternatively, we can write the translation vectors as
\begin{equation}
    \vec{T}^{\E{5}}_{\A{1}} = L_{A} \begin{pmatrix} 1 \\ 0 \\ 0 \end{pmatrix}, \quad
    \vec{T}^{\E{5}}_{\A{2}} = L_{A} \begin{pmatrix} -1/2 \\ \sqrt{3}/2 \\ 0 \end{pmatrix}, \quad
    \vec{T}^{\E{5}}_{B} = L_B \begin{pmatrix} \cos\beta\cos\gamma \\ \cos\beta\sin\gamma \\ \sin\beta \end{pmatrix} .
\end{equation}
\\
\noindent \textit{Associated \E{1}}: In addition to $\vec{T}^{\E{5}}_1$ and $\vec{T}^{\E{5}}_2$ defined above, a third independent translation follows from $(\mat{M}^{\E{5}}_B)^6=\identity$:
\begin{equation}
    g^{\E{5}}_3 \equiv (g^{\E{5}}_B )^6:\vec{x}\to \vec{x} + \vec{T}^{\E{5}}_3,
\end{equation}
for
\begin{equation}
    \label{eqn:E5assocE1}
    \vec{T}^{\E{5}}_3 \equiv \begin{pmatrix} 0 \\ 0 \\ 6 L_{Bz} \end{pmatrix}
      = 6 L_B \begin{pmatrix} 0 \\ 0 \\ \sin\beta \end{pmatrix}.
\end{equation}
\\
\noindent \textit{Volume}:
\begin{equation}
    \label{eqn:VE5}
    V_{\E{5}}= \frac{1}{6}
    \vert(\vec{T}^{\E{5}}_{{1}}\times\vec{T}^{\E{5}}_{{2}})\cdot\vec{T}^{\E{5}}_{3}\vert 
    = \frac{\sqrt{3}}{2} L_A^2 L_B |\sin\beta|.
\end{equation}
\\
\noindent \textit{Tilts versus origin position}: As in \E{2}, a shift of origin will change $L_{Bx}$ and $L_{By}$ but will not affect $L_{Bz}$.
This again leads to a tilt being equivalent to a shift of origin, which allows us to choose $\vec{T}^{\E{5}}_B = L_B \unitvec{e}_z$ at the expense of the observer no longer being at the origin. \\

\noindent \textit{Real parameters (4 independent)}: There are 4 independent parameters required to fully define \E{5}.
As in \E{2}, some are redundant with shifting the origin.
Thus we have:
\begin{itemize}
    \item $L_A$ and $L_A/L_B$ are intrinsic parameters of the manifold;
    \item $L_{Bx}$ and $L_{By}$, or equivalently $\beta$ and $\gamma$, can be traded for ${x}^{\E{5}}_{0x}$ and ${x}^{\E{5}}_{0y}$;
    \item the standard (special origin, i.e., ``untilted'') form is $0 = L_{Bx} = L_{By}$, or equivalently $\beta=\pi/2$ (with $\gamma$ irrelevant).
\end{itemize}

\noindent \textit{Parameter ranges}: We want to ensure that we do not double-count parameter choices that appear different but actually generate the same lattice of clones.
Similar to \E{1}, we therefore require:
\begin{enumerate}
    \item $0 < L_A$ and $0 \neq L_{Bz}$, i.e., choice of orientation;
    \item $\vert\vec{T}^{\E{5}}_{\A{1}}\vert/\vert\vec{T}^{\E{5}}_{\A{2}}\vert = 1$, automatically enforced by parametrization (note that we cannot constrain $\vert\vec{T}^{\E{5}}_B\vert$ since it is identified not as the longest vector but as the one associated with $\mat{M}^{\E{5}}_B\neq\identity$);
    \item $\vert\vec{T}^{\E{5}}_{\A{2}}\cdot\vec{T}^{\E{5}}_{\A{1}}\vert = \frac{1}{2} \vert\vec{T}^{\E{5}}_{\A{1}}\vert^2$, automatically enforced by parametrization;
    \item $\vert\vec{T}^{\E{5}}_B\cdot\vec{T}^{\E{5}}_{\A{1}}\vert \leq \frac{1}{2}\vert\vec{T}^{\E{5}}_{\A{1}}\vert^2$, i.e., $\vec{T}^{\E{5}}_B$ cannot be shortened by adding or subtracting $\vec{T}^{\E{5}}_{\A{1}}$;
    \item $\vert\vec{T}^{\E{5}}_B\cdot\vec{T}^{\E{5}}_{\A{2}}\vert \leq \frac{1}{2}\vert\vec{T}^{\E{5}}_{\A{2}}\vert^2$, i.e., $\vec{T}^{\E{5}}_B$ cannot be shortened by adding or subtracting $\vec{T}^{\E{5}}_{\A{2}}$;
    \item  $\vert\vec{T}^{\E{5}}_B\cdot(\vec{T}^{\E{5}}_{\A{1}}\pm\vec{T}^{\E{5}}_{\A{2}})\vert \leq \frac{1}{2}\vert\vec{T}^{\E{5}}_{\A{1}}\pm\vec{T}^{\E{5}}_{\A{2}}\vert^2$, i.e., $\vec{T}^{\E{5}}_B$ cannot be shortened by adding or subtracting $\vec{T}^{\E{5}}_{\A{1}}\pm\vec{T}^{\E{5}}_{\A{2}}$;
    \item  $\vert\vec{T}^{\E{5}}_{\A{1}}\cdot(\vec{T}^{\E{5}}_{\A{2}}\pm\vec{T}^{\E{5}}_3)\vert \leq \frac{1}{2}\vert\vec{T}^{\E{5}}_{\A{2}}\pm\vec{T}^{\E{5}}_3\vert^2$, automatically enforced given that $\vec{T}^{\E{5}}_{\A{1}}\cdot\vec{T}^{\E{5}}_3=0$ and $\vec{T}^{\E{5}}_{\A{2}}\cdot\vec{T}^{\E{5}}_3=0$ (note that it is $\vec{T}^{\E{5}}_3$ that appears here, not $\vec{T}^{\E{5}}_B$);
    \item  $\vert\vec{T}^{\E{5}}_{\A{2}}\cdot(\vec{T}^{\E{5}}_3\pm\vec{T}^{\E{5}}_{\A{1}})\vert \leq \frac{1}{2}\vert\vec{T}^{\E{5}}_3\pm\vec{T}^{\E{5}}_{\A{1}}\vert^2$, automatically enforced given that $\vec{T}^{\E{5}}_{\A{1}}\cdot\vec{T}^{\E{5}}_3=0$ and $\vec{T}^{\E{5}}_{\A{2}}\cdot\vec{T}^{\E{5}}_3=0$ (note that it is $\vec{T}^{\E{5}}_3$ that appears here, not $\vec{T}^{\E{5}}_B$).
\end{enumerate}

In terms of the parameters these conditions are the same as those for \E{4}, so the necessary conditions become:
\begin{enumerate}
    \item $0 < L_A$, $0 < \vert \beta \vert \leq\pi/2$ so that $0 \neq \sin\beta$, and $0\leq\gamma<2\pi$;
    \item $0< L_A/L_B < \infty $;
    \stepcounter{enumi}
    \item $\cos\beta \vert \cos\gamma\vert \leq \frac{1}{2} L_A / L_B$;
    \item $\cos\beta \vert \cos(\gamma+\pi/3)\vert \leq \frac{1}{2} L_A / L_B$;
    \item $\cos\beta \vert \cos(\gamma-\pi/3)\vert \leq \frac{1}{2} L_A / L_B$
    and
    $\cos\beta \vert \cos(\gamma+\pi/6) \vert \leq \frac{\sqrt{3}}{2} L_A / L_B$.
\end{enumerate}

These conditions can be written compactly as $0 < L_A$, $0 \neq L_{Bz}$, and
\begin{enumerate}
    \item $-\frac{1}{2} L_A < L_{Bx} \leq 0$ and $-L_A + L_{Bx} < \sqrt{3} L_{By} \leq L_A + 3 L_{Bx}$, or
    \item $0 < L_{Bx} \leq \frac{1}{2} L_A$ and $-L_A + 3 L_{Bx} < \sqrt{3} L_{By} \leq L_A + L_{Bx}$.
\end{enumerate}

\subsection{\E{6}: Hantzsche-Wendt space}
\label{secn:topologyE6}

\textit{Properties}: As listed in \cref{tab:properties}, this manifold is compact, orientable, inhomogeneous, and anisotropic (for more information see \rcite{Aurich:2014sea}). \\

\noindent \textit{Generators}: In general (see \cref{app:E6}) the generators of \E{6} can be written as
\begin{align}
    \label{eqn:E6generalT}
    & \mat{M}^{\E{6}}_{A} = 
    \begin{pmatrix}
        1 & \hphantom{-}0 & \hphantom{-}0\\
        0 & -1 & \hphantom{-}0\\
        0 & \hphantom{-}0 & -1
    \end{pmatrix}, \quad
        \mat{M}^{\E{6}}_{B} = 
    \begin{pmatrix}
        -1 & \hphantom{-}0 & \hphantom{-}0\\
        \hphantom{-}0 & \hphantom{-}1 & \hphantom{-}0\\
        \hphantom{-}0 & \hphantom{-}0 & -1
    \end{pmatrix} , \quad
        \mat{M}^{\E{6}}_{C} = 
    \begin{pmatrix}
        -1 & \hphantom{-}0 & \hphantom{-}0\\
        \hphantom{-}0 & -1 & \hphantom{-}0\\
        \hphantom{-}0 & \hphantom{-}0 & \hphantom{-}1
    \end{pmatrix}, \quad \mbox{with} \nonumber \\
    & \vec{T}^{\E{6}}_{A} = \begin{pmatrix} L_{Ax} \\ L_{By} + L_{Cy} \\ L_{Az} \end{pmatrix} , \quad
    \vec{T}^{\E{6}}_{B} = \begin{pmatrix} L_{Bx} \\ L_{By} \\ L_{Cz} + L_{Az} \end{pmatrix} , \quad
    \vec{T}^{\E{6}}_{C} = \begin{pmatrix} L_{Ax} + L_{Bx} \\ L_{Cy} \\ L_{Cz} \end{pmatrix} .
\end{align}
Alternatively, we can reparametrize some of the lengths via
\begin{equation}
    L_{Bx} = \left( r_x - \frac{1}{2} \right) L_{Ax}, \quad
    L_{Cy} = \left( r_y - \frac{1}{2} \right) L_{By}, \quad
    L_{Az} = \left( r_z - \frac{1}{2} \right) L_{Cz},
\end{equation}
so that the translation vectors become
\begin{equation}
    \label{eqn:E6generalTalt}
    \vec{T}^{\E{6}}_{A} = \begin{pmatrix} L_{Ax} \\ (r_y + \frac{1}{2})L_{By} \\ (r_z- \frac{1}{2})L_{Cz} \end{pmatrix} , \quad
    \vec{T}^{\E{6}}_{B} = \begin{pmatrix} (r_x- \frac{1}{2})L_{Ax} \\ L_{By} \\ (r_z + \frac{1}{2})L_{Cz} \end{pmatrix} , \quad
    \vec{T}^{\E{6}}_{C} = \begin{pmatrix} (r_x + \frac{1}{2})L_{Ax} \\ (r_y- \frac{1}{2})L_{By} \\ L_{Cz} \end{pmatrix} .
\end{equation}
\\
\noindent \textit{Associated \E{1}}: Since $(\mat{M}^{\E{6}}_A)^2 = (\mat{M}^{\E{6}}_B)^2 = (\mat{M}^{\E{6}}_C)^2 = \identity$ we define three independent translations as
\begin{align}
    \label{eqn:E6assocE1}
    g^{\E{6}}_1 & \equiv (g^{\E{6}}_A)^2: \vec{x} \to \vec{x} + \vec{T}^{\E{6}}_1, \nonumber \\
    g^{\E{6}}_2 & \equiv (g^{\E{6}}_B)^2: \vec{x} \to \vec{x} + \vec{T}^{\E{6}}_2, \\
    g^{\E{6}}_3 & \equiv (g^{\E{6}}_C)^2: \vec{x} \to \vec{x} + \vec{T}^{\E{6}}_3, \nonumber
\end{align}
for
\begin{equation}
    \vec{T}^{\E{6}}_1 \equiv \begin{pmatrix} 2 L_{Ax} \\ 0 \\ 0 \end{pmatrix}, \quad
    \vec{T}^{\E{6}}_2 \equiv \begin{pmatrix} 0 \\ 2 L_{By} \\ 0 \end{pmatrix}, \quad
    \vec{T}^{\E{6}}_3 \equiv \begin{pmatrix} 0 \\ 0 \\ 2 L_{Cz} \end{pmatrix}.
\end{equation}
\\
\noindent \textit{Volume}:
\begin{equation}
    \label{eqn:VE6}
    V_{\E{6}}= \frac{1}{4}
    \vert(\vec{T}^{\E{6}}_{{1}}\times\vec{T}^{\E{6}}_{{2}})\cdot\vec{T}^{\E{6}}_{3}\vert 
    = 2 \vert L_{Ax} L_{By} L_{Cz} \vert .
\end{equation}
\\
\noindent \textit{Tilts versus origin position}: Similar to \E{2}, a shift of the origin affects the tilts of the translation vectors.
Here since there are rotations around all three axes, a shift of origin will affect all three of the translation vectors.
In this case, $\vec{x}^{\E{6}}_0$ can replace $L_{Bx}$, $L_{Cy}$, and $L_{Az}$, or equivalently $r_x$, $r_y$, and $r_z$.
This leads to the translation vectors with respect to the special origin given by
\begin{equation}
    \vec{T}^{\E{6}}_A = \begin{pmatrix} L_{Ax} \\ L_{By} \\ 0 \end{pmatrix}, \quad
    \vec{T}^{\E{6}}_B = \begin{pmatrix} 0 \\ L_{By} \\ L_{Cz} \end{pmatrix}, \quad
    \vec{T}^{\E{6}}_C = \begin{pmatrix} L_{Ax} \\ 0 \\ L_{Cz} \end{pmatrix}, \quad \mbox{[special origin]}
\end{equation}
at the expense of the observer no longer being at the origin. \\

\noindent \textit{Real parameters (6 independent)}: There are 6 independent parameters required to fully define \E{6}.
As in \E{2}, some are redundant with shifting the origin.
Thus we have:
\begin{itemize}
    \item $L_{Ax}$, $L_{By}$, and $L_{Cz}$ are intrinsic parameters of the manifold;
    \item $L_{Bx}$, $L_{Cy}$, and $L_{Az}$, or equivalently $r_{x}$, $r_{y}$, and $r_{z}$, can be traded for $\vec{x}^{\E{6}}_0$;
    \item the standard (special origin, i.e., ``untilted'') form is $0 = L_{Bx} = L_{Cy} = L_{Az}$, or equivalently $1/2 = r_x = r_y = r_z$.
\end{itemize}

\noindent \textit{Parameter ranges}: We want to ensure that we do not double-count parameter choices that appear different but actually generate the same lattice of clones.
Although most of the rotational freedom was used to set the three orthogonal rotation axes as the coordinate axes, there remains the freedom to perform a half turn (rotation by $\pi$) about any two of the axes.
With this freedom, we can use a rotation by $\pi$ around the $y$-axis to always have $0 < L_{Ax}$ and a rotation by $\pi$ around the $x$-axis to always have $0 < L_{By}$.
Further, we have the freedom to order the axes by the lengths of the associated \E{1} vectors.
Similar to \E{1}, we therefore require:\footnote{
   Note that because $\vec{T}^{\E{6}}_1$, $\vec{T}^{\E{6}}_2$, and $\vec{T}^{\E{6}}_3$ are orthogonal to one another, we do not need to require that $\vec{T}^{\E{6}}_a$ ($a=A,B,C$) cannot be shortened by adding/subtracting linear combinations of $\vec{T}^{\E{6}}_i$ ($i=1,2,3$).
}
\begin{enumerate}
    \item $0 < L_{A x}$, $0 < L_{By}$, and $0 \neq L_{Cz}$, i.e., choice of orientation;
    \item $0 < \vert \vec{T}^{\E{6}}_1\vert/\vert \vec{T}^{\E{6}}_3\vert  
        \leq \vert \vec{T}^{\E{6}}_1\vert/\vert \vec{T}^{\E{6}}_2\vert \leq 1$, i.e., choice of ordering;
    \item $\vert\vec{T}^{\E{6}}_A\cdot\vec{T}^{\E{6}}_2\vert \leq \frac{1}{2}\vert\vec{T}^{\E{6}}_2\vert^2$, i.e., $\vec{T}^{\E{6}}_A$ cannot be shortened by adding or subtracting $\vec{T}^{\E{6}}_2$;
    \item $\vert\vec{T}^{\E{6}}_A\cdot\vec{T}^{\E{6}}_3\vert \leq \frac{1}{2}\vert\vec{T}^{\E{6}}_3\vert^2$ , i.e., $\vec{T}^{\E{6}}_A$ cannot be shortened by adding or subtracting $\vec{T}^{\E{6}}_3$;
    \item $\vert\vec{T}^{\E{6}}_B\cdot\vec{T}^{\E{6}}_1\vert \leq \frac{1}{2}\vert\vec{T}^{\E{6}}_1\vert^2$, i.e., $\vec{T}^{\E{6}}_B$ cannot be shortened by adding or subtracting $\vec{T}^{\E{6}}_1$;
    \item $\vert\vec{T}^{\E{6}}_B\cdot\vec{T}^{\E{6}}_3\vert \leq \frac{1}{2}\vert\vec{T}^{\E{6}}_3\vert^2$, i.e., $\vec{T}^{\E{6}}_B$ cannot be shortened by adding or subtracting $\vec{T}^{\E{6}}_3$;
    \item $\vert\vec{T}^{\E{6}}_C\cdot\vec{T}^{\E{6}}_1\vert \leq \frac{1}{2}\vert\vec{T}^{\E{6}}_1\vert^2$, i.e., $\vec{T}^{\E{6}}_C$ cannot be shortened by adding or subtracting $\vec{T}^{\E{6}}_1$;
    \item $\vert\vec{T}^{\E{6}}_C\cdot\vec{T}^{\E{6}}_2\vert \leq \frac{1}{2}\vert\vec{T}^{\E{6}}_2\vert^2$, i.e., $\vec{T}^{\E{6}}_C$ cannot be shortened by adding or subtracting $\vec{T}^{\E{6}}_2$.
\end{enumerate}
In terms of the parameters, a direct application of these conditions gives:
\begin{enumerate}
    \item $0 < L_{Ax} \leq L_{By}$ and $0 \neq L_{Cz}$ (including the next condition);
    \item $0 < L_{Ax} / \vert L_{Cz} \vert \leq L_{Ax} / L_{By} \leq 1$;
    \item $-2 L_{By} < L_{Cy} \leq 0$;
    \item $\vert L_{Az} \vert \leq \vert L_{Cz} \vert$;
    \item $\vert L_{Bx} \vert \leq L_{Ax}$;
    \item $-2 \vert L_{Cz} \vert < L_{Az} < 2 \vert L_{Cz} \vert$;
    \item $-2 L_{Ax} < L_{Bx} \leq 0$;
    \item $\vert L_{Cy} \vert \leq L_{By}$.
\end{enumerate}
These conditions are not all independent and can be written more compactly as:
\begin{enumerate}
    \item $0 < L_{Ax} \leq L_{By} \leq \vert L_{Cz} \vert$;
    \item $-L_{Ax} < L_{Bx} \leq 0$;
    \item $-L_{By} < L_{Cy} \leq 0$;
    \item $\vert L_{Az} \vert \leq \vert L_{Cz} \vert$.
\end{enumerate}
Finally, in terms of the alternative form parameters these conditions (2)--(4) are equivalent to
\begin{equation}
    \{ r_x, r_y, r_z \} \in (-1/2, 1/2].
\end{equation}

\subsection{\E{11}: Chimney space}
\label{secn:topologyE11}

The chimney space \E{11} is the basis for all the Euclidean manifolds with two compact dimensions (i.e., compact cross-sections), in much the same way as \E{1} is for all the compact manifolds.
It can be thought of \E{1} with one non-compact dimension.
All of \E{12}--\E{15} are roots of \E{11}. \\

\noindent \textit{Properties}:
As listed in \cref{tab:properties}, this manifold has compact cross-sections and is orientable, homogeneous, and anisotropic. \\

\noindent \textit{Generators}: Since \E{11} only has two compact dimensions, it is described by two generators.
In general (see \cref{app:E11}) the $z$ direction is chosen to be non-compact and the generators of \E{11} are given by
\begin{align} 
    \label{eqn:E11generalT}
    & \mat{M}^{\E{11}}_A = \identity, \quad\mbox{with} \nonumber \\
    & \vec{T}^{\E{11}}_{\A{1}} = \begin{pmatrix} L_{\A{1}x} \\ 0 \\ 0 \end{pmatrix} \equiv \vec{T}^{\E{11}}_{1} , \quad
    \vec{T}^{\E{11}}_{\A{2}} = \begin{pmatrix} L_{\A{2}x} \\ L_{\A{2}y} \\ 0 \end{pmatrix} \equiv \vec{T}^{\E{11}}_{2}, \\
    \intertext{or alternately}
    & \vec{T}^{\E{11}}_{\A{1}} = L_{\A{1}} \begin{pmatrix} 1 \\ 0 \\ 0 \end{pmatrix}, \quad
    \vec{T}^{\E{11}}_{\A{2}} = L_{\A{2}} \begin{pmatrix} \cos\alpha \\ \sin\alpha \\ 0 \end{pmatrix}. \nonumber
\end{align}
Similar to \E{1}, it is often convenient to simplify notation by dropping the $A$ label and instead use
\begin{align}
    L_{iw} &\equiv L_{\A{i}w}, \quad \mbox{for } i\in \{1,2\}, \, w\in\{x,y,z\}; \nonumber \\
    L_{i} &\equiv L_{\A{i}}, \quad \mbox{for } i\in \{1,2\} .
\end{align}
\\
\noindent\textit{Cross-sectional area}: 
Since the chimney spaces have two compact dimensions their volumes are infinite, but their cross-sections perpendicular to the non-compact direction are finite:
\begin{equation}
    \label{eqn:AE11}
    A_{\E{11}} = \vert \vec{T}^{\E{11}}_1 \times \vec{T}^{\E{11}}_2 \vert = L_1 L_2 \vert \sin\alpha \vert.
\end{equation}
\\
\noindent\textit{Origin}:
Since $\mat{M}^{\E{11}}_A=\identity$ the manifold is homogeneous and the lattice of an observer is independent of the location of the observer. \\

\noindent \textit{Real parameters (3 independent)}: There are 3 independent parameters to fully define \E{11}.
Since \E{11} is homogeneous they are all required, none can be traded for shifts of the origin. \\

\noindent\textit{Parameter ranges}:  We want to ensure that we do not double-count parameter choices that appear different but actually generate the same lattice of clones.
The constraints on the parameter ranges are similar to those in \E{1}, though simplified due to \E{11} having one less compact dimension.
The two translation vectors $\vec{T}^{\E{11}}_{\A{j}}$ can be used to define the $xy$-plane.
They can be ordered such that $\vert \vec{T}^{\E{11}}_{\A{1}} \vert \leq \vert \vec{T}^{\E{11}}_{\A{2}} \vert$.
Finally, we can rotate around the $x$-axis to always choose $0 < L_{\A{2}y}$.
With this we have:
\begin{enumerate}
    \item $0< L_{\A{1}x} = L_{\A{1}}$ and $0 < L_{\A{2}y}$, i.e., choice of orientation;
    \item $0 \leq \vert\vec{T}^{\E{11}}_{\A{1}}\vert/\vert\vec{T}^{\E{11}}_{\A{2}}\vert \leq 1$, i.e., choice of ordering;
    \item $\vert\vec{T}^{\E{11}}_{\A{2}}\cdot\vec{T}^{\E{11}}_{\A{1}}\vert \leq \frac{1}{2}\vert\vec{T}^{\E{11}}_{\A{1}}\vert^2$, i.e., $\vec{T}^{\E{11}}_{\A{2}}$ cannot be shortened by adding or subtracting $\vec{T}^{\E{11}}_{\A{1}}$.
\end{enumerate}
In terms of the parameters the conditions become:
\begin{enumerate}
    \item $0 < \alpha \leq \pi/2$ so that $0 < \sin\alpha \leq 1$; 
    \item $0<L_1 / L_2 \leq 1$;
    \item $\vert\cos\alpha\vert \leq \frac{1}{2} L_1 / L_2$.
\end{enumerate}

\noindent \textit{Convenient fundamental domain:}
A convenient choice of FD in the $xy$-plane is an infinite ``cylinder'' with a parallelogram cross-section in any constant-$z$ plane, the vertices of which are any base point and three clones.

There are two convenient choices of base point:
\begin{enumerate}[(A)]
    \item Origin-centered FD.
    \begin{itemize}
        \item Base point:
        $\vec{x}_A\equiv\frac{1}{2}(-\vec{T}^{\E{11}}_{\A{1}}-\vec{T}^{\E{11}}_{\A{2}})$;
        \item Three other corners of the face:
        \begin{itemize}
            \item $\vec{x}_B=g^{\E{11}}_{\A{1}}\vec{x}_A=\frac{1}{2}(\vec{T}^{\E{11}}_{\A{1}}-\vec{T}^{\E{11}}_{\A{2}});$
           \item $\vec{x}_C=g^{\E{11}}_{\A{2}}g^{\E{11}}_{\A{1}}\vec{x}_A=\frac{1}{2}(\vec{T}^{\E{11}}_{\A{1}}+\vec{T}^{\E{11}}_{\A{2}});$
            \item $\vec{x}_D=g^{\E{11}}_{\A{2}}\vec{x}_A=\frac{1}{2}(-\vec{T}^{\E{11}}_{\A{1}}+\vec{T}^{\E{11}}_{\A{2}})
            $.
        \end{itemize}
    \end{itemize}
\item Origin-rooted FD.
    \begin{itemize}
        \item Base point:
        $\vec{x}_A\equiv\vec{0}$;
        \item Three other corners of the face:
        \begin{itemize}
            \item $\vec{x}_B=g^{\E{11}}_{\A{1}}\vec{x}_A=\vec{T}^{\E{11}}_{\A{1}}$;
           \item $\vec{x}_C=g^{\E{11}}_{\A{2}}g^{\E{11}}_{\A{1}}\vec{x}_A=\vec{T}^{\E{11}}_{\A{1}}+\vec{T}^{\E{11}}_{\A{2}}$;
            \item $\vec{x}_D=g^{\E{11}}_{\A{2}}\vec{x}_A=\vec{T}^{\E{11}}_{\A{2}}
            $.
        \end{itemize}
    \end{itemize}
\end{enumerate}

\subsection{\E{12}: Chimney space with half turn}
\label{secn:topologyE12}

The chimney space with half turn is a root of \E{11} and can be thought of as \E{2} with one non-compact dimension. \\

\noindent \textit{Properties}:
As listed in \cref{tab:properties}, this manifold has compact cross-sections and is orientable, inhomogeneous, and anisotropic. \\

\noindent \textit{Generators}:
Similar to \E{11}, there are two generators, and similar to \E{2}, one of the matrices is a rotation by $\pi$.
Conventionally this rotation is chosen to be around the $y$-axis (cf.\ \rcite{Riazuelo2004:prd}).
Here we instead choose the rotation to be around the $z$-axis, as is done in \E{2}.
This makes it clear that \E{12} is the limit of \E{2} with $ \vert\vec{T}^{\E{2}}_{\A{2}} \vert \to \infty$.
In general (see \cref{app:E12}), the generators of \E{12} can be written as
\begin{align} 
    \label{eqn:E12generalT}
    & \mat{M}^{\E{12}}_A = \identity, \quad \mat{M}^{\E{12}}_B = \mat{R}_{\unitvec{z}}(\pi) = \diag(-1, -1, 1), \quad \mbox{with} \nonumber \\
    & \vec{T}^{\E{12}}_A = \begin{pmatrix} L_{A x} \\ 0 \\ 0 \end{pmatrix} \equiv \vec{T}^{\E{12}}_{1} , \quad
    \vec{T}^{\E{12}}_{B} = \begin{pmatrix} L_{Bx} \\ L_{By} \\ L_{Bz} \end{pmatrix},
\end{align}
or alternately,
\begin{equation}
    \vec{T}^{\E{12}}_{A} = L_{A} \begin{pmatrix} 1 \\ 0 \\ 0 \end{pmatrix}, \quad
    \vec{T}^{\E{12}}_{B} = L_{B} \begin{pmatrix} \cos\beta\cos\gamma \\ \cos\beta\sin\gamma \\ \sin\beta \end{pmatrix}.
\end{equation}
\\
\noindent \textit{Associated \E{11}}: In addition to $\vec{T}^{\E{12}}_1$ defined above, a second independent translation is
\begin{equation}
    ( g^{\E{12}}_B )^2: \vec{x} \to \vec{x} + \vec{T}^{\E{12}}_2,
\end{equation}
for
\begin{equation}
    \label{eqn:E12assocE11}
    \vec{T}^{\E{12}}_2 \equiv \begin{pmatrix} 0 \\ 0 \\ 2 L_{Bz} \end{pmatrix}
        = 2 L_B \begin{pmatrix} 0 \\ 0 \\ \sin\beta \end{pmatrix}\,.
\end{equation}
\\
\noindent\textit{Cross-sectional area}:
\begin{equation}
    \label{eqn:AE12}
    A_{\E{12}} = \frac{1}{2} \vert \vec{T}^{\E{12}}_1 \times \vec{T}^{\E{12}}_2 \vert = L_A L_B \vert \sin\beta \vert.
\end{equation}
\\
\noindent\textit{Tilts versus origin position}: A shift of origin will change $L_{Bx}$ and $L_{By}$ (the two components of $\vec{T}^{\E{12}}_B$ perpendicular to the axis of rotation) but will not affect $L_{Bz}$.
By special choice of origin on the axis of rotation, the tilt can be traded for a shift and we can choose $\vec{T}^{\E{12}}_B = L_B \unitvec{e}_z$ at the expense of the observer no longer being at the origin. \\

\noindent \textit{Real parameters (4 independent)}: There are 4 independent parameters required to fully define \E{12} with some being redundant with shifting the origin.
Thus we have:
\begin{itemize}
    \item $L_A$ and $L_A / L_B$ are intrinsic parameters of the manifold;
    \item $L_{Bx}$ and $L_{By}$, or equivalently $\beta$ and $\gamma$, can be traded for $x^{\E{12}}_{0x}$ and $x^{\E{12}}_{0y}$;
    \item the standard (special origin, i.e., ``untilted'') form is $0 = L_{Bx} = L_{By}$, or equivalently, $\beta = \pi/2$ (with $\gamma$ irrelevant).
\end{itemize}

\noindent \textit{Parameter ranges}:  We want to ensure that we do not double-count parameter choices that appear different but actually generate the same lattice of clones.
Similar to \E{11}, we therefore require:
\begin{enumerate}
    \item $0< L_A$ and $0 < L_{Bz}$, i.e., choice of orientation;
    \item the lengths of $\vec{T}^{\E{12}}_A$ and $\vec{T}^{\E{12}}_B$ are unconstrained;
    \item $\vert\vec{T}^{\E{12}}_B\cdot\vec{T}^{\E{12}}_A\vert \leq \frac{1}{2}(\vec{T}^{\E{12}}_A)^2$, i.e., $\vec{T}^{\E{12}}_B$ cannot be shortened by adding or subtracting $\vec{T}^{\E{12}}_A$.
\end{enumerate}
In terms of the parameters, the necessary conditions become: 
\begin{enumerate}
    \item $0 < \beta \leq \pi/2$ so that $0 < \cos\beta \leq 1$;
    \stepcounter{enumi}
    \item $\vert L_{Bx} \vert \leq \frac{1}{2} L_{Ax}$, or equivalently, $\cos\beta \vert\cos\gamma\vert \leq \frac{1}{2} L_A / L_B$.
\end{enumerate}

\subsection{\E{16}: Slab space including rotation}
\label{secn:topologyE16}

The slab space \E{16} is the basis for all Euclidean three-manifolds with one compact dimension (i.e., compact lengths), in much the same way as \E{1} and \E{11} are for all compact and two compact dimensions, respectively.
The possibility of having a corkscrew (as opposed to a pure translation) in the slab space 
appears to be new, at least in the cosmology literature. 
While topologically the corkscrew is continuously deformable to the unrotated slab space, physically the corkscrew leads to a distinguishable pattern of clones.  
Due to this, we split the description of \E{16} into two cases: \slabh\ and \slabi. \\

\subsubsection{\slabh: Conventional unrotated slab space}

The conventional definition of \E{16} only includes a translation.
Here we call this choice \slabh, because the space is homogeneous.\\

\noindent \textit{Properties:} As listed in \cref{tab:properties}, this manifold has a compact length and is orientable, homogeneous, and anisotropic. \\

\noindent \textit{Generators}: In general, since \E{16} has one compact dimension it is described by one generator, which we may take to be a translation in the $z$ direction (see \cref{app:E16h}), so the generator of \slabh\ is
\begin{equation} 
    \label{eqn:E16hgeneralT}
    \mat{M}^{\slabh}_A = \identity, \quad \mbox{with} \quad \vec{T}^{\slabh}_A = L\begin{pmatrix} 0 \\ 0 \\ 1 \end{pmatrix}.
\end{equation}
Even though there is only one generator, since this generator is a pure translation, we follow the convention of using $A$ to label it.\\

\noindent \textit{Length}: Since the slab spaces have only one compact dimension their volumes and cross-sectional areas are infinite.
The shortest path length around the manifold at any point is $L$. \\

\noindent \textit{Origin}: Since $\mat{M}^{\slabh}_A = \identity$ the manifold is homogeneous and the lattice of an observer is independent of the location of the observer. \\

\noindent \textit{Real parameters (1 independent)}: There is 1 independent parameter required to fully define \slabh: the length of the compact dimension. \\

\noindent \textit{Parameter ranges}: We want to ensure that we do not double-count parameter choices that appear different but actually generate the same lattice of clones.
In this case, we can always choose $0 < L$ through the orientation of the coordinate axes.

\subsubsection{\slabi: General rotated slab space}

The orientable slab space also allows for a corkscrew motion.
Physically this corkscrew is distinguishable and must be treated as a separate case. As discussed in \cref{app:E16i}, the rotation angle must be a rational multiple of $2\pi$ in order for the eigenmodes of the Laplacian not to have azimuthal symmetry around the corkscrew axis.
Such a symmetry would exclude them as a basis for general smooth functions on the manifold.
\\

\noindent \textit{Properties}: Due to the corkscrew motion this differs from \slabh\ in that it is inhomogeneous.
As listed in \cref{tab:properties}, this manifold has a compact length and is orientable, inhomogeneous, and anisotropic. \\

\noindent \textit{Generators}: Similar to \slabh, there is one generator.
In general the generator of \slabi\ (see \cref{app:E16i}) can be written as
\begin{align} 
    \label{eqn:E16igeneralT}
    &\mat{M}^{\slabi}_B = \mat{R}_{\unitvec{z}}(2\pi p/q)
     = \begin{pmatrix} \cos(2\pi p/q) & -\sin(2\pi p/q) & 0 \\
       \sin(2\pi p/q) & \hphantom{-}\cos(2\pi p/q) & 0 \\
       0 & \hphantom{-}0 & 1 \end{pmatrix} ,
       \quad \mbox{with}  \nonumber \\
     &\vec{T}^{\slabi}_B = \begin{pmatrix} L_{x} \\ 0 \\ L_{z} \end{pmatrix} = L\begin{pmatrix} \cos\beta \\ 0 \\ \sin\beta \end{pmatrix} ,
\end{align}
and $p\in \integers^{\neq 0}$, $q\in \integers^{>0}$, and $|p|$ and $q$ relatively prime.
As in \slabh, here since the generator is a rotation we use $B$ to label it.
\\

\noindent \textit{Associated \slabh}: A pure translation can be defined for \slabi\ as
\begin{equation}
    \label{eqn:E16iassocE16h}
    g^{\slabi}_1 \equiv (g^{\slabi}_B)^q: \vec{x} \to \vec{x} + \vec{T}^{\slabi}_1, \quad \mbox{for }
    \vec{T}^{\slabi}_1 \equiv \begin{pmatrix} 0 \\ 0 \\ q L_{z} \end{pmatrix} = q L \sin\beta \begin{pmatrix} 0 \\ 0 \\ 1 \end{pmatrix} .
\end{equation}
\\
\noindent \textit{Length}: The length of the associated \slabh\ is 
\begin{equation}
    \label{eqn:LE16i}
    L_{\slabi} = q L \vert \sin\beta \vert.
\end{equation}
\\
\noindent \textit{Tilts versus origin position}: A shift of origin will change $L_x$ and $L_y$.
Rotational freedom can always be used to restore $L_y=0$.
By special choice of origin on the axis of rotation, the tilt can be traded for a shift and we can choose $\vec{T}^{\slabi}_B = L \unitvec{e}_z$ at the expense of the observer no longer being at the origin. \\

\noindent \textit{Real parameters (2 independent)}: There are 2 independent parameters required to fully define \slabi\ with 1 parameter interchangeable with a shift of origin.
Thus we have:
\begin{itemize}
    \item $L$ is an intrinsic parameter of the manifold;
    \item $L_x$, or equivalently $\beta$, can be traded for $x^{\slabi}_{0x}$;
    \item $L_y=0$ is not a parameter, even though it can be traded for $x^{\slabi}_{0y}$; it is redundant with the orientation of the coordinate system, and thus the rotation of an observer's coordinate system, about the topology rotation axis ($\unitvec{e}_z$);
    \item the standard (special origin, i.e., ``untilted'') form is $0 = L_x$, or equivalently $\beta=\pi/2$.
\end{itemize}

\noindent \textit{Parameter ranges}: We want to ensure that we do not double-count parameter choices that appear different but actually generate the same lattice of clones. Similar to \slabh, we can always require $0 < L_x$ and $0 < L_z$, or equivalently $0 < \beta \leq \pi/2$, through orientation of the coordinate system.

\subsection{\E{18}: The covering space of \Espace}
\label{secn:topologyE18}

Three-dimensional Euclidean space \E{18} is the covering space of the \Espace\ geometry.
It is infinite in all directions and has no generators.
As listed in \cref{tab:properties}, it has no compact dimensions and is orientable, homogeneous, and the only Euclidean topology that is isotropic.

\section{Eigenmodes of the scalar Laplacian and correlation matrices}
\label{secn:eigenmodes}

A key ingredient of cosmological perturbation theory is the set of the scalar (and tensor) eigenmodes of the Laplacian.
Characteristically, it is the amplitudes of these modes for which theories give statistical predictions \cite{Lachieze-rey2005,Nelson2022EigenmodeAO}.

In this section, we present the scalar eigenmodes for the orientable Euclidean manifolds in their full generality.
While such eigenmodes have been presented before \cite{Inoue1999:cqg,Lehoucq2002:cqg, Lachieze-rey2005, Riazuelo2004:prd,Weeks2006:cqg}, it has been in a context where the full topology parameter space has not been included, even when its existence has been hinted at.  Note that we are not faithful to the notational conventions of those works, so any comparisons should be made carefully.

In the covering space, \E{18}, the eigenmodes of the Laplacian are\footnote{
    There are many conventions for normalizing the eigenmodes.
    Here we choose not to include any additional factors and will discuss the implications of this for each manifold below.
}
\begin{equation}
    \label{eqn:EuclideanFourierBasis}
    \Upsilon^{\E{18}}_{\vec{k}}(\vec{x}) = e^{i\vec{k}\cdot(\vec{x}-\vec{x}_0)}.
\end{equation}
Here $\vec{x}_0$ is the position of an arbitrary origin relative to the observer's coordinate system\footnote{
    The inclusion of $\vec{x}_0$ has no particular role for the covering space, for \E{1}, for \E{11}, or for \slabh, but is crucial
    for \E{2}--\E{6}, \E{12}, and \slabi.
    }
and $\vec{k}=\transpose{(k_x,k_y,k_z)}$, referred to as the wavevector, is any triplet of real numbers with its magnitude $k$ referred to as the wavenumber.
Since 
\begin{equation}
   \nabla^2 \Upsilon^{\E{18}}_{\vec{k}}(\vec{x}) = -\vert\vec{k}\vert^2 \Upsilon^{\E{18}}_{\vec{k}}(\vec{x}),
\end{equation}
the eigenvalue associated with $\Upsilon^{\E{18}}_{\vec{k}}$ is $-\vert\vec{k}\vert^2\equiv -k^2$.
It can assume any non-positive real value.

In standard inflationary cosmological theory, the adiabatic curvature perturbation field $\delta^{\mathcal{R}}$ is the sum of the eigenmodes $\Upsilon^{\E{18}}_{\vec{k}}(\vec{x})$ with amplitudes $\delta^{\mathcal{R}}(\vec{k})$ that are described by Gaussian random variables of zero mean and dimensionless power spectrum ${\mathcal{P}}^{\mathcal{R}}(k)$.\footnote{
   Some small amount of non-Gaussianity is often predicted, but we reserve such considerations in a topological context for future work.
}
We can write the resulting three-dimensional scalar field as
\begin{equation}
    \label{eqn:scalarfieldinE18Cartesian}
    \delta^{\mathcal{R}}(\vec{x}) = \int \frac{\mathrm{d}^3 k}{(2\pi)^3}\; \delta^{\mathcal{R}}(\vec{k}) \Upsilon^{\E{18}}_{\vec{k}}(\vec{x})\,.
\end{equation}
However, we will be interested in other scalar fields $\delta^X$ that are linearly related to $\delta^{\mathcal{R}}$
by a transfer function that we should write as $\Delta^{\mathcal{R}X}$, but we will drop the $\mathcal{R}$.
The expectation value of any pair of $\delta^X(\vec{k})$ is
\begin{equation}
    \label{eqn:CE18XY}
    C^{\E{18};XY}_{\vec{k} \vec{k}'} 
        \equiv \langle \delta^X(\vec{k}) \delta^{Y*}(\vec{k}')\rangle 
        = (2\pi)^3\frac{2\pi^2}{k^3} {\mathcal{P}}^{\mathcal{R}}(k) \Delta^{X}(k) \DeltaYstar(k')
        \Ddelta(\vec{k}-\vec{k}')\; ,
\end{equation}
where $\Ddelta (\vec{k}-\vec{k}')$ is the three-dimensional Dirac delta function and we have assumed that the transfer function depends only on the magnitude of $\vec{k}$.\footnote{
    For three-dimensional scalar quantities $X$ and $Y$ the transfer function will typically depend only on the magnitude of $\vec{k}$.  However, it may be useful to consider a more general dependence on $\vec{k}$, for example when deriving a transfer function of CMB temperature and polarization on the sky. We will continue to write $\Delta^{X}(k)$.
    }
For the adiabatic curvature $\delta^{\mathcal{R}}$, the primordial power spectrum is often written as\footnote{
    The normalization by ${2\pi^2}/{k^3}$ in \eqref{eqn:CE18XY} is a common convention, but not universal. In this convention, ${\mathcal{P}}^X(k)$ is the contribution to the variance per logarithmic interval of wavenumber: the total variance of $\delta^X$ is $\int \mathrm{d}(\ln k)\; {\mathcal{P}}^X(k)$. In the large-scale structure literature, the matter power spectrum is usually denoted by the quantity $P(k)=2\pi^2 {\mathcal{P}}(k)/k^3$.} 
\begin{equation}
{\mathcal{P}}^{\mathcal{R}}(k) = A_\mathrm{s} \left(\frac{k}{k_*}\right)^{n_\mathrm{s}-1}\;,
\end{equation}
with the scalar amplitude $A_\mathrm{s}$ defined at the fiducial wavenumber $k_*$, and the scalar spectral tilt $n_\mathrm{s}$.
We  assume throughout that ${\mathcal{P}}^{\mathcal{R}}$ is the same function for \E{1}--\E{18}, 
as might be expected to result, for example, from an epoch of inflation, and so do not add a topology label to ${\mathcal{P}}^{\mathcal{R}}$. 

Since Euclidean geometry has both translational and rotational isometry, there are other natural bases of the eigenmodes of the Laplacian.
Given the nature of cosmological observations, in particular those of the cosmic microwave background, it is more convenient to work in spherical coordinates $(r,\theta,\phi)$ with the plane waves expanded in terms of spherical harmonics as
\begin{equation}
    \label{eqn:planewaveinsphericalcoords}
    e^{i\vec{k}\cdot(\vec{x}-\vec{x}_0)} 
        = 4\pi e^{-i\vec{k}\cdot\vec{x}_0}\sum_{\ell m} i^\ell j_\ell(kr) Y_{\ell m}^*(\unitvec{k}) Y_{\ell m}(\theta, \phi),  
\end{equation}
where $j_\ell$ are the spherical Bessel functions and $Y_{\ell m}$ are the (scalar) spherical harmonics.
This allows us to always expand the eigenmode as
\begin{equation}
    \Upsilon^{\E{i}}_{\vec{k}}(\vec{x}) =  4\pi \sum_{\ell m} j_\ell(kr) \xi^{\E{i};\unitvec{k}}_{k \ell m} Y_{\ell m}(\theta, \phi).
\end{equation}
For \E{18}
\begin{equation}
    \xi^{\E{18};\unitvec{k}}_{k \ell m} = e^{-i\vec{k}\cdot\vec{x}_0} i^\ell  Y_{\ell m}^*(\unitvec{k}).
\end{equation}

In certain cases, e.g., CMB fluctuations, observations project the scalar field $\delta^X$ onto the sphere of the sky, integrating along the line of sight with an appropriate transfer function, so
\begin{equation}
    \label{eqn:observable}
    \delta^X (\theta, \phi) = \sum_{\ell m} a^{\E{18};X}_{\ell m} Y_{\ell m} (\theta, \phi),
\end{equation}
where
\begin{align}
    \label{eqn:almE18}
    a^{\E{18};X}_{\ell m} 
    &=  \frac{4\pi}{(2\pi)^3}
    \int \mathrm{d}^3k ~\delta^{\mathcal{R}}({\vec{k}})
    \xi^{\E{18};\unitvec{k}}_{k \ell m}
     \Delta^X_\ell(k).
\end{align}
Here $\Delta^X_\ell(k)$ is the spherical-harmonic transfer function from $\mathcal{R}$ to $X$,
and which, relative to $\Delta^X(\vec{k})$  absorbs the $j_\ell(kr)$ that contributed to the integrand of the radial integral.\footnote{As with $\Delta^X(k)$, $\Delta^X_\ell(k)$ is typically only a function of the wavevector magnitude $k$, but in more generality we can have $\Delta^X_\ell(\vec{k})$ dependent on the full three-dimensional wavevector.}

If, as usual, $X$ represents a real scalar quantity, then the spherical-harmonic coefficients satisfy $a_{\ell m}^* = (-1)^m a_{\ell -m}$ and we only obtain unique physical information about the $a_{\ell m}$ from $m\geq 0$.
In this paper, we will be particularly interested in the properties of the CMB temperature fluctuations, i.e., in $X=T$.

It is surprising, but easily proved, that the isotropy of \E{18} means that if the $\delta^X(\vec{k})$ are independent Gaussian random variables of zero mean, with variance only dependent upon the magnitude $k$, then the coefficients $a^X_{\ell m}$ are independent with variance only dependent upon $\ell$.
This leads to the customary statement of statistical isotropy,
\begin{equation}
    \label{eqn:SI-2d}
    C^{\E{18};XY}_{\ell m \ell' m'} \equiv 
    \langle a^X_{\ell m}a^{Y*}_{\ell' m'} \rangle = C^{XY}_{\ell} \Kdelta_{\ell\ell'}\Kdelta_{mm'},
\end{equation}
where $\Kdelta_{ij}$ is the Kronecker delta.

Non-trivial topological boundary conditions have two important effects on the eigenmodes of the Laplacian:
\begin{enumerate}
    \item Only certain wavevectors $\vec{k}$ are ``allowed'' by the boundary conditions.
    For the fully compact topologies \E{1}--\E{10}, the allowed wavevectors form a discrete lattice.
    We write $\delta^X_{\vec{k}}$, not $\delta^X(\vec{k})$.
    Thus the correlator $\langle\delta^X_{\vec{k}}\delta^{X*}_{\vec{k}'}\rangle$ contains terms involving ${\mathcal{P}}^X(k) \Kdelta_{\vec{k} \vec{k}'}$ (i.e., a Kronecker, rather than Dirac, delta, although there can be a mix of the two for the chimney and slab spaces with a mix of finite and infinite directions).
    \item Except for \E{1}, \E{11}, and \slabh, the eigenmodes are not single covering-space eigenmodes but instead linear combinations thereof, with different $\vec{k}$ of the same magnitude. This induces extra terms in the correlator coupling $\vec{k}$ to the generator's rotations of $\vec{k}$ with Kronecker or Dirac deltas.
\end{enumerate}
Each of these effects encodes the violation of statistical isotropy,
and each of them breaks the surprising connection presented above between the statistics of $\delta^X(\vec{k})$ and the statistics of $a^X_{\ell m}$.
Equations \eqref{eqn:CE18XY} and \eqref{eqn:SI-2d} no longer hold.
Instead of $C^{\E{18};XY}_{\vec{k} \vec{k}'}\equiv \langle \delta^X(\vec{k}) \delta^{Y*}(\vec{k}')\rangle $ being proportional to a Dirac delta function of $\vec{k}$ and $\vec{k}'$, it vanishes except for certain allowed $\vec{k}$ and generically connects all pairs of allowed $\vec{k}$ with correlations of equal magnitude and location-dependent phase.
Meanwhile, $\langle a^X_{\ell m} a^{Y*}_{\ell' m'}\rangle$ is also not diagonal:
\begin{equation}
    \label{eqn:noSI-2d}
    \langle a^X_{\ell m} a^{Y*}_{\ell' m'}\rangle = C^{XY}_{\ell m \ell' m'}.
\end{equation}
Despite the reality condition on the spherical-harmonic coefficients themselves, the quantity $C^{XY}_{\ell -m \ell'm'}=\langle a^X_{\ell -m} a^{Y*}_ {\ell' m'} \rangle =  (-1)^m \langle a^{X*}_ {\ell m} a^{Y*}_{\ell' m'} \rangle$ does contain independent information. Rather, the $C^{XY}_{\ell m\ell'm'}$ matrix is Hermitian in the $(X, \ell m),(Y, \ell'm')$ index sets.

In the subsections below, we present the eigenmodes and eigenspectra of the orientable Euclidean manifolds as functions of their topological parameters in their full generality.
Assuming that it is the amplitudes of these eigenmodes that are Gaussian random variables of zero mean and dimensionless power spectrum ${\mathcal{P}}^{\mathcal{R}}(k)$, we present the correlation matrices for Fourier-mode amplitudes $C^{\E{i};XY}_{\vec{k} \vec{k}'}$ and spherical-harmonic amplitudes $C^{\E{i};XY}_{\ell m \ell' m'}$.
The important results for each topology are boxed.
The generality of the results employs the orientation and other choices described in \cref{secn:topologiesmanifolds}, but also includes both an arbitrary origin for the definition of the manifold parameters and an arbitrary location for the observer.
As discussed in \cref{secn:topologiesmanifolds}, there are redundancies in these choices.
Any comprehensive search over parameters must take care to avoid overweighting some parts of parameter space.
In practice it is convenient to make one of two choices when employing the results below, either
\begin{enumerate}
    \item choose the observer to be at the origin, $\vec{x}_0 = \vec{0}$, and use the ``tilted'' parameters of the manifold, or
    \item choose the ``special'' origin for the coordinate system, in which case the manifold parameters are simplified, but some of the components of observer location, $\vec{x}_0$, become significant.
\end{enumerate}

\subsection{General considerations for eigenmodes}

In each of the manifolds, the eigenmodes of the scalar Laplacian $\Upsilon^{\E{i}}_{\vec{k}}(\vec{x})$ must be invariant under every possible  group transformation $G_\alpha \in \Gamma^{\E{i}}$:
\begin{equation}
    \label{eqn:generaleigenmodeinvariance}
    \Upsilon^{\E{i}}_{\vec{k}}(G_\alpha\vec{x}) = \Upsilon^{\E{i}}_{\vec{k}}(\vec{x}) \,.
\end{equation}
Formally, the solution is that $\Upsilon^{\E{i}}_{\vec{k}}(\vec{x})$ is a simple linear combination of all covering-space eigenmodes related by the group transformations 
\begin{equation}
    \Upsilon^{\E{i}}_{\vec{k}}(\vec{x}) \propto \sum_{G_\alpha \in \Gamma^{\E{i}}} e^{i\vec{k}\cdot G_\alpha\vec{x}} \,.
\end{equation}
More practically, we can limit the sum to a small, finite set of group elements $G_\alpha \in \mathcal{G}^{\E{i}}$,
\begin{equation}
    \label{eqn:generaleigenmodeformula}
    \Upsilon^{\E{i}}_{\vec{k}}(\vec{x}) = \frac{1}{\sqrt{N(\mathcal{G}^{\E{i}})}}\sum_{G_\alpha \in \mathcal{G}^{\E{i}}} e^{i\vec{k}\cdot G_\alpha\vec{x}} \,,
\end{equation}
where $N(\mathcal{G}^{\E{i}})$ is the number of elements in $\mathcal{G}^{\E{i}}$.
$\mathcal{G}^{\E{i}}$ includes one group element for each of the $SO(3)$ matrices $\mat{M}^{(G_\alpha)}$ that appears when we explicitly write the action of the group elements,
\begin{equation}
    \label{eqn:actionofGalpha}
    G_\alpha:(\vec{x}-\vec{x}_0)\to \mat{M}^{(G_\alpha)} (\vec{x}-\vec{x}_0) + \vec{v}^{(G_\alpha)}\,.
\end{equation}
These $\mat{M}^{(G_\alpha)}$ are then just the matrices $\mat{M}^{\E{i}}_a$ that appear in the generators, as described in \cref{secn:topologiesmanifolds-general}, plus all non-identical $SO(3)$ matrices that can be built from arbitrary products of those $\mat{M}^{\E{i}}_a$.
Below, we will present the $\mathcal{G}^{\E{i}}$ for each \E{i}.

Equation \eqref{eqn:generaleigenmodeinvariance} must still be satisfied for every group element $G_\alpha \in \Gamma^{\E{i}}$.
Among those group elements are a subgroup of pure translations, which are all the integer linear combinations of the $\vec{T}^{\E{i}}_j$, i.e., the translations of the associated homogeneous space (\E{1}, \E{11}, or \slabh) of that manifold.
Considering the invariance of $\Upsilon^{\E{i}}_{\vec{k}}(\vec{x})$ under the translation by $\vec{T}^{\E{i}}_j$, and recognizing that 
$\mathcal{G}^{\E{i}}$ always includes the identity matrix, we learn that one must have
\begin{equation}
      e^{i\vec{k}\cdot[(\vec{x}-\vec{x}_0)+\vec{T}^{\E{i}}_j]} = e^{i\vec{k}\cdot(\vec{x}-\vec{x}_0)}
\end{equation}
or more compactly
\begin{equation}
    \label{eqn:associatedhomogeneousdiscretization}
    \vec{k}\cdot\vec{T}^{\E{i}}_j = 2\pi n_j\,, \quad \mbox{for } n_j\in\integers.
\end{equation}
This is exactly the discretization condition that we get with an \E{1}, \E{11}, or \slabh.
In other words, the eigenmodes of the Laplacian on an \E{i} manifold are linear combinations of the Fourier modes that are eigenmodes of the associated homogeneous space.
For each \E{i} below, we present those discretization conditions.

Equation \eqref{eqn:generaleigenmodeformula} satisfies the invariance condition \eqref{eqn:generaleigenmodeinvariance} for all $\vec{k}$ allowed by 
\eqref{eqn:associatedhomogeneousdiscretization}, however in some cases the sum over $G_\alpha \in \mathcal{G}^{\E{i}}$ yields more than one identical term.
This occurs when $\mat{M}^{(G_\alpha)}\vec{k}=\vec{k}$ for certain $\vec{k}$ allowed by \eqref{eqn:associatedhomogeneousdiscretization}.
More specifically, for the manifolds in question, where $\mat{M}^{(G_\alpha)}$ are matrices $\mat{M}^{\E{i}}_a$  representing rotations by $2\pi p_a/q_a$ ($p_a\in\integers^{\neq 0}$, $q_a\in\integers^{> 0}$, $\vert p \vert$ and $q$ relatively prime) about one of the three coordinate axes, this occurs when $(\mat{M}^{\E{i}}_a)^{N_a}\vec{k} = \vec{k}$ has a solution for $N_a < q_a$.  We will consider those cases explicitly for each \E{i}.

\subsection{\E{1}: 3-torus}
\label{secn:eigenmodesE1}

The 3-torus is the simplest of the compact Euclidean topologies and will serve as a model for determining the eigenspectrum and eigenmodes of all the Euclidean three-manifolds.
In this subsection we determine which of the eigenvalues and eigenmodes of the scalar Laplacian acting on the  covering space \E{18} are preserved by the isometries of the topology.
We then use that information to present the Fourier space and spherical-harmonic space correlation matrices of any fluctuations that are linearly related to independent Gaussian random fluctuations of the amplitudes of those eigenmodes.

We begin with the covering-space (\E{18}) eigenmodes \eqref{eqn:EuclideanFourierBasis}.
Though in general the eigenmodes of \E{i} can be linear combinations of the \E{18} eigenmodes, 
the \E{1} eigenmodes are the subset of the \E{18} eigenmodes that respect the \E{1} symmetries, 
\begin{equation}
    \label{eqn:E1eigenmodeinvariance}
    \Upsilon^{\E{1}}_{\vec{k}}(g^{\E{1}}_{\A{j}}\vec{x}) = \Upsilon^{\E{1}}_{\vec{k}}(\vec{x}).
\end{equation}
This is because all the group elements of $\Gamma^{\E{1}}$ are pure translations, i.e.,
$\mat{M}^{(G_\alpha)}=\identity$ for all $G_\alpha\in \mathcal{G}^{\E{1}}$, 
so $\mat{M}^{(G_\alpha)}\vec{k}=\vec{k}$ trivially.

As discussed above in general (cf.\ \eqref{eqn:associatedhomogeneousdiscretization}),
the symmetry condition \eqref{eqn:E1eigenmodeinvariance} leads to the discretization of the allowed $\vec{k}$ in \E{1}:
\begin{align}
    2\pi n_1 &= (\vec{k}_{\vec{n}})_{x} L_1 , \nonumber \\
    2\pi n_2 &= (\vec{k}_{\vec{n}})_{x} L_2 \cos\alpha + (\vec{k}_{\vec{n}})_{y} L_2 \sin\alpha,  \\
    2\pi n_3 &= (\vec{k}_{\vec{n}})_{x} L_3 \cos\beta\cos\gamma
        + (\vec{k}_{\vec{n}})_{y} L_3 \cos\beta\sin\gamma 
        + (\vec{k}_{\vec{n}})_{z} L_3 \sin\beta .  \nonumber
\end{align}
Since the wavenumbers are now discretized, they are labeled by integers $n_i \in \integers$ and we denote this explicitly by writing the wavevector as $\vec{k}_{\vec{n}}$ for $\vec{n} = (n_1, n_2, n_3)$.
Here and below we will use either the $n_i$ or $(\vec{k}_{\vec{n}})_i$ labels as convenient for the situation.
Inverting these requirements, the components of the wavevectors are 
\begin{align}
    \label{eqn:E1_ki}
    (\vec{k}_{\vec{n}})_{x} &= \frac{2\pi n_1}{L_1} , \nonumber \\
    (\vec{k}_{\vec{n}})_{y} &= \frac{2\pi n_2}{L_2\sin\alpha} - \frac{2\pi n_1}{L_1}\frac{\cos\alpha}{\sin\alpha},\\
    (\vec{k}_{\vec{n}})_{z} &= \frac{2\pi n_3}{L_3\sin\beta}
        - 
        \frac{2\pi n_2}{L_2}
          \frac{\cos\beta\sin\gamma}{\sin\alpha\sin\beta} 
        - 
        \frac{2\pi n_1}{L_1}
          \frac{\cos\beta
            (\sin\alpha\cos\gamma-\cos\alpha\sin\gamma)
          }{\sin\alpha\sin\beta}.
          \nonumber    
\end{align}
Clearly the eigenvalues $k_{\vec{n}} = \vert\vec{k}_{\vec{n}}\vert$ are complicated functions of $\vec{n}$.

Thus
\begin{empheq}[box=\fbox]{equation}
    \Upsilon^{\E{1}}_{\vec{k}_{\vec{n}}}(\vec{x}) = e^{i\vec{k}_{\vec{n}}\cdot(\vec{x}-\vec{x}_0)}, \quad \mbox{for } \vec{n}\in \setN^{\E{1}},
\end{empheq}
where
\begin{empheq}[box=\fbox]{equation}
    \setN^{\E{1}} \equiv \{(n_1,n_2,n_3) \vert n_i\in\integers\}\setminus (0,0,0).
\end{empheq}

Following \eqref{eqn:CE18XY} the Fourier-mode correlation matrix for \E{1} is
\begin{empheq}[box=\fbox]{align}
    \label{eqn:CE1XY}
    C^{\E{1};XY}_{\vec{k}_{\vec{n}} \vec{k}_{\vecnp}} 
    &= V_{\E{1}}\frac{2\pi^2}{k_{\vec{n}}^3}{\mathcal{P}}^{\mathcal{R}}(k_{\vec{n}}) 
    \Delta^X(k_{\vec{n}}) \DeltaYstar(k_{\vec{n}}) 
    \Kdelta_{\vec{k}_{\vec{n}}\vec{k}_{\vecnp}}\;.
\end{empheq}
In transitioning from the covering space \E{18} we have replaced $(2\pi)^3\Ddelta(\vec{k}-\vec{k}')$ with $V_{\E{1}}\Kdelta_{\vec{k}_{\vec{n}}\vec{k}_{\vecnp}}$, where the volume factor $V_{\E{1}}$ is given by \eqref{eqn:E1volume}.

As for \E{18} above, we can project the field $\delta^X$ onto the sky by performing a radial integral with suitable weight function and transfer function, giving
\begin{align}
    \label{eqn:almE1}
    a^{\E{1};X}_{\ell m} 
    &=  \frac{4\pi}{V_{\E{1}}}
    \sum_{\vec{n}\in\setN^{\E{1}} } \delta^{\mathcal{R}}_{\vec{k}_{\vec{n}}} 
     \xi^{\E{1};\unitvec{k}_{\vec{n}}}_{k_{\vec{n}} \ell m}
     \Delta^X_\ell(k_{\vec{n}}),
\end{align}
with
\begin{empheq}[box=\fbox]{equation}
    \xi^{\E{1};\unitvec{k}_{\vec{n}}}_{k_{\vec{n}} \ell m} \equiv e^{-i\vec{k}_{\vec{n}}\cdot\vec{x}_0} i^\ell  Y_{\ell m}^*(\unitvec{k}_{\vec{n}}).
\end{empheq}
Because $\setN^{\E{1}}$ labels only a discrete set of $\vec{k}_{\vec{n}}$, the integral over $\mathrm{d}^3 k$ in \cref{eqn:almE18} is replaced by a sum over $\vec{n}\in\setN^{\E{1}}$.

For the compact topologies \E{i} with $i \in \{1, \ldots, 6\}$, the spherical-harmonic space covariance matrix has the general form\footnote{
    Note that, while $\Delta^Y(\vec{k})$ is complex, $\Delta^Y_{\ell}$ is real for the usual cases of CMB temperature and polarization; nevertheless, we retain the complex conjugate for generic $Y$.}
\begin{empheq}[box=\fbox]{equation}
    \label{eqn:HarmonicCovariance}
    C^{\E{i};XY}_{\ell m\ell'm'}  =
     \frac{(4\pi)^2}{{V_{\E{i}}}}
    \sum_{\vec{n}\in \setN^{\E{i}}} 
    \Delta^X_{\ell}(k_{\vec{n}})
    \DeltaYstarforell_{\ell'}(k_{\vec{n}})
    \frac{2\pi^2 \mathcal{P}^{\mathcal{R}}(k_{\vec{n}})}{k_{\vec{n}}^3}
    \xi^{\E{i};\unitvec{k}_{\vec{n}}}_{k_{\vec{n}} \ell m}
    \xi^{\E{i};\unitvec{k}_{\vec{n}}*}_{k_{\vec{n}} \ell' m'} .
\end{empheq}

\subsection{\E{2}: Half-turn space}
\label{secn:eigenmodesE2}

The eigenspectrum and eigenmodes of the half-turn space can be determined in a manner analogous to that of the 3-torus.
We could begin from the covering space, but it is more expedient to recognize that \E{2} is \E{1} with extra symmetries imposed.
With this, the eigenspectrum of \E{2} will be discretized with wavevectors $\vec{k}_{\vec{n}}$ and the eigenfunctions $\Upsilon^{\E{2}}_{\vec{k}_{\vec{n}}}(\vec{x})$ will be linear combinations of $\Upsilon^{\E{1}}_{\vec{k_{\vec{n}}}}(\vec{x})$.
For \E{2}, the discretization condition \eqref{eqn:associatedhomogeneousdiscretization} from the translation vectors $\vec{T}^{\E{2}}_j$ leads to the components of the allowed wavevectors,
\begin{align}
    \label{eqn:E2_kni}
    (\vec{k}_{\vec{n}})_{x} &= \frac{2\pi n_1}{L_1}, \nonumber \\
    (\vec{k}_{\vec{n}})_{y} &= \frac{2\pi n_2}{L_2\sin\alpha} - \frac{2\pi n_1}{L_1}\frac{\cos\alpha}{\sin\alpha},\\
    (\vec{k}_{\vec{n}})_{z} &= \frac{2\pi n_3}{2L_B\sin\beta}. \nonumber
\end{align}

Unlike in \E{1}, the eigenmodes of \E{2} can include a linear combination of two \E{1} eigenmodes.
This follows in the application of \cref{eqn:generaleigenmodeformula} since the condition $(\mat{M}^{\E{2}}_B)^N\vec{k}_{\vec{n}} = \vec{k}_{\vec{n}}$ has more than one solution for the minimum positive $N$, depending on $\vec{k}_{\vec{n}}$,
namely, $N=1$ for $(\vec{k}_{\vec{n}})_{x} = (\vec{k}_{\vec{n}})_{y} = 0$ and $N=2$ otherwise.
Written explicitly,
\begin{description}
    \item[\textbf{$N=1$ eigenmodes: }]  $\vec{k}_{\vec{n}}=\transpose{(0,0,(\vec{k}_{\vec{n}})_z)}$, 
        i.e., $\vec{n}=(0,0,n_3)$, 
        $n_3 \in 2\integers^{\neq 0}$,\footnote{
            Odd $n_3$  is excluded because when $n_3$ is odd
            $ e^{i\vec{k}_{\vec{n}}\cdot\vec{T}^{\E{2}}_B}=-1$,
            and so $\Upsilon^{\E{2}}_{(0,0,n_3)}(\vec{x})$ does not then satisfy the boundary conditions. 
            Similar conditions apply to the $N=1$ eigenmodes of other manifolds for similar reasons.}  with
    \begin{empheq}[box=\fbox]{equation}
        \Upsilon^{\E{2}}_{\vec{k}_{\vec{n}}}(\vec{x}) = e^{i\vec{k}_{\vec{n}} \cdot (\vec{x}-\vec{x}_0)} = e^{i(\vec{k}_{\vec{n}})_{z} (z-z_0)}, 
    \end{empheq}
    \item[\textbf{$N=2$ eigenmodes: }] $((\vec{k}_{\vec{n}})_{x}, (\vec{k}_{\vec{n}})_{y}) \neq (0,0)$, 
        i.e., $(n_1,n_2) \neq (0,0)$, and per \cref{eqn:generaleigenmodeformula},
    \begin{empheq}[box=\fbox]{equation}
        \Upsilon^{\E{2}}_{\vec{k}_{\vec{n}}}(\vec{x}) = \frac{1}{\sqrt{2}} \left( e^{i\vec{k}_{\vec{n}}\cdot(\vec{x}-\vec{x}_0)} + e^{i\vec{k}_{\vec{n}}\cdot(\mat{M}^{\E{2}}_B(\vec{x}-\vec{x}_0)+\vec{T}^{\E{2}}_B)} \right).
    \end{empheq}
\end{description}
The linear combination in the $N=2$ modes requires some care.
Notice that $\transpose{\vec{k}}_{\vec{n}}\mat{M}^{\E{2}}_B=(-(\vec{k}_{\vec{n}})_{x},-(\vec{k}_{\vec{n}})_{y},(\vec{k}_{\vec{n}})_{z})$, i.e., $\mat{M}^{\E{2}}_B$ maps $(n_1,n_2,n_3)\to(-n_1,-n_2,n_3)$.
One implication of this is that summing over $(n_1,n_2,n_3)$ would double-count eigenmodes if all $n_1\in\integers$ and all $n_2\in\integers$ were included.
Hence, we define two sets of allowed modes, one for $N=1$ and another for $N=2$,
\begin{empheq}[box=\fbox]{align}
    \setN^{\E{2}}_1 &= \{(0,0,n_3)|n_3\in 2\integers^{\neq0} \} , \nonumber\\
    \setN^{\E{2}}_2 &= \{(n_1,n_2,n_3)| n_1 \in \integers^{>0}, n_2 \in \integers, n_3 \in \integers\} 
                    \cup \{(0,n_2,n_3)| n_2 \in \integers^{>0}, n_3 \in \integers\} , \\
    \setN^{\E{2}} &= \setN^{\E{2}}_1 \cup \setN^{\E{2}}_2 . \nonumber
\end{empheq}
With these the Fourier-mode correlation matrix can now be expressed as
\begin{empheq}[box=\fbox]{align}
\label{eqn:E2FourierCovarianceStandardConvention}
    C^{\E{2};XY}_{\vec{k}_{\vec{n}}\vec{k}_{\vecnp}} 
     = {} & V_{\E{2}} \frac{2\pi^2}{k^3_{\vec{n}}}
        {\mathcal{P}}^{\mathcal{R}}(k_{\vec n})\Delta^X(k_{\vec{n}})\DeltaYstar(k_{\vec{n}})
        e^{i(\vec{k}_{\vecnp}-\vec{k}_{\vec{n}})\cdot\vec{x}_0}
        \left[
            \sum_{\vec{\tilde{n}} \in \setN^{\E{2}}_1}
                 \Kdelta_{\vec{k}_{\vec{n}}\vec{k}_{\vec{\tilde{n}}}}
                 \Kdelta_{\vec{k}_{\vecnp}\vec{k}_{\vec{\tilde{n}}}} + {}
        \right.
            \\
            & \left. \quad
                {} + \frac{1}{2}
                \sum_{\vec{\tilde{n}}\in \setN^{\E{2}}_2}
                \sum_{a=0}^1\sum_{b=0}^1 
                e^{i\vec{k}_{\vec{\tilde{n}}}\cdot(\vec{T}^{(a)}-\vec{T}^{(b)})}
                 \Kdelta_{\vec{k}_{\vec{n}}([(\mat{M}^{\E{2}}_B){}^T]^a\vec{k}_{\vec{\tilde{n}}})}
                 \Kdelta_{\vec{k}_{\vecnp}([(\mat{M}^{\E{2}}_B){}^T]^b\vec{k}_{\vec{\tilde{n}}})} 
         \right], 
         \nonumber
\end{empheq}
where $V_{\E{2}}$ is given in \eqref{eqn:VE2}, $\vec{T}^{(0)}\equiv\vec{0}$, and $\vec{T}^{(1)}\equiv\vec{T}^{\E{2}}_B$.\footnote{
    In  \cref{eqn:E2FourierCovarianceStandardConvention}, $\vec{k}_{\vec{n}}$ and $\vec{k}_{\vecnp}$ are wavevectors of the associated \E{1}, as specified above in \eqref{eqn:E2_kni}.
    $C^{\E{2};XY}_{\vec{k}_{\vec{n}}\vec{k}_{\vecnp}}$ describes correlations between amplitudes of the plane waves that comprise the eigenmodes of a specific manifold --- i.e., of a specific topology, with specific values of its parameters.
    It is this object that would be used, for example, in creating realizations of initial conditions for large-scale structure simulations.
    If one was, instead, constructing a likelihood function to compare data with expectations from \E{2} manifolds, one would need to convolve $C^{\E{2};XY}_{\vec{k}_{\vec{n}}\vec{k}_{\vecnp}}$ with a kernel characterizing the Fourier structure of the survey of interest.
    }
Note that $C^{\E{2};XY}_{\vec{k}_{\vec{n}}\vec{k}_{\vecnp}}=0$ for $\vert k_{\vecnp}\vert\neq\vert k_{\vec{n}}\vert$, so ${\mathcal{P}}^{\mathcal{R}}(k_{\vec n})$, $\Delta^X(k_{\vec{n}})$, and $\DeltaYstar(k_{\vec{n}})$ are each a function only of $k_{\vec{n}}$.\footnote{
    The terms in the square brackets of \cref{eqn:E2FourierCovarianceStandardConvention} encode the correlations between Fourier modes with wavevectors related by application of $\mat{M}^{\E{2}}_B$. 
    Similar terms will appear in the expressions for the correlation matrices of other topologies.
    }

Another implication of the rotation in $\mat{M}^{\E{2}}_B$ comes when representing the eigenmodes in the harmonic basis.
Here we will combine modes with the same eigenvalue $k_{\vec{n}}$ and orientations $\unitvec{k}_{\vec{n}}$ and $(\mat{M}^{\E{2}}_B){}^T \unitvec{k}_{\vec{n}}$.
Since the half turn is a rotation around the $z$-axis by $\pi$ we can use the rotation properties of the spherical harmonics to simplify our expressions.
In particular
\begin{equation}
    Y^*_{\ell m} ((\mat{M}^{\E{2}}_B){}^T \unitvec{k}_{\vec{n}}) = e^{i m \pi} Y^*_{\ell m} (\unitvec{k}_{\vec{n}}) = (-1)^m Y^*_{\ell m} (\unitvec{k}_{\vec{n}}).
\end{equation}
This gives for the eigenmodes in the harmonic basis
\begin{empheq}[box=\fbox]{align}
    \xi^{\E{2};\unitvec{k}_{\vec{n}}}_{k_{\vec{n}}\ell m} 
    &= e^{-i \vec{k}_{\vec{n}}\cdot\vec{x}_0} i^\ell  Y^*_{\ell m} (\unitvec{k}_{\vec{n}}), \quad \mbox{for } \vec{n} \in \setN^{\E{2}}_1 ,\\
    \xi^{\E{2};\unitvec{k}_{\vec{n}}}_{k_{\vec{n}}\ell m} 
     &= \frac{1}{\sqrt{2}} i^\ell Y^*_{\ell m}(\unitvec{k}_{\vec{n}})
        \sum_{j=0}^1 (-1)^{j m}
        e^{-i\vec{k}_{\vec{n}}\cdot(\mat{M}^{\E{2}}_B)^j\vec{x}_0}
        e^{i \vec{k}_{\vec{n}}\cdot\mat{M}^{\E{2}}_{0j}\vec{T}^{\E{2}}_B},\quad \mbox{for } \vec{n}\in \setN^{\E{2}}_2 , \nonumber
\end{empheq}
where it will prove useful for many of the topologies to define
\begin{equation}
    \label{eqn:M0jdef}
    \mat{M}^{\E{i}}_{00} \equiv\mathbb{0}, \quad \mat{M}^{\E{i}}_{01} \equiv\identity, \quad
    \mbox{and} \quad \mat{M}^{\E{i}}_{0j} \equiv \sum_{r=0}^{j-1} (\mat{M}^{\E{i}}_B)^r \quad \mbox{for} \quad j>1.
\end{equation}
Finally, the spherical-harmonic space covariance matrix has the form \eqref{eqn:HarmonicCovariance}.

\subsection{\E{3}: Quarter-turn space}
\label{secn:eigenmodesE3}

The eigenspectrum and eigenmodes of the quarter-turn space can be determined in a manner analogous to that for \E{2}.
For \E{3}, the discretization condition \eqref{eqn:associatedhomogeneousdiscretization} from the translation vectors $\vec{T}^{\E{3}}_j$ leads to the components of the allowed wavevectors,
\begin{equation}
    (\vec{k}_{\vec{n}})_{x} = \frac{2\pi n_1}{L_A }, \quad
    (\vec{k}_{\vec{n}})_{y} = \frac{2\pi n_2}{L_A},  \quad
     (\vec{k}_{\vec{n}})_{z} = \frac{2\pi n_3}{4 L_B \sin\beta}.
\end{equation}

As in \E{2}, the eigenmodes of \E{3} can include linear combinations of \E{1} eigenmodes since $(\mat{M}^{\E{3}}_B)^N \vec{k}_{\vec{n}} = \vec{k}_{\vec{n}}$ has more than one solution for the minimum positive $N$, depending on $\vec{k}_{\vec{n}}$.
Here we have
\begin{description}
    \item[\textbf{$N=1$ eigenmodes: }]  $\vec{k}_{\vec{n}}=\transpose{(0,0,(\vec{k}_{\vec{n}})_z)}$, 
        i.e., $\vec{n}=(0,0,n_3)$, 
        $n_3 \in 4\integers^{\neq 0}$, with
    \begin{empheq}[box=\fbox]{align}
        \Upsilon^{\E{3}}_{\vec{k}_{\vec{n}}}(\vec{x}) = e^{i\vec{k}_{\vec{n}} \cdot (\vec{x}-\vec{x}_0)}, 
    \end{empheq}
    \item[\textbf{$N=4$ eigenmodes: }] $((\vec{k}_{\vec{n}})_{x}, (\vec{k}_{\vec{n}})_{y}) \neq (0,0)$, 
        i.e., $(n_1,n_2) \neq (0,0)$, with
    \begin{empheq}[box=\fbox]{align}
        \Upsilon^{\E{3}}_{\vec{k}_{\vec{n}}}(\vec{x}) = \frac{1}{\sqrt{4}} \sum_{j=0}^3 
        e^{i\vec{k}_{\vec{n}}\cdot ((\mat{M}^{\E{3}}_B)^j(\vec{x}-\vec{x}_0)+\mat{M}^{\E{3}}_{0j}\vec{T}^{\E{3}}_B)},
    \end{empheq}
\end{description}
and $\mat{M}^{\E{3}}_{0j}$ defined in \eqref{eqn:M0jdef}.

As in \E{2}, the cyclic properties of $\mat{M}^{\E{3}}_B$ would lead to repeated counting of eigenmodes if all $n_1\in\integers$ and $n_2\in\integers$ were included.
In this case, under the repeated action of $\mat{M}^{\E{3}}_B$ we have the mappings $(n_1, n_2) \to (n_2, -n_1) \to (-n_1, -n_2) \to (-n_2, n_1)$.
To avoid this, we define two sets of allowed modes, now for $N=1$ and $N=4$,
\begin{empheq}[box=\fbox]{align}
    \setN^{\E{3}}_1 &= \{(0,0,n_3)|n_3\in 4\integers^{\neq0} \} , \nonumber \\
    \setN^{\E{3}}_4 &= \{(n_1,n_2,n_3)| n_1 \in \integers^{\geq 0}, n_2 \in \integers^{>0}, n_3 \in \integers\} , \\
    \setN^{\E{3}} &= \setN^{\E{3}}_1 \cup \setN^{\E{3}}_4 . \nonumber
\end{empheq}
With these the Fourier-mode correlation matrix can now be expressed as
\begin{empheq}[box=\fbox]{align}
\label{eqn:E3FourierCovarianceStandardConvention}
    C^{\E{3};XY}_{\vec{k}_{\vec{n}}\vec{k}_{\vecnp}} 
     = {} & V_{\E{3}} \frac{2\pi^2}{k^3_{\vec{n}}}
        {\mathcal{P}}^{\mathcal{R}}(k_{\vec n})\Delta^X(k_{\vec{n}})\DeltaYstar(k_{\vec{n}})
        e^{i(\vec{k}_{\vecnp}-\vec{k}_{\vec{n}})\cdot\vec{x}_0}
        \left[
            \sum_{\vec{\tilde{n}} \in \setN^{\E{3}}_1}
                 \Kdelta_{\vec{k}_{\vec{n}}\vec{k}_{\vec{\tilde{n}}}}
                 \Kdelta_{\vec{k}_{\vecnp}\vec{k}_{\vec{\tilde{n}}}} + {}
        \right.
            \\
            & \left. \quad
                {} + \frac{1}{4}
                \sum_{\vec{\tilde{n}}\in \setN^{\E{3}}_4}
                \sum_{a=0}^3\sum_{b=0}^3 
                e^{i\vec{k}_{\vec{\tilde{n}}}\cdot(\vec{T}^{(a)}-\vec{T}^{(b)})}
                 \Kdelta_{\vec{k}_{\vec{n}}([(\mat{M}^{\E{3}}_B){}^T]^a\vec{k}_{\vec{\tilde{n}}})}
                 \Kdelta_{\vec{k}_{\vecnp}([(\mat{M}^{\E{3}}_B){}^T]^b\vec{k}_{\vec{\tilde{n}}})} 
         \right] ,
         \nonumber
\end{empheq}
where $V_{\E{3}}$ is given in \eqref{eqn:VE3}, $\vec{T}^{(0)} \equiv \vec{0}$, and $\vec{T}^{(a)} \equiv \mat{M}^{\E{3}}_{0a} \vec{T}^{\E{3}}_B$ for $a \in \{ 1,2,3 \}$.

Also as in \E{2}, we can use the rotation properties of the spherical harmonics along with the fact that $\mat{M}^{\E{3}}_B$ is a rotation around the $z$-axis by $\pi/2$ to note that
\begin{equation}
    Y^*_{\ell m}([(\mat{M}^{\E{3}}_B){}^T]^j \unitvec{k}_{\vec{n}}) = e^{i m j\pi/2} Y^*_{\ell m}(\unitvec{k}_{\vec{n}}) = i^{m j} Y^*_{\ell m}(\unitvec{k}_{\vec{n}}).
\end{equation}
This gives for the eigenmodes in the harmonic basis
\begin{empheq}[box=\fbox]{align}
    \xi^{\E{3};\unitvec{k}_{\vec{n}}}_{k_{\vec{n}}\ell m} 
    &= e^{-i \vec{k}_{\vec{n}}\cdot\vec{x}_0} i^\ell  Y^*_{\ell m} (\unitvec{k}_{\vec{n}}), \quad \mbox{for } \vec{n} \in \setN^{\E{3}}_1 ,\\
    \xi^{\E{3};\unitvec{k}_{\vec{n}}}_{k_{\vec{n}}\ell m} 
     &= \frac{1}{\sqrt{4}} i^\ell Y^*_{\ell m}(\unitvec{k}_{\vec{n}})
        \sum_{j=0}^3 i^{m j}
        e^{-i\vec{k}_{\vec{n}}\cdot(\mat{M}^{\E{3}}_B)^j\vec{x}_0}
        e^{i \vec{k}_{\vec{n}}\cdot\mat{M}^{\E{3}}_{0j}\vec{T}^{\E{3}}_B},\quad \mbox{for } \vec{n}\in \setN^{\E{3}}_4 , \nonumber
\end{empheq}
and the spherical-harmonic space covariance matrix has the form \eqref{eqn:HarmonicCovariance}.

\subsection{\E{4}: Third-turn space}
\label{secn:eigenmodesE4}

The eigenspectrum and eigenmodes of the third-turn space can be determined in a manner analogous to those from above.
For \E{4}, the discretization condition \eqref{eqn:associatedhomogeneousdiscretization} from the translation vectors $\vec{T}^{\E{4}}_j$ leads to the components of the allowed wavevectors,
\begin{equation}
    (\vec{k}_{\vec{n}})_{x} = \frac{2\pi n_1}{L_A}, \quad
    (\vec{k}_{\vec{n}})_{y} = \frac{2\pi }{\sqrt{3} L_A}(n_1 + 2n_2), \quad
    (\vec{k}_{\vec{n}})_{z} = \frac{2\pi n_3}{3 L_B \sin\beta}.
\end{equation}

As in \E{2}, the eigenmodes of \E{4} can include linear combinations of \E{1} eigenmodes since $(\mat{M}^{\E{4}}_B)^N \vec{k}_{\vec{n}} = \vec{k}_{\vec{n}}$ has more than one solution for the minimum positive $N$, depending on $\vec{k}_{\vec{n}}$.
Here we have
\begin{description}
    \item[\textbf{$N=1$ eigenmodes: }]  $\vec{k}_{\vec{n}}=\transpose{(0,0,(\vec{k}_{\vec{n}})_z)}$, 
        i.e., $\vec{n}=(0,0,n_3)$, 
        $n_3 \in 3\integers^{\neq 0}$, with
    \begin{empheq}[box=\fbox]{align}
        \Upsilon^{\E{4}}_{\vec{k}_{\vec{n}}}(\vec{x}) = e^{i\vec{k}_{\vec{n}} \cdot (\vec{x}-\vec{x}_0)}, 
    \end{empheq}
    \item[\textbf{$N=3$ eigenmodes: }] $((\vec{k}_{\vec{n}})_{x}, (\vec{k}_{\vec{n}})_{y}) \neq (0,0)$, 
        i.e., $(n_1,n_2) \neq (0,0)$, with
    \begin{empheq}[box=\fbox]{align}
        \Upsilon^{\E{4}}_{\vec{k}_{\vec{n}}}(\vec{x}) = \frac{1}{\sqrt{3}} \sum_{j=0}^2 
        e^{i\vec{k}_{\vec{n}}\cdot ((\mat{M}^{\E{4}}_B)^j(\vec{x}-\vec{x}_0)+\mat{M}^{\E{4}}_{0j}\vec{T}^{\E{4}}_B)},
    \end{empheq}
\end{description}
and $\mat{M}^{\E{4}}_{0j}$ defined in \eqref{eqn:M0jdef}.

As in \E{2}, the cyclic properties of $\mat{M}^{\E{4}}_B$ would lead to repeated counting of eigenmodes if all $n_1\in\integers$ and $n_2\in\integers$ were included.
To avoid this, we define two sets of allowed modes, now for $N=1$ and $N=3$,
\begin{empheq}[box=\fbox]{align}
    \setN^{\E{4}}_1 &= \{(0,0,n_3)| n_3\in 3\integers^{\neq0} \} , \nonumber \\
    \setN^{\E{4}}_3 &= \{(n_1,n_2,n_3)| n_1 \in \integers^{\neq 0}, n_2 \in \integers, n_1 n_2 \geq 0,  n_3 \in \integers\},  \\
    \setN^{\E{4}} &= \setN^{\E{4}}_1 \cup \setN^{\E{4}}_3 . \nonumber
\end{empheq}
With these the Fourier-mode correlation matrix can now be expressed as
\begin{empheq}[box=\fbox]{align}
\label{eqn:E4FourierCovarianceStandardConvention}
    C^{\E{4};XY}_{\vec{k}_{\vec{n}}\vec{k}_{\vecnp}} 
     = {} & V_{\E{4}} \frac{2\pi^2}{k^3_{\vec{n}}}
        {\mathcal{P}}^{\mathcal{R}}(k_{\vec n})\Delta^X(k_{\vec{n}})\DeltaYstar(k_{\vec{n}})
        e^{i(\vec{k}_{\vecnp}-\vec{k}_{\vec{n}})\cdot\vec{x}_0}
        \left[
            \sum_{\vec{\tilde{n}} \in \setN^{\E{4}}_1}
                 \Kdelta_{\vec{k}_{\vec{n}}\vec{k}_{\vec{\tilde{n}}}}
                 \Kdelta_{\vec{k}_{\vecnp}\vec{k}_{\vec{\tilde{n}}}} + {}
        \right.
            \\
            & \left. \quad
                {} + \frac{1}{3}
                \sum_{\vec{\tilde{n}}\in \setN^{\E{4}}_3}
                \sum_{a=0}^2\sum_{b=0}^2 
                e^{i\vec{k}_{\vec{\tilde{n}}}\cdot(\vec{T}^{(a)}-\vec{T}^{(b)})}
                 \Kdelta_{\vec{k}_{\vec{n}}([(\mat{M}^{\E{4}}_B){}^T]^a\vec{k}_{\vec{\tilde{n}}})}
                 \Kdelta_{\vec{k}_{\vecnp}([(\mat{M}^{\E{4}}_B){}^T]^b\vec{k}_{\vec{\tilde{n}}})} 
         \right] ,
         \nonumber
\end{empheq}
where $V_{\E{4}}$ is given in \eqref{eqn:VE4}, $\vec{T}^{(0)} \equiv \vec{0}$, and $\vec{T}^{(a)} \equiv \mat{M}^{\E{4}}_{0a} \vec{T}^{\E{4}}_B$ for $a \in \{ 1,2 \}$.

Also as in \E{2}, we can use the rotation properties of the spherical harmonics along with the fact that $\mat{M}^{\E{4}}_B$ is a rotation around the $z$-axis by $2\pi/3$ to note that
\begin{equation}
    Y^*_{\ell m}([(\mat{M}^{\E{4}}_B){}^T]^j \unitvec{k}_{\vec{n}}) = e^{i m j 2\pi/3} Y^*_{\ell m} (\unitvec{k}_{\vec{n}}).
\end{equation}
This gives for the eigenmodes in the harmonic basis
\begin{empheq}[box=\fbox]{align}
    \xi^{\E{4};\unitvec{k}_{\vec{n}}}_{k_{\vec{n}}\ell m} 
    &= e^{-i \vec{k}_{\vec{n}}\cdot\vec{x}_0} i^\ell  Y^*_{\ell m} (\unitvec{k}_{\vec{n}}), \quad \mbox{for } \vec{n} \in \setN^{\E{4}}_1 ,\\
    \xi^{\E{4};\unitvec{k}_{\vec{n}}}_{k_{\vec{n}}\ell m} 
     &= \frac{1}{\sqrt{3}} i^\ell Y^*_{\ell m}(\unitvec{k}_{\vec{n}})
        \sum_{j=0}^2 e^{i m j 2\pi/3}
        e^{-i\vec{k}_{\vec{n}}\cdot(\mat{M}^{\E{4}}_B)^j\vec{x}_0}
        e^{i \vec{k}_{\vec{n}}\cdot\mat{M}^{\E{4}}_{0j}\vec{T}^{\E{4}}_B},\quad \mbox{for } \vec{n}\in \setN^{\E{4}}_3 , \nonumber
\end{empheq}
and the spherical-harmonic space covariance matrix has the form \eqref{eqn:HarmonicCovariance}.

\subsection{\E{5}: Sixth-turn space}
\label{secn:eigenmodesE5}

The eigenspectrum and eigenmodes of the sixth-turn space can be determined in a manner analogous to those from above, in particular it is very similar to \E{4}.
For \E{5}, the discretization condition \eqref{eqn:associatedhomogeneousdiscretization} from the translation vectors $\vec{T}^{\E{5}}_j$ leads to the components of the allowed wavevectors,
\begin{equation}
    (\vec{k}_{\vec{n}})_{x} = \frac{2\pi n_1}{L_A}, \quad
    (\vec{k}_{\vec{n}})_{y} = \frac{2\pi}{\sqrt{3} L_A}(n_1 + 2n_2), \quad
    (\vec{k}_{\vec{n}})_{z} = \frac{2\pi n_3}{6 L_B \sin\beta}.
\end{equation}

As in \E{2}, the eigenmodes of \E{5} can include linear combinations of \E{1} eigenmodes since $(\mat{M}^{\E{5}}_B)^N \vec{k}_{\vec{n}} = \vec{k}_{\vec{n}}$ has more than one solution for the minimum positive $N$, depending on $\vec{k}_{\vec{n}}$.
Here we have
\begin{description}
    \item[\textbf{$N=1$ eigenmodes: }]  $\vec{k}_{\vec{n}}=\transpose{(0,0,(\vec{k}_{\vec{n}})_z)}$, 
        i.e., $\vec{n}=(0,0,n_3)$, 
        $n_3 \in 6\integers^{\neq 0}$, with
    \begin{empheq}[box=\fbox]{align}
        \Upsilon^{\E{5}}_{\vec{k}_{\vec{n}}}(\vec{x}) = e^{i\vec{k}_{\vec{n}} \cdot (\vec{x}-\vec{x}_0)}, 
    \end{empheq}
    \item[\textbf{$N=6$ eigenmodes: }] $((\vec{k}_{\vec{n}})_{x}, (\vec{k}_{\vec{n}})_{y}) \neq (0,0)$, 
        i.e., $(n_1,n_2) \neq (0,0)$, with
    \begin{empheq}[box=\fbox]{align}
        \Upsilon^{\E{5}}_{\vec{k}_{\vec{n}}}(\vec{x}) = \frac{1}{\sqrt{6}} \sum_{j=0}^5 
        e^{i\vec{k}_{\vec{n}}\cdot ((\mat{M}^{\E{5}}_B)^j(\vec{x}-\vec{x}_0)+\mat{M}^{\E{5}}_{0j}\vec{T}^{\E{5}}_B)},
    \end{empheq}
\end{description}
and $\mat{M}^{\E{5}}_{0j}$ defined in \eqref{eqn:M0jdef}.

As in \E{2}, the cyclic properties of $\mat{M}^{\E{5}}_B$ would lead to repeated counting of eigenmodes if all $n_1\in\integers$ and $n_2\in\integers$ were included.
To avoid this, we define two sets of allowed modes, now for $N=1$ and $N=6$,
\begin{empheq}[box=\fbox]{align}
    \setN^{\E{5}}_1 &= \{(0,0,n_3)| n_3\in 6\integers^{\neq0} \} , \nonumber \\
    \setN^{\E{5}}_6 &= \{(n_1,n_2,n_3)| n_1 \in \integers^{> 0}, n_2 \in \integers^{\geq 0}, n_3 \in \integers\}
    , \\
    \setN^{\E{5}} &= \setN^{\E{5}}_1 \cup \setN^{\E{5}}_6 . \nonumber
\end{empheq}
With these the Fourier-mode correlation matrix can now be expressed as
\begin{empheq}[box=\fbox]{align}
\label{eqn:E5FourierCovarianceStandardConvention}
    C^{\E{5};XY}_{\vec{k}_{\vec{n}}\vec{k}_{\vecnp}} 
     = {} & V_{\E{5}} \frac{2\pi^2}{k^3_{\vec{n}}}
        {\mathcal{P}}^{\mathcal{R}}(k_{\vec n})\Delta^X(k_{\vec{n}})\DeltaYstar(k_{\vec{n}})
        e^{i(\vec{k}_{\vecnp}-\vec{k}_{\vec{n}})\cdot\vec{x}_0}
        \left[
            \sum_{\vec{\tilde{n}} \in \setN^{\E{5}}_1}
                 \Kdelta_{\vec{k}_{\vec{n}}\vec{k}_{\vec{\tilde{n}}}}
                 \Kdelta_{\vec{k}_{\vecnp}\vec{k}_{\vec{\tilde{n}}}} + {}
        \right.
            \\
            & \left. \quad
                {} + \frac{1}{6}
                \sum_{\vec{\tilde{n}}\in \setN^{\E{5}}_6}
                \sum_{a=0}^5\sum_{b=0}^5 
                e^{i\vec{k}_{\vec{\tilde{n}}}\cdot(\vec{T}^{(a)}-\vec{T}^{(b)})}
                 \Kdelta_{\vec{k}_{\vec{n}}([(\mat{M}^{\E{5}}_B){}^T]^a\vec{k}_{\vec{\tilde{n}}})}
                 \Kdelta_{\vec{k}_{\vecnp}([(\mat{M}^{\E{5}}_B){}^T]^b\vec{k}_{\vec{\tilde{n}}})} 
         \right] ,
         \nonumber
\end{empheq}
where $V_{\E{5}}$ is given in \eqref{eqn:VE5}, $\vec{T}^{(0)} \equiv \vec{0}$, and $\vec{T}^{(a)} \equiv \mat{M}^{\E{5}}_{0a} \vec{T}^{\E{5}}_B$ for $a \in \{ 1,\ldots, 5 \}$.

Also as in \E{2}, we can use the rotation properties of the spherical harmonics along with the fact that $\mat{M}^{\E{5}}_B$ is a rotation around the $z$-axis by $\pi/3$ to note that
\begin{equation}
    Y^*_{\ell m}([(\mat{M}^{\E{5}}_B){}^T]^j \unitvec{k}_{\vec{n}}) = e^{i m j \pi/3} Y^*_{\ell m}(\unitvec{k}_{\vec{n}}) .
\end{equation}
This gives for the eigenmodes in the harmonic basis
\begin{empheq}[box=\fbox]{align}
    \xi^{\E{5};\unitvec{k}_{\vec{n}}}_{k_{\vec{n}}\ell m} 
    &= e^{-i \vec{k}_{\vec{n}}\cdot\vec{x}_0} i^\ell  Y^*_{\ell m} (\unitvec{k}_{\vec{n}}), \quad \mbox{for } \vec{n} \in \setN^{\E{5}}_1 ,\\
    \xi^{\E{5};\unitvec{k}_{\vec{n}}}_{k_{\vec{n}}\ell m} 
     &= \frac{1}{\sqrt{6}} i^\ell Y^*_{\ell m}(\unitvec{k}_{\vec{n}})
        \sum_{j=0}^5 e^{i m j \pi/3}
        e^{-i\vec{k}_{\vec{n}}\cdot(\mat{M}^{\E{5}}_B)^j\vec{x}_0}
        e^{i \vec{k}_{\vec{n}}\cdot\mat{M}^{\E{5}}_{0j}\vec{T}^{\E{5}}_B},\quad \mbox{for } \vec{n}\in \setN^{\E{5}}_6 , \nonumber
\end{empheq}
and the spherical-harmonic space covariance matrix has the form \eqref{eqn:HarmonicCovariance}.

\subsection{\E{6}: Hantzsche-Wendt space}
\label{secn:eigenmodesE6}

The eigenspectrum and eigenmodes of the Hantzsche-Wendt space can be determined in a manner analogous to those from above.
Complications arise from the fact that \E{6} contains rotations around multiple axes so a more careful discussion is warranted.
The discretization condition \eqref{eqn:associatedhomogeneousdiscretization} from the translation vectors $\vec{T}^{\E{6}}_j$ is still straightforward and leads to the components of the allowed wavevectors,
\begin{equation}
    (\vec{k}_{\vec{n}})_{x} = \frac{2\pi n_1}{2 L_{Ax}}, \quad
    (\vec{k}_{\vec{n}})_{y} = \frac{2\pi n_2}{2 L_{By}}, \quad
    (\vec{k}_{\vec{n}})_{z} = \frac{2\pi n_3}{2 L_{Cz}}.
\end{equation}

As in \E{2}, the eigenmodes of \E{6} can include linear combinations of \E{1} eigenmodes since $(\mat{M}^{\E{6}}_a)^{N_a} \vec{k}_{\vec{n}} = \vec{k}_{\vec{n}}$ has more than one solution for the minimum positive $N_a$, depending on $\vec{k}_{\vec{n}}$, for $a\in\{A, B, C\}$.
Since all the rotations are half turns, i.e., all $(\mat{M}^{\E{6}}_a)^2 = \identity$, there are linear combinations with the $N_a=1$ and $N_a=2$.
At first glance \cref{eqn:generaleigenmodeformula} seems to suggest that the eigenmodes with $N_a=1$ will be a linear combination of four eigenmodes of \E{1}.
However, since $\mat{M}^{\E{6}}_A \mat{M}^{\E{6}}_B = \mat{M}^{\E{6}}_C$ (and all permutations of $\{A, B, C\}$) along with the invariance of the eigenmodes under the group action \eqref{eqn:generaleigenmodeinvariance}, a linear combination of only two eigenmodes of \E{1} is required in this case.
Based on this we can choose
\begin{description}
    \item[\textbf{$N_A=1$ eigenmodes: }]  $\vec{k}_{\vec{n}}=\transpose{((\vec{k}_{\vec{n}})_x, 0, 0)}$, 
        i.e., $\vec{n}=(n_1,0,0)$, 
        $n_1 \in 2\integers^{\neq 0}$, with
    \begin{empheq}[box=\fbox]{align}
        \Upsilon^{\E{6}}_{\vec{k}_{\vec{n}}}(\vec{x}) = \frac{1}{\sqrt{2}} \left[ e^{i\vec{k}_{\vec{n}} \cdot (\vec{x}-\vec{x}_0)}
          + e^{-i\vec{k}_{\vec{n}} \cdot (\vec{x}-\vec{x}_0)} e^{i\vec{k}_{\vec{n}}\cdot\vec{T}^{\E{6}}_B}  \right], 
    \end{empheq}
    \item[\textbf{$N_B=1$ eigenmodes: }]  $\vec{k}_{\vec{n}}=\transpose{(0,(\vec{k}_{\vec{n}})_y, 0)}$, 
        i.e., $\vec{n}=(0,n_2,0)$, 
        $n_2 \in 2\integers^{\neq 0}$, with
    \begin{empheq}[box=\fbox]{align}
        \Upsilon^{\E{6}}_{\vec{k}_{\vec{n}}}(\vec{x}) = \frac{1}{\sqrt{2}} \left[ e^{i\vec{k}_{\vec{n}} \cdot (\vec{x}-\vec{x}_0)}
          + e^{-i\vec{k}_{\vec{n}} \cdot (\vec{x}-\vec{x}_0)} e^{i\vec{k}_{\vec{n}}\cdot\vec{T}^{\E{6}}_C}  \right], 
    \end{empheq}
    \item[\textbf{$N_C=1$ eigenmodes: }]  $\vec{k}_{\vec{n}}=\transpose{(0,0,(\vec{k}_{\vec{n}})_z)}$, 
        i.e., $\vec{n}=(0,0,n_3)$, 
        $n_3 \in 2\integers^{\neq 0}$, with
    \begin{empheq}[box=\fbox]{align}
        \Upsilon^{\E{6}}_{\vec{k}_{\vec{n}}}(\vec{x}) = \frac{1}{\sqrt{2}} \left[ e^{i\vec{k}_{\vec{n}} \cdot (\vec{x}-\vec{x}_0)}
          + e^{-i\vec{k}_{\vec{n}} \cdot (\vec{x}-\vec{x}_0)} e^{i\vec{k}_{\vec{n}}\cdot\vec{T}^{\E{6}}_A} \right], 
    \end{empheq}
    \item[\textbf{$N_a=2$ eigenmodes: }] at most one component of $\vec{k}_{\vec{n}}$ zero, i.e., at most one of the $n_i$ equal to zero, with
    \begin{empheq}[box=\fbox]{align}
        \Upsilon^{\E{6}}_{\vec{k}_{\vec{n}}}(\vec{x}) = \frac{1}{\sqrt{4}} \left[
        e^{i\vec{k}_{\vec{n}} \cdot (\vec{x}-\vec{x}_0)} +
        \sum_{a \in\{A, B, C\}} e^{i\vec{k}_{\vec{n}} \cdot \mat{M}^{\E{6}}_a (\vec{x}-\vec{x}_0)} e^{i \vec{k}_{\vec{n}} \cdot \vec{T}^{\E{6}}_a} \right].
    \end{empheq}
\end{description}

The cyclic properties of the $\mat{M}^{\E{6}}_a$ would lead to repeated counting of eigenmodes.
To avoid this, we define the sets of allowed modes as
\begin{empheq}[box=\fbox]{align}
    \setN^{\E{6}}_{1A} &= \{(n_1,0,0)| n_1\in 2\integers^{\neq0} \} , \nonumber \\
    \setN^{\E{6}}_{1B} &= \{(0,n_2,0)| n_2\in 2\integers^{\neq0} \} , \nonumber \\
    \setN^{\E{6}}_{1C} &= \{(0,0,n_3)| n_3\in 2\integers^{\neq0} \} , \\
    \setN^{\E{6}}_2 &= \{(n_1,n_2,n_3)| n_1 \in \integers^{> 0}, n_2 \in \integers^{> 0}, n_3 \in \integers\} \nonumber \\
    & {} \qquad \cup \{(0, n_2, n_3)| n_2 \in \integers^{>0}, n_3 \in \integers^{>0} \} \nonumber \\
    & {} \qquad \cup \{(n_1, 0, n_3)| n_1 \in \integers^{>0}, n_3 \in \integers^{>0} \}, \nonumber \\
    \setN^{\E{6}} &= \setN^{\E{6}}_{1A} \cup \setN^{\E{6}}_{1B} \cup \setN^{\E{6}}_{1C} \cup \setN^{\E{6}}_2 . \nonumber
\end{empheq}
\newpage
\newgeometry{textwidth=0.75\paperwidth, textheight=0.8\paperheight}
With these the Fourier-mode correlation matrix can now be expressed as
\begin{empheq}[box=\fbox]{align}
    \label{eqn:E6FourierCovarianceStandardConvention}
    C^{\E{6};XY}_{\vec{k}_{\vec{n}}\vec{k}_{\vecnp}} 
     = {} & V_{\E{6}} \frac{2\pi^2}{k^3_{\vec{n}}}
        {\mathcal{P}}^{\mathcal{R}}(k_{\vec n})\Delta^X(k_{\vec{n}})\DeltaYstar(k_{\vec{n}})
        e^{i(\vec{k}_{\vecnp}-\vec{k}_{\vec{n}})\cdot\vec{x}_0} \times {} \\
        & {} \quad \times \left[
            \frac{1}{2} \sum_{\vec{\tilde{n}} \in \setN^{\E{6}}_{1A}}
                \sum_{a,b\in\{0, B\}} e^{i\vec{k}_{\vec{n}} \cdot (\vec{T}^{(a)} - \vec{T}^{(b)})}
                 \Kdelta_{\vec{k}_{\vec{n}}((\mat{M}^{\E{6}}_a){}^T \vec{k}_{\tilde{\vec{n}}})}
                 \Kdelta_{\vec{k}_{\vecnp}((\mat{M}^{\E{6}}_b){}^T \vec{k}_{\tilde{\vec{n}}})} + {}
            \right. \nonumber \\
        & {} \qquad + \frac{1}{2} \sum_{\vec{\tilde{n}} \in \setN^{\E{6}}_{1B}}
                \sum_{a,b\in\{0, C\}} e^{i\vec{k}_{\vec{n}} \cdot (\vec{T}^{(a)} - \vec{T}^{(b)})}
                 \Kdelta_{\vec{k}_{\vec{n}}((\mat{M}^{\E{6}}_a){}^T \vec{k}_{\tilde{\vec{n}}})}
                 \Kdelta_{\vec{k}_{\vecnp}((\mat{M}^{\E{6}}_b){}^T \vec{k}_{\tilde{\vec{n}}})} + {} \nonumber \\
        & {} \qquad + \frac{1}{2} \sum_{\vec{\tilde{n}} \in \setN^{\E{6}}_{1C}}
                \sum_{a,b\in\{0, A\}} e^{i\vec{k}_{\vec{n}} \cdot (\vec{T}^{(a)} - \vec{T}^{(b)})}
                 \Kdelta_{\vec{k}_{\vec{n}}((\mat{M}^{\E{6}}_a){}^T \vec{k}_{\tilde{\vec{n}}})}
                 \Kdelta_{\vec{k}_{\vecnp}((\mat{M}^{\E{6}}_b){}^T \vec{k}_{\tilde{\vec{n}}})} + {} \nonumber \\
        & \left. \qquad
                {} + \frac{1}{4}
                \sum_{\vec{\tilde{n}}\in \setN^{\E{6}}_2}
                \sum_{a,b\in\{0, A, B, C\}} e^{i\vec{k}_{\vec{n}} \cdot (\vec{T}^{(a)} - \vec{T}^{(b)})}
                 \Kdelta_{\vec{k}_{\vec{n}}((\mat{M}^{\E{6}}_a){}^T \vec{k}_{\tilde{\vec{n}}})}
                 \Kdelta_{\vec{k}_{\vecnp}((\mat{M}^{\E{6}}_b){}^T \vec{k}_{\tilde{\vec{n}}})} 
         \right] ,
         \nonumber
\end{empheq}
where $V_{\E{6}}$ is given in \eqref{eqn:VE6}, $\vec{T}^{(0)} \equiv \vec{0}$, and $\vec{T}^{(a)} \equiv \vec{T}^{\E{6}}_a$ for $a \in \{ A, B, C \}$.

The rotation properties of the spherical harmonics can again be used to simplify the eigenmodes of \E{6} in the harmonic basis.
Though there are multiple axes of rotation, the fact that they are half turns allows by direct computation to show that
\begin{align}
    Y_{\ell m}^*((\mat{M}^{\E{6}}_A){}^T \unitvec{k}_{\vec{n}})
        &= (-1)^\ell Y^*_{\ell ~-m}(\unitvec{k}_{\vec{n}}),
    \nonumber\\
    Y_{\ell m}^*((\mat{M}^{\E{6}}_B){}^T \unitvec{k}_{\vec{n}})
        &= (-1)^{\ell+m} Y^*_{\ell ~-m}(\unitvec{k}_{\vec{n}}), \\
    Y_{\ell m}^*((\mat{M}^{\E{6}}_C){}^T \unitvec{k}_{\vec{n}})
        &= (-1)^m Y_{\ell m}^*(\unitvec{k}_{\vec{n}}) . \nonumber
\end{align}
With these the eigenmodes in the harmonic basis can be written in a number of useful forms.
We have general expressions patterned after the expressions given for \E{1}--\E{5},
\begin{empheq}[box=\fbox]{align}
    \xi^{\E{6};\unitvec{k}_{\vec{n}}}_{k_{\vec{n}}\ell m} 
    &= \frac{1}{\sqrt{2}} i^\ell Y^*_{\ell m}(\unitvec{k}_{\vec{n}}) \left[ e^{-i \vec{k}_{\vec{n}}\cdot\vec{x}_0} + (-1)^\ell e^{i \vec{k}_{\vec{n}} \cdot \vec{x}_0} e^{i \vec{k}_{\vec{n}} \cdot \vec{T}^{\E{6}}_B} \right] , \quad \mbox{for } \vec{n} \in \setN^{\E{6}}_{1A} , \nonumber \\
    \xi^{\E{6};\unitvec{k}_{\vec{n}}}_{k_{\vec{n}}\ell m} 
    &= \frac{1}{\sqrt{2}} i^\ell Y^*_{\ell m}(\unitvec{k}_{\vec{n}}) \left[ e^{-i \vec{k}_{\vec{n}}\cdot\vec{x}_0} + (-1)^\ell e^{i \vec{k}_{\vec{n}} \cdot \vec{x}_0} e^{i \vec{k}_{\vec{n}} \cdot \vec{T}^{\E{6}}_C} \right] , \quad \mbox{for } \vec{n} \in \setN^{\E{6}}_{1B} , \nonumber \\
    \xi^{\E{6};\unitvec{k}_{\vec{n}}}_{k_{\vec{n}}\ell m} 
    &= \frac{1}{\sqrt{2}} i^\ell Y^*_{\ell m}(\unitvec{k}_{\vec{n}}) \left[ e^{-i \vec{k}_{\vec{n}}\cdot\vec{x}_0} + (-1)^\ell e^{i \vec{k}_{\vec{n}} \cdot \vec{x}_0} e^{i \vec{k}_{\vec{n}} \cdot \vec{T}^{\E{6}}_A} \right] , \quad \mbox{for } \vec{n} \in \setN^{\E{6}}_{1C} , \\
    \xi^{\E{6};\unitvec{k}_{\vec{n}}}_{k_{\vec{n}}\ell m} 
     &= \frac{1}{\sqrt{4}} i^\ell\left[ Y^*_{\ell m}(\unitvec{k}_{\vec{n}})
        \left( 1 + (-1)^m e^{-i \vec{k}_{\vec{n}} \cdot \mat{M}^{\E{6}}_C \vec{x_0}} e^{i\vec{k}_{\vec{n}} \cdot \vec{T}^{\E{6}}_C} \right) + {} \right. \nonumber \\
      & {} \qquad + (-1)^\ell Y^*_{\ell -m}(\unitvec{k}_{\vec{n}}) \left(
        e^{-i \vec{k}_{\vec{n}} \cdot \mat{M}^{\E{6}}_A \vec{x_0}} e^{i\vec{k}_{\vec{n}} \cdot \vec{T}^{\E{6}}_A}
        \right. + {} \nonumber \\
      & {} \left. \left. \qquad \qquad \qquad \qquad {} 
        + (-1)^m e^{-i \vec{k}_{\vec{n}} \cdot \mat{M}^{\E{6}}_B \vec{x_0}} e^{i\vec{k}_{\vec{n}} \cdot \vec{T}^{\E{6}}_B} \right) \right], \quad \mbox{for } \vec{n}\in \setN^{\E{6}}_2 , \nonumber
\end{empheq}
and the spherical-harmonic space covariance matrix has the form \eqref{eqn:HarmonicCovariance}.

\subsection{\E{11}: Chimney space}
\label{secn:eigenmodesE11}

The chimney space is similar to the 3-torus though with only two compact dimensions and thus a finite ``cross-sectional area'' \eqref{eqn:AE11}.
As in \E{1}, we begin with the covering-space (\E{18}) eigenmodes \eqref{eqn:EuclideanFourierBasis} and restrict ourselves to the set that respect the \E{11} symmetries,
\begin{equation}
    \Upsilon^{\E{11}}_{\vec{k}} (g^{\E{11}}_{\A{j}} \vec{x}) = \Upsilon^{\E{11}}_{\vec{k}} (\vec{x}).
\end{equation}
\restoregeometry
Also as in \E{1}, since all the group elements of $\Gamma^{\E{11}}$ are pure translations, only one \E{18} eigenmode will contribute for each allowed $\vec{k}$.
The two generators of \E{11} \eqref{eqn:E11generalT} lead to two (rather than three as in \E{1}) discretization conditions following \eqref{eqn:associatedhomogeneousdiscretization},
\begin{align}
    \label{eqn:E11discretization}
    2\pi n_1 &= (\vec{k}_{\vec{n}})_{x} L_1, \\
    2\pi n_2 &= (\vec{k}_{\vec{n}})_{x} L_2 \cos\alpha + (\vec{k}_{\vec{n}})_{y} L_2 \sin\alpha . \nonumber
\end{align}
These can be inverted to determine the components of the allowed wavevectors,
\begin{align}
    \label{eqn:E11_ki}
    (\vec{k}_{\vec{n}})_{x} &= \frac{2\pi n_1}{L_1}, \\
    (\vec{k}_{\vec{n}})_{y} &= \frac{2\pi n_2}{L_2\sin\alpha} - \frac{2\pi n_1}{L_1}\frac{\cos\alpha}{\sin\alpha} , \nonumber
\end{align}
while $(\vec{k}_{\vec{n}})_{z}\equiv k_z$ is unconstrained.
We will write $\vec{k}_{\vec{n}}$ as a shorthand for the wavevector parametrized by $(n_1,n_2; k_z)$, i.e., by the integers $n_1$ and $n_2$ and the real variable $k_z$.
The usual (untilted) results are recovered for $\alpha=\pi/2$.

Thus
\begin{empheq}[box=\fbox]{equation}
    \Upsilon^{\E{11}}_{\vec{k}_{\vec{n}}}(\vec{x}) = e^{i\vec{k}_{\vec{n}}\cdot(\vec{x}-\vec{x}_0)}\,, \quad \mbox{for } \vec{n}\in \setN^{\E{11}}\,,
\end{empheq}
where
\begin{empheq}[box=\fbox]{equation}
    \setN^{\E{11}} \equiv \{(n_1,n_2) \vert n_i\in\integers\}\,.
\end{empheq}

Following \eqref{eqn:CE18XY} the Fourier-mode correlation matrix for \E{11} is
\begin{empheq}[box=\fbox]{align}
    \label{eqn:CE11XY}
    C^{\E{11};XY}_{\vec{k}_{\vec{n}} \vec{k}_{\vecnp}} 
    &= 2\pi A_{\E{11}}\frac{2\pi^2}{k_{\vec{n}}^3}{\mathcal{P}}^{\mathcal{R}}(k_{\vec{n}}) 
    \Delta^X(k_{\vec{n}}) \DeltaYstar(k_{\vec{n}}) 
    \Kdelta_{(\vec{k}_{\vec{n}})_x (\vec{k}_{\vecnp})_x}
    \Kdelta_{(\vec{k}_{\vec{n}})_y (\vec{k}_{\vecnp})_y}
    \Ddelta(k_z - k'_z)
    \,.
\end{empheq}
In transitioning from \E{1} (compare \cref{eqn:CE11XY} to \cref{eqn:CE1XY}) we have replaced $V_{\E{1}}$ with $2\pi A_{\E{11}}$ (the cross-sectional area given by \eqref{eqn:AE11}) and  $\Kdelta_{\vec{k}_{\vec{n}}\vec{k}_{\vecnp}}$ by a Kronecker delta for the $x$ and $y$ components of $\vec{k}_{\vec{n}}$ and a Dirac delta function for the $z$ component.

We can project the field $\delta^X$ onto the sky by performing a radial integral with a suitable weight function and transfer function, giving
\begin{align}
    \label{eqn:almE11}
    a^{\E{11};X}_{\ell m} 
    &= \frac{4\pi}{2\pi A_{\E{11}}}
    \sum_{(n_1,n_2)\in\setN^{\E{11}} } \int_{-\infty}^{\infty} \mathrm{d} k_z \,
    \delta^{\mathcal{R}}_{\vec{k}_{\vec{n}}} 
     \xi^{\E{11};\unitvec{k}_{\vec{n}}}_{k_{\vec{n}} \ell m}
     \Delta^X_\ell(k_{\vec{n}}),
\end{align}
with
\begin{empheq}[box=\fbox]{equation}
    \xi^{\E{11};\unitvec{k}_{\vec{n}}}_{k_{\vec{n}} \ell m} \equiv e^{-i\vec{k}_{\vec{n}}\cdot\vec{x}_0} i^\ell  Y_{\ell m}^*(\unitvec{k}_{\vec{n}})\,,
\end{empheq}
where a similar transition to that above was performed in starting from \cref{eqn:almE1} and replacing the sum over the $(\vec{k}_{\vec{n}})_z$ with the integral over $k_z$.

For the chimney spaces \E{i} with $i=11$ or $i=12$, the spherical-harmonic space covariance matrix now has the form
\begin{empheq}[box=\fbox]{equation}
    \label{eqn:HarmonicCovarianceE11}
    C^{\E{i};XY}_{\ell m\ell'm'}  =
     \frac{(4\pi)^2}{{2\pi A_{\E{i}}}}
    \sum_{(n_1,n_2)\in \setN^{\E{i}}} \int_{-\infty}^{\infty} \mathrm{d} k_z \,
    \Delta^X_{\ell}(k_{\vec{n}})
    \DeltaYstarforell_{\ell'}(k_{\vec{n}})
    \frac{2\pi^2 \mathcal{P}^{\mathcal{R}}(k_{\vec{n}})}{k_{\vec{n}}^3}
    \xi^{\E{i};\unitvec{k}_{\vec{n}}}_{k_{\vec{n}} \ell m}
    \xi^{\E{i};\unitvec{k}_{\vec{n}}*}_{k_{\vec{n}} \ell' m'} .
\end{empheq}

\subsection{\E{12}: Chimney space with half turn}
\label{secn:eigenmodesE12}

The eigenspectrum and eigenmodes of the chimney space with half turn can be determined in a manner analogous to that of the chimney space in much the same way that these quantities for $\E{2}$ were determined from \E{1}.
For \E{12}, the discretization condition \eqref{eqn:associatedhomogeneousdiscretization} leads to the components of the allowed wavevectors,
\begin{equation}
    (\vec{k}_{\vec{n}})_{x} = \frac{2\pi n_1}{L_{Ax}} , \quad (\vec{k}_{\vec{n}})_{z} = \frac{\pi n_2}{L_{Bz}} ,
\end{equation}
with $k_y$ again unconstrained.
As in \E{11}, we will write $\vec{k}_{\vec{n}}$ as a shorthand for the wavevector parametrized by the integer array $\vec{n}=(n_1,n_2)$ and the real variable $k_y$.

Unlike in \E{11}, the eigenmodes of \E{12} can include a linear combination of two \E{11} eigenmodes.
Here $(\mat{M}^{\E{12}}_B)^N \vec{k}_{\vec{n}} =\vec{k}_{\vec{n}}$ has two solutions, $N=1$ and $N=2$.
Written explicitly,
\begin{description}
    \item[\textbf{$N=1$ eigenmodes: }]  $\vec{k}_{\vec{n}}=\transpose{(0,0,(\vec{k}_{\vec{n}})_z)}$, 
        i.e., $\vec{n}=(0,n_2)$, $n_2 \in 2\integers^{\neq 0}$, $k_y = 0$, with
    \begin{empheq}[box=\fbox]{equation}
        \Upsilon^{\E{12}}_{\vec{k}_{\vec{n}}}(\vec{x}) = e^{i\vec{k}_{\vec{n}} \cdot (\vec{x}-\vec{x}_0)}
        = e^{i \pi n_2 (z - z_0) / L_{Bz}},
    \end{empheq}
    \item[\textbf{$N=2$ eigenmodes: }] $((\vec{k}_{\vec{n}})_{x}, (\vec{k}_{\vec{n}})_{z}) \neq (0,0)$, 
        i.e., $(n_1,n_2) \neq (0,0)$, with
    \begin{empheq}[box=\fbox]{equation}
        \Upsilon^{\E{12}}_{\vec{k}_{\vec{n}}}(\vec{x}) = \frac{1}{\sqrt{2}} \left( e^{i\vec{k}_{\vec{n}}\cdot(\vec{x}-\vec{x}_0)} + e^{i\vec{k}_{\vec{n}}\cdot(\mat{M}^{\E{12}}_B(\vec{x}-\vec{x}_0)+\vec{T}^{\E{12}}_B)} \right).
    \end{empheq}
\end{description}
Here the two sets of allowed modes are defined by
\begin{empheq}[box=\fbox]{align}
     \setN^{\E{12}}_1 &\equiv \{(0,n_2) \vert n_2\in 2\integers^{\neq0}\}\,, \nonumber\\
    \setN^{\E{12}}_2 &\equiv \{(n_1,n_2) \vert n_1 \in \integers^{>0}, n_2 \in \integers\} , \\
    \setN^{\E{12}} &\equiv \setN^{\E{12}}_1 \cup \setN^{\E{12}}_2 \,.\nonumber
\end{empheq}

With these the Fourier-mode correlation matrix can now be expressed as
\begin{empheq}[box=\fbox]{align}
\label{eqn:E12FourierCovarianceStandardConvention}
    C^{\E{12};XY}_{\vec{k}_{\vec{n}}\vec{k}_{\vecnp}} 
     = {} & 2\pi A_{\E{12}} \frac{2\pi^2}{k^3_{\vec{n}}}
        {\mathcal{P}}^{\mathcal{R}}(k_{\vec n})\Delta^X(k_{\vec{n}})\DeltaYstar(k_{\vec{n}})
        e^{i(\vec{k}_{\vecnp}-\vec{k}_{\vec{n}})\cdot\vec{x}_0} \times {} \nonumber \\
        & {} \; \times
            \frac{1}{2} \!\!\!\!\!\!\!\!\!\!\! 
                \sum_{(\tilde{n}_1, \tilde{n}_2)\in \setN^{\E{12}}_2}
                \int_{-\infty}^{\infty} \mathrm{d}\tilde{k}_y \, \sum_{a=0}^1\sum_{b=0}^1
                e^{i\vec{k}_{\vec{\tilde{n}}}\cdot(\vec{T}^{(a)}-\vec{T}^{(b)})} 
                \Kdelta_{(\vec{k}_{\vec{n}})_x (\vec{k}^{(a)}_{\vec{\tilde{n}}})_x}
                \Kdelta_{(\vec{k}_{\vecnp})_x (\vec{k}^{(b)}_{\vec{\tilde{n}}})_x}
                \\
            & \qquad\quad {} \times
                 \Kdelta_{(\vec{k}_{\vec{n}})_z (\vec{k}^{(a)}_{\vec{\tilde{n}}})_z}
                 \Kdelta_{(\vec{k}_{\vecnp})_z (\vec{k}^{(b)}_{\vec{\tilde{n}}})_z}
                 \Ddelta(k_y - \tilde{k}^{(a)}_y)
                 \Ddelta(k'_y - \tilde{k}^{(b)}_y)
         , 
         \nonumber
\end{empheq}
where the terms with $\vec{n} \in \setN^{\E{12}}_1$ are of measure zero and have been dropped, $A_{\E{12}}$ is given in \eqref{eqn:AE12}, $\vec{T}^{(0)}\equiv\vec{0}$, $\vec{T}^{(1)}\equiv\vec{T}^{\E{12}}_B$, $\vec{k}^{(a)}_{\tilde{\vec{n}}} \equiv [(\mat{M}^{\E{12}}_B){}^T]^a \vec{k}_{\vec{\tilde{n}}}$, and $\tilde{k}^{(a)}_y \equiv (\vec{k}^{(a)}_{\tilde{\vec{n}}})_y$.

In the harmonic basis we have
\begin{empheq}[box=\fbox]{align}
    \xi^{\E{12};\unitvec{k}_{\vec{n}}}_{k_{\vec{n}}\ell m} 
    &= e^{-i \vec{k}_{\vec{n}}\cdot\vec{x}_0} i^\ell Y^*_{\ell m} (\unitvec{k}_{\vec{n}}), \quad \mbox{for } \vec{n} \in \setN^{\E{12}}_1 \,,\\
    \xi^{\E{12};\unitvec{k}_{\vec{n}}}_{k_{\vec{n}}\ell m} 
     &= \frac{1}{\sqrt{2}} i^\ell Y^*_{\ell m}(\unitvec{k}_{\vec{n}})
        \sum_{j=0}^1 i^{m j}
        e^{-i\vec{k}_{\vec{n}}\cdot(\mat{M}^{\E{12}}_B)^j\vec{x}_0}
        e^{i \vec{k}_{\vec{n}}\cdot\mat{M}^{\E{12}}_{0j}\vec{T}^{\E{12}}_B}\,,\quad \mbox{for } \vec{n}\in \setN^{\E{12}}_2 , \nonumber
\end{empheq}
and the spherical-harmonic space covariance matrix has the form \eqref{eqn:HarmonicCovarianceE11}.

\subsection{\E{16}: Slab space including rotation}
\label{secn:eigenmodesE16}

As discussed above (\cref{secn:topologyE16}), there are two physically distinct cases for \E{16}, which will be discussed separately.

\subsubsection{\slabh: Conventional unrotated slab space}

Similar to \E{1} and \E{11}, \slabh\ is homogeneous.
It is compact in only one dimension so there is only one generator of the topology.
The one discretization condition following from \eqref{eqn:associatedhomogeneousdiscretization} leads to
\begin{equation}
     \label{eqn:E16h_ki}
    (\vec{k}_{\vec{n}})_{z} = \frac{2\pi n}{L}\,,
\end{equation}
while $(\vec{k}_{\vec{n}})_x \equiv k_x$ and $(\vec{k}_{\vec{n}})_y \equiv k_y$ are unconstrained.
As in the chimney spaces, we will again write $\vec{k}_{\vec{n}}$ as a shorthand for the wavevector characterized by the integer $n$ and the real variables $k_x$ and $k_y$.

Thus
 \begin{empheq}[box=\fbox]{align}
    \Upsilon^{\slabh}_{\vec{k}_{\vec{n}}}(\vec{x}) = 
        e^{i\vec{k}_{\vec{n}}\cdot(\vec{x}-\vec{x}_0)}, \quad \mbox{for } n\in \setN^{\slabh},
\end{empheq}
where
\begin{empheq}[box=\fbox]{equation}
    \setN^{\slabh} = \{n\in \integers \} .
\end{empheq}

Following \eqref{eqn:CE18XY} the Fourier-mode correlation matrix for \slabh\ is
\begin{empheq}[box=\fbox]{align}
     \label{eqn:CE16hXY}
    C^{\slabh;XY}_{\vec{k}_{\vec{n}}\vec{k}_{\vecnp}} 
        &= (2\pi)^2 L \frac{2\pi^2}{k^3}
        \mathcal{P}^{\mathcal{R}}(k) \Delta^X(k_{\vec{n}}) \DeltaYstar(k_{\vec{n}}) \times {} \\
        &\qquad \quad {} \times
        \Ddelta(k_x - k'_x) \Ddelta(k_y - k'_y)
        \Kdelta_{(\vec{k}_{\vec{n}})_z (\vec{k}_{\vecnp})_z}.
        \nonumber
\end{empheq}
In transitioning from \E{1} (compare \cref{eqn:CE16hXY} to \cref{eqn:CE1XY}) we have replaced $V_{\E{1}}$ with $(2\pi)^2 L$ and $\Kdelta_{\vec{k}_{\vec{n}}\vec{k}_{\vecnp}}$ by a Kronecker delta for the $z$ component of $\vec{k}_{\vec{n}}$ and Dirac delta functions for the $x$ and $y$ components.

We can project the field $\delta^X$ onto the sky by performing a radial integral with a suitable weight function and transfer function, giving
\begin{equation}
    \label{eqn:almE16h}
    a^{\slabh;X}_{\ell m} 
     = \frac{4\pi}{(2\pi)^2 L}
        \sum_{n\in \setN^{\slabh}}
        \int_{-\infty}^\infty \mathrm{d} k_x\, \int_{-\infty}^\infty \mathrm{d} k_y \,
        \delta^{\mathcal{R}}_{\vec{k}_{\vec{n}}}
        \xi^{\slabh;\unitvec{k}_{\vec{n}}}_{k_{\vec{n}} \ell m}
         \Delta^X_\ell(k_{\vec{n}}), 
\end{equation}
with
\begin{empheq}[box=\fbox]{equation}
    \label{eqn:xiE16h}
    \xi_{k_{\vec{n}} \ell m}^{\slabh;\unitvec{k}_{\vec{n}}}
    \equiv  e^{-i\vec{k}_{\vec{n}}\cdot\vec{x_0}} i^\ell Y^*_{\ell m}(\unitvec{k}_{\vec{n}})\,,\quad n\in \setN^{\slabh} ,
\end{empheq}
where a similar transition to that above was performed in starting from \cref{eqn:almE1}.

For the slab spaces the spherical-harmonic space covariance matrix now has the form
\begin{empheq}[box=\fbox]{equation}
    \label{eqn:HarmonicCovarianceE16h}
    C^{\E{16}^{(a)};XY}_{\ell m\ell'm'} 
    = \frac{(4\pi)^2}{(2\pi)^2 L}
    \sum_{n\in \setN^{\E{16}^{(a)}}}
    \int_{-\infty}^\infty \mathrm{d} k_x\, \int_{-\infty}^\infty \mathrm{d} k_y \, 
    \Delta^X_{\ell}(k_{\vec{n}})
    \DeltaYstar_{\ell'}(k_{\vec{n}})
    \frac{2\pi^2 \mathcal{P}^{\mathcal{R}}(k)}{k_{\vec{n}}^3} 
    \xi^{\E{16}^{(a)};\unitvec{k}_{\vec{n}}}_{k_{\vec{n}} \ell m}
    \xi^{\E{16}^{(a)};\unitvec{k}_{\vec{n}}*}_{k_{\vec{n}} \ell' m'} \,,
\end{empheq}
where $a \in \{\mathrm{h},\mathrm{i}\}$ labels the homogeneous (\slabh) and inhomogeneous (\slabi) slab spaces.

\newpage
\newgeometry{textwidth=0.8\paperwidth, textheight=0.8\paperheight}
\subsubsection{\slabi: General rotated slab space}

The eigenspectrum and eigenmodes of \slabi\ can be determined from \slabh\ in a manner similar to that for \E{12} from \E{11} since $\mat{M}^{\slabi}_B$ is a rotation around the $z$-axis.
For \slabi, the discretization condition \eqref{eqn:associatedhomogeneousdiscretization} leads to the component of the allowed wavevectors,
\begin{equation}
    \label{eqn:kzinE16i}
    (\vec{k}_{\vec{n}})_{z} = \frac{2\pi n}{q L\sin\beta} ,
\end{equation}
while $(\vec{k}_{\vec{n}})_x \equiv k_x$ and $(\vec{k}_{\vec{n}})_y \equiv k_y$ are again unconstrained.
As in \slabh, we will again write $\vec{k}_{\vec{n}}$ as a shorthand for the wavevector characterized by the integer $n$ and the real variables $k_x$ and $k_y$.

Unlike in \slabh, the eigenmodes of \slabi\ can include linear combinations of \slabh\ eigenmodes.
Here $(\mat{M}^{\slabi}_B)^N \vec{k}_{\vec{n}} = \vec{k}_{\vec{n}}$ has two solutions, $N=1$ and $N=q$. Written explicitly,
\begin{description}
    \item[\textbf{$N=1$ eigenmodes: }]  $\vec{k}_{\vec{n}}=\transpose{(0, 0, (\vec{k}_{\vec{n}})_z)}$, 
        i.e., $n \in q\integers^{\neq 0}$, $k_x = k_y = 0$, with
    \begin{empheq}[box=\fbox]{equation}
        \Upsilon^{\slabi}_{\vec{k}_{\vec{n}}}(\vec{x}) = e^{i\vec{k}_{\vec{n}} \cdot (\vec{x}-\vec{x}_0)},
    \end{empheq}
    \item[\textbf{$N=q$ eigenmodes: }] $(k_x, k_y) \neq (0,0)$, $n \in \integers$, with
    \begin{empheq}[box=\fbox]{equation}
        \Upsilon^{\slabi}_{\vec{k}_{\vec{n}}}(\vec{x}) = \frac{1}{\sqrt{q}} \sum_{j=0}^{q-1} e^{i\vec{k}_{\vec{n}}\cdot(\mat{M}^{\slabi}_B)^j(\vec{x}-\vec{x}_0)} e^{i \vec{k}_{\vec{n}} \cdot \mat{M}^{\slabi}_{0j} \vec{T}^{\slabi}_B} ,
    \end{empheq}
\end{description}
and $\mat{M}^{\slabi}_{00} \equiv 0$ and $\mat{M}^{\slabi}_{0j}$ defined in \eqref{eqn:M0jdef}.
Here the two sets of allowed modes are defined by
\begin{empheq}[box=\fbox]{align}
    \setN^{\slabi}_1 &= \{n\in q\integers^{\neq0} \} , \nonumber \\
    \setN^{\slabi}_q &= \{n\in \integers \} , \\
    \setN^{\slabi} &= \setN^{\slabi}_1 \cup \setN^{\slabi}_q . \nonumber
\end{empheq}

With these the Fourier-mode correlation matrix can now be expressed as
\begin{empheq}[box=\fbox]{align}
     \label{eqn:E16iFourierCovarianceStandardConvention}
    C^{\slabi;XY}_{\vec{k}_{\vec{n}}\vec{k}_{\vecnp}}  
     &= (2\pi)^2 L_{\slabi}
        \frac{2\pi^2}{k_{\vec{n}}^3} \mathcal{P}^{\mathcal{R}}(k_{\vec{n}}) \Delta^X(k_{\vec{n}}) \DeltaYstar(k_{\vec{n}})
            e^{i(\vec{k}_{\vecnp}-\vec{k}_{\vec{n}})\cdot\vec{x}_0} \times {} \\
     & \quad {} \times \frac{1}{q} \!\!\!\!\!\!
     \sum_{\tilde{n} \in \setN^{\slabi}_q} \int_{-\infty}^{\infty} \mathrm{d} k_x\, \int_{-\infty}^{\infty} \mathrm{d} k_y\, \sum_{a=0}^{q-1}\sum_{b=0}^{q-1}
        e^{i\vec{k}_{\tilde{\vec{n}}} \cdot (\vec{T}^{(a)} - \vec{T}^{(b)})} 
        \Kdelta_{(\vec{k}_{\vec{n}})_z (\vec{k}_{\tilde{\vec{n}}}^{(a)})_z} \Kdelta_{(\vec{k}_{\vecnp})_z (\vec{k}_{\tilde{\vec{n}}}^{(b)})_z}
        \times {} \nonumber \\
     & \qquad {} \times
        \Ddelta(k_x - \tilde{k}^{(a)}_{x}) \Ddelta(k'_x - \tilde{k}^{(b)}_{x})
        \Ddelta(k_y - \tilde{k}^{(a)}_{y}) \Ddelta(k'_y - \tilde{k}^{(b)}_{y}) , \nonumber
\end{empheq}
where the terms with $n\in\setN^{\slabi}_1$ are of measure zero and have been dropped, $L_{\slabi}$ is given in \eqref{eqn:LE16i}, $\vec{T}^{(0)} \equiv \vec{0}$, $\vec{T}^{(a)} \equiv \mat{M}^{\slabi}_{0a} \vec{T}^{\slabi}_B$ for $a \in \{ 1,\ldots, q-1 \}$, $\vec{k}^{(a)}_{\tilde{\vec{n}}} \equiv [(\mat{M}^{\slabi}_B){}^T]^a \vec{k}_{\vec{\tilde{n}}}$, and $\tilde{k}^{(a)}_w \equiv (\vec{k}^{(a)}_{\tilde{\vec{n}}})_w$ for $w \in \{x, y\}$.

In the harmonic basis we have
\begin{empheq}[box=\fbox]{align}
    \xi^{\slabi;\unitvec{k}_{\vec{n}}}_{k_{\vec{n}}\ell m} 
    &= e^{-i \vec{k}_{\vec{n}}\cdot\vec{x}_0} i^\ell Y^*_{\ell m} (\unitvec{k}_{\vec{n}}), \quad \mbox{for } n \in \setN^{\slabi}_1 ,\\
    \xi^{\slabi;\unitvec{k}_{\vec{n}}}_{k_{\vec{n}}\ell m} 
     &= \frac{1}{\sqrt{q}} i^\ell Y^*_{\ell m}(\unitvec{k}_{\vec{n}})
        \sum_{j=0}^{q-1} e^{i\pi m j p / q}
        e^{-i\vec{k}_{\vec{n}}\cdot(\mat{M}^{\slabi}_B)^j\vec{x}_0}
        e^{i \vec{k}_{\vec{n}}\cdot\mat{M}^{\slabi}_{0j}\vec{T}^{\slabi}_B}, \; \mbox{for } n\in \setN^{\slabi}_q , \nonumber
\end{empheq}
and the spherical-harmonic space covariance matrix has the form \eqref{eqn:HarmonicCovarianceE16h}.
\restoregeometry

\subsection{\E{18}: The covering space of \Espace}
\label{secn:eigenmodesE18}

The covering space itself has all of the usual Fourier and spherical-harmonic mode functions, along with a simple isotropic correlation structure. The eigenmodes are
\begin{empheq}[box=\fbox]{align}
    \Upsilon^{\E{18}}_{\vec{k}}(\vec{x}) 
        &= e^{i \vec{k} \cdot (\vec{x}-\vec{x}_0)} \,,
\end{empheq}
the Fourier-mode correlation matrix is
\begin{empheq}[box=\fbox]{align}
     \label{eqn:E18FourierCovarianceStandardConvention}
    C^{\E{18};XY}_{\vec{k}\vec{k}'} 
        &= 
        \frac{2\pi^2}{k^3} \mathcal{P}^{\mathcal{R}}(k) \Delta^X(k)\Delta^{Y*}(k)\Ddelta(\vec{k}-\vec{k}'),
\end{empheq}
and the spherical-harmonic space covariance matrix has the form
\begin{empheq}[box=\fbox]{align}
    \label{eqn:E18HarmonicCovarianceStandardConvention}
    C^{\E{18};XY}_{\ell m\ell'm'} 
    &=
     \frac{2}{\pi}
    \int \mathrm{d}^3\vec{k} \,
    \Delta^X_{\ell}(\vec{k}) \DeltaYstarforell_{\ell'}(\vec{k})
    \frac{2\pi^2 \mathcal{P}^{\mathcal{R}}(k)}{k^3}
    \xi_{\ell m}^{\E{18};\vec{k}}
    \xi_{\ell' m'}^{\E{18};\vec{k}*},
\end{empheq}
with
\begin{empheq}[box=\fbox]{align}
    \xi^{\E{18};\vec{k}}_{\ell m}
     &= 
     e^{-i\vec{k}\cdot\vec{x}_0}
     i^\ell
     Y^*_{\ell m}(\unitvec{k}).
\end{empheq}
In the typical case (e.g., CMB temperature and polarization) where $\Delta^X_{\ell}(\vec{k})=\Delta^X_{\ell}(k)$, \cref{eqn:E18HarmonicCovarianceStandardConvention} gives the isotropic
\begin{equation}
C^{\E{18};XY}_{\ell m\ell'm'} = C^{\E{18};XY}_\ell\Kdelta_{\ell\ell'}\Kdelta_{mm'},
\end{equation}
with the angular power spectrum given by
\begin{equation}
	C^{\E{18};XY}_\ell = 4\pi\int \frac{\mathrm{d}k}{k}\mathcal{P}^{\mathcal{R}}(k)\Delta^X_{\ell}(k) \DeltaYstarforell_{\ell}(k)\;.
\end{equation}
\section{Numerical analysis}
\label{secn:numerical_results}

As described above, to the extent that the CMB is Gaussian, all the information about the temperature anisotropies is given by the 2-point correlation matrix of the spherical-harmonic coefficients, namely the covariance matrix $C_{\ell m\ell'm'}^{\E{i};XY} = \langle a_{\ell m}^{\E{i};X}  a_{\ell' m'}^{\E{i};Y*} \rangle$ \cite{Hu2002}.
In the familiar case of the isotropic covering space, we have the usual formula  $\langle  a_{\ell m}^{\E{18};X}  a_{\ell' m'}^{\E{18};Y*} \rangle = C_\ell^{\E{18};XY} \Kdelta_{\ell\ell'}\Kdelta_{mm'}$, where only the diagonal elements are non-zero and they are independent of $m$.
But a non-trivial topology breaks the assumption of isotropy (see, e.g., \rcite{Riazuelo2004:prd}), inducing non-zero values for off-diagonal components of the correlation matrix.
In this section, we numerically compute the correlation matrix elements for sample manifolds of each of the compact, orientable, Euclidean topologies \E{1}--\E{6}.\footnote{
    The non-compact cases \E{11}, \E{12}, and \slabh\ can be regarded as limiting cases of \E{1} and  \E{2}.
    We reserve detailed investigation of \slabi\ for future specific work.}
We then estimate the detectability of this off-diagonal signal of topology given the cosmic variance associated with the primary CMB temperature anisotropies.

\subsection{Evaluation of CMB temperature covariance matrices}
\label{subsecn:numerical_covariances}

The covariance matrices for the fully compact, orientable, Euclidean topologies can generally be written like Eq.~\eqref{eqn:HarmonicCovariance}, which we rewrite here for auto-correlation of the CMB temperature (T) fluctuations:
\begin{align}
\label{eqn:general_covariance_matrix}
    C^{\E{i};\, \mathrm{TT}}_{\ell m\ell'm'} = 
   \frac{(4\pi)^2}{{V_{\E{i}}}}
    \sum_{\vec{n}\in \setN^{\E{i}}} 
    \Delta^{\mathrm{T}}_{\ell}(k_{\vec{n}})
    \Delta^{\mathrm{T}}_{\ell'}(k_{\vec{n}})
    \frac{2\pi^2 \mathcal{P}^{\mathcal{R}}(k_{\vec{n}})}{k_{\vec{n}}^3}
    \xi^{\E{i};\unitvec{k}_{\vec{n}}}_{k_{\vec{n}} \ell m}
    \xi^{\E{i};\unitvec{k}_{\vec{n}}*}_{k_{\vec{n}} \ell' m'}.
\end{align}
For brevity, we will drop the $\mathrm{T}$ and $\mathrm{TT}$ labels in this section. The covering space, \E{18}, reduces this three-dimensional summation to a one-dimensional integral over the magnitude of the wavevector, $\vert\vec{k}\vert$ (or $k$), which is computationally fast to evaluate numerically.
However, for non-trivial topologies, we must perform computationally costly summations.
In principle, we need to perform the sum over an infinite set of wavevectors, but in practice, we must choose a maximum value of $\vert\vec{k}\vert$, $\vert\vec{k}_{\textrm{max}}(\ell)\vert$, at which we stop the summation.
As we are interested in probing Dirichlet domains that fully contain the last scattering surface, we are forced to prioritize large topology scales (heuristically, the various $L_i$) over the computational accuracy that might be achieved by calculating the summation up to a large $|\vec{k}|$.

As an example, consider the cubic \E{1}, the 3-torus with equal lengths $L_{\A{1}}=L_{\A{2}}=L_{\A{3}}=L$. 
Here, the wavevector is
\begin{equation}
    \vec{k} = \frac{2\pi}{L}\vec{n}.
\end{equation}
In making  $L$ larger, we are forced to go to larger values of $\vert\vec{n}\vert$ to achieve a given accuracy for a given element of $C^{\E{1}}_{\ell m\ell'm'}$,
meaning we must execute a larger number of summations.
Since we live in three spatial dimensions, the number of operations we must perform for a given $\{\ell,m, \ell',m'\}$ scales as $\mathcal{O}(L^3)$. Therefore, we need to make a clear definition for our multipole-dependent cutoff $|\vec{k}_{\textrm{max}}(\ell)|$. We define the ratio function
\begin{equation}
    \label{eqn:cut_off_definition}
    R_\ell(|\vec{k}|) = \frac{C^{|\vec{k}|}_\ell}{C^{\Lambda\textrm{CDM}}_\ell},
\end{equation}
where
\begin{equation}
    C^{|\vec{k}|}_\ell = 4\pi \int^{|\vec{k}|}_0 \mathrm{d} k'\,\frac{\mathcal{P}^{\mathcal{R}}(k')}{k'}\Delta_\ell(k')^2,
\end{equation}
and $C^{\Lambda\textrm{CDM}}_\ell$ is the standard $\Lambda$CDM angular power spectrum produced by \texttt{CAMB} \cite{Lewis:1999bs,2011ascl.soft02026L} for \E{18}.
We then find $|\vec{k}_{\textrm{max}}(\ell)|$ such that wavevectors with $|\vec{k}|\leq|\vec{k}_{\textrm{max}}(\ell)|$
would contribute at least 99\% of $C^{\Lambda\textrm{CDM}}_\ell$.
In the limit of $L\rightarrow \infty$, we would find a power spectrum that is 1\% smaller than the actual power spectrum.
For computing the off-diagonal elements $\ell \neq \ell'$, we perform the sum in \cref{eqn:general_covariance_matrix} up to $|\vec{k}_{\textrm{max}}(\max(\ell, \ell'))|$.
Increasing the precision to a higher number than $99\%$ has a negligible impact on the KL divergence we define later.
\begin{figure}
  \centering
\includegraphics[width=\linewidth]{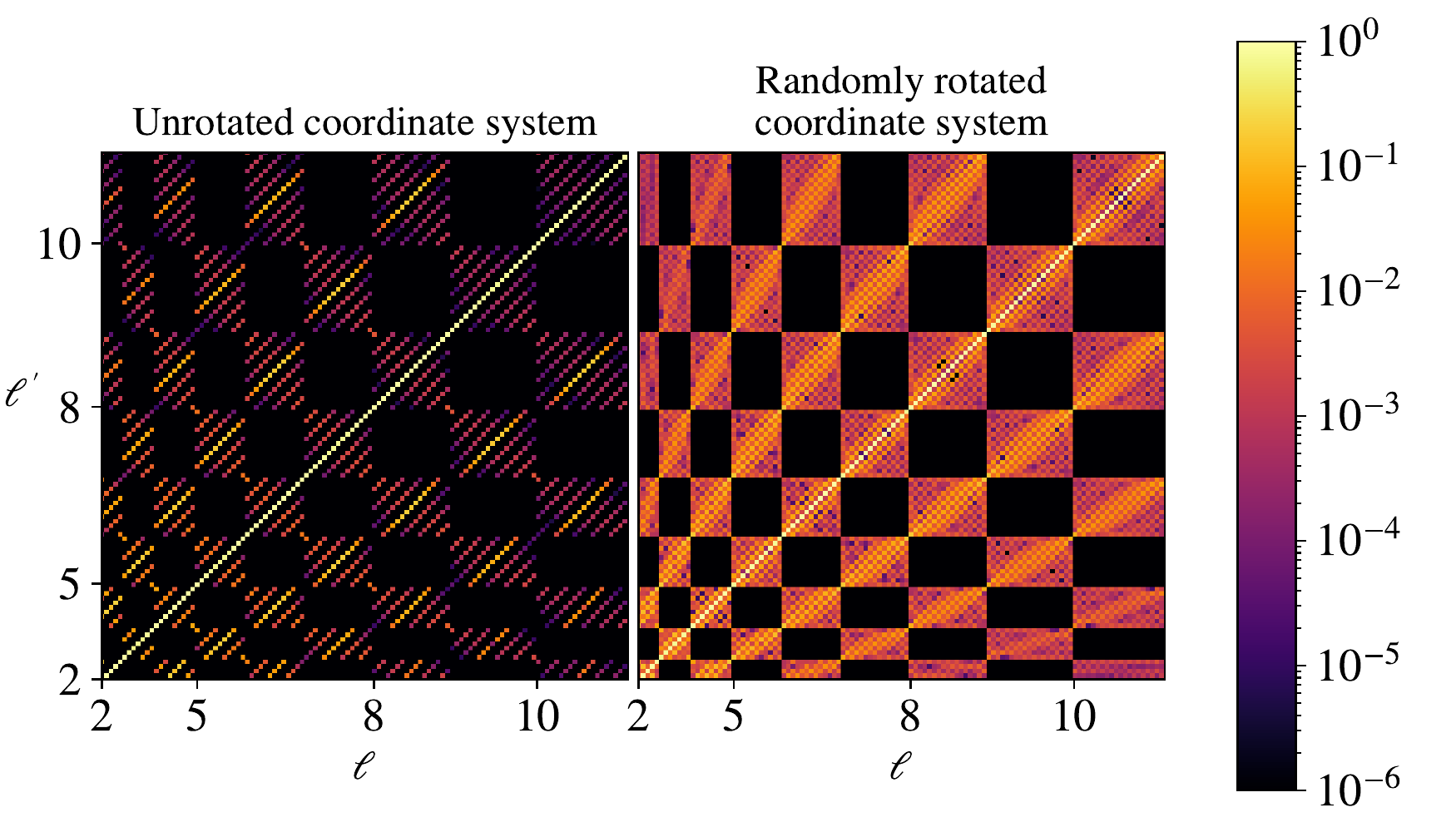}
\caption{Effect of rotating the coordinate system on the CMB temperature covariance matrix. Here, the topology is the untilted \E{2} with $L_{{A_1}}=L_{{A_2}}=1.4L_{\mathrm{LSS}}$ and $L_B = L_{\mathrm{LSS}}$, where $L_{\mathrm{LSS}}$ is the diameter of the last scattering surface, and with an on-axis observer.
    The left panel shows the absolute value of the rescaled covariance matrix $\Xi^{\E{2};\, \mathrm{TT}}_{\ell m \ell' m'}$ (at low multipoles $\ell$) in the coordinate system used for this work as opposed to a random rotation of the coordinate system in the right panel.
    For each $(\ell, \ell')$ block the matrix elements show the corresponding $(m, m')$ in increasing order, i.e., $-\ell \leq m \leq \ell$.
    Although the right panel seems brighter, both plots contain an equal amount of information about the topology of the Universe, as confirmed by both resulting in the same KL divergence.}
\label{fig:rot_C}
\end{figure}

We solve these equations by developing a special-purpose Python code.
To obtain the transfer function $\Delta_\ell(k)$ and the primordial power spectrum $ \mathcal{P}^{\mathcal{R}}(k)$, we use \texttt{CAMB} with the {\it Planck} 2018 best-fit $\Lambda\textrm{CDM}$ cosmological parameters \cite{Planck:2018vyg}.
Before we present the covariance matrices for all the compact, orientable topologies, we note that the choice of coordinate system will affect the appearance and symmetries of the covariance matrix.
We will, in this paper, plot the \emph{rescaled} covariance matrix, which we define as
\begin{equation}
    \Xi^{\E{i}}_{\ell m\ell'm'} \equiv \frac{C^{\E{i}}_{\ell m\ell'm'}} {\sqrt{C^{\Lambda \mathrm{CDM}}_{\ell}C^{\Lambda \mathrm{CDM}}_{\ell'}}},
\end{equation}
and will be of importance later when we discuss the KL divergence.

In \cref{fig:rot_C}, we show absolute values of the rescaled covariance matrix for the unshifted and untilted \E{2} in the coordinate system used for this work.
In the presented plots, we order the $\{\ell,m\}$ elements as $s = \ell(\ell+1) + m$, with $m$ in the range $-\ell \leq m \leq\ell$.
In the right panel of the figure, we randomly rotate the coordinate system by applying the Wigner D-matrices $D^{\ell}_{m\bar{m}}(\theta_0, \phi_0)$ (see, e.g., \rcite{wigner2013group}).
Here, the new coordinates are $\theta' =\theta+\theta_0$ and $\phi'=\phi+\phi_0$ so that the covariance matrix in the new coordinates becomes
\begin{equation}
    C^{\prime\E{i}}_{\ell m\ell'm'} = \sum^{\ell}_{\bar{m}=-\ell}\sum^{\ell'}_{\bar{m}'=-\ell'} D^{\ell}_{m \bar{m}}(\theta_0, \phi_0)D^{*\ell'}_{m' \bar{m}'}(\theta_0, \phi_0)C^{\E{i}}_{\ell \bar{m}\ell'\bar{m}'}.
\end{equation}
The rotation operation only mixes the $m$ and $m'$ elements in each $(\ell,\ell')$ block, meaning that all the dark, uncorrelated elements in the $\ell-\ell' \equiv 1\Mod{2}$ of the left panel of \cref{fig:rot_C} stay uncorrelated after the rotation.
Unlike this configuration for \E{2}, some topologies will induce correlations in all $(\ell, \ell')$ blocks, meaning that a random rotation of the coordinate system can induce correlations between all combinations of $\{\ell,m\}$ and $\{\ell',m'\}$.

\begin{figure}
  \centering
\includegraphics[width=0.49\linewidth]{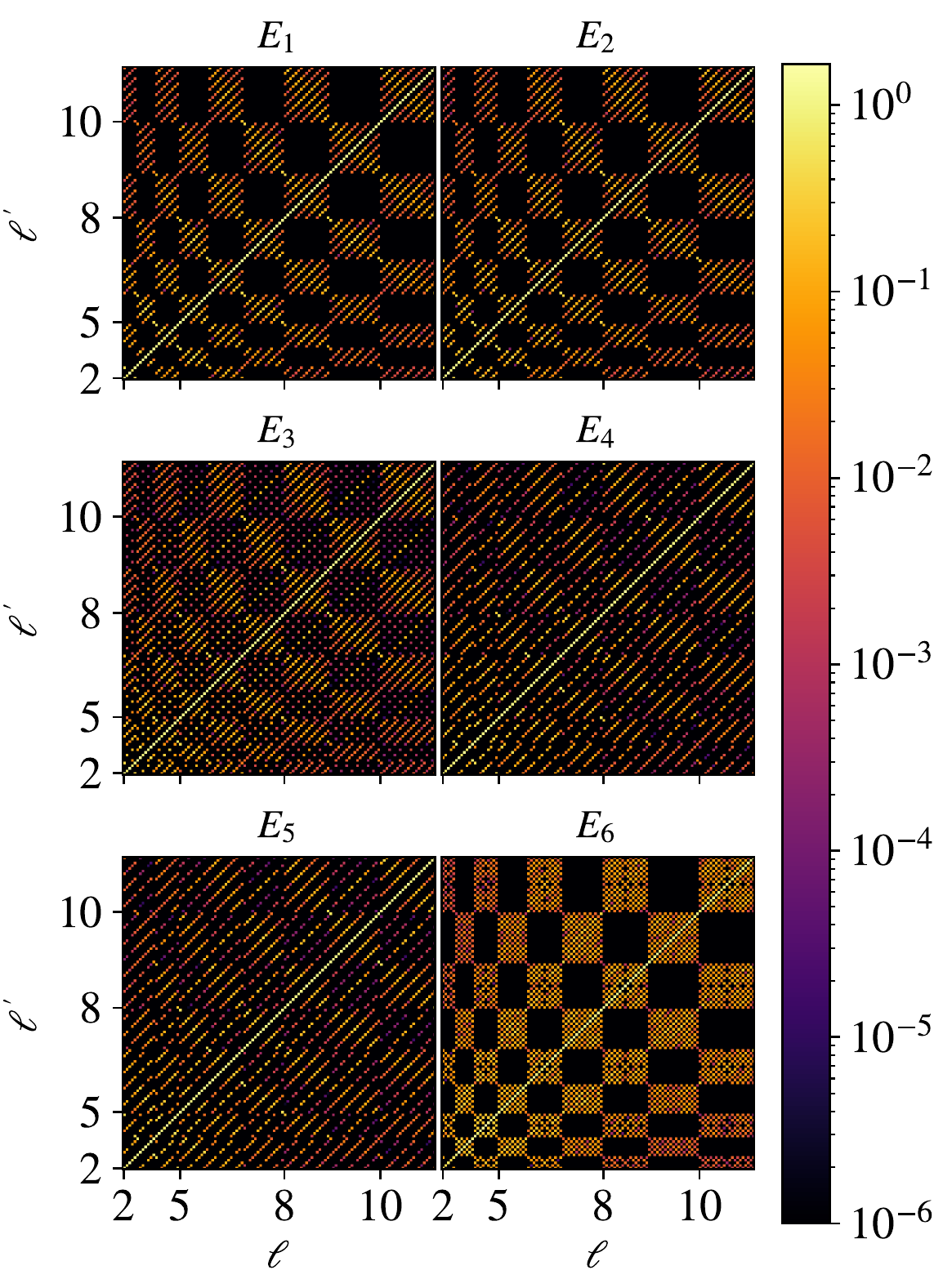}
\includegraphics[width=0.49\linewidth]{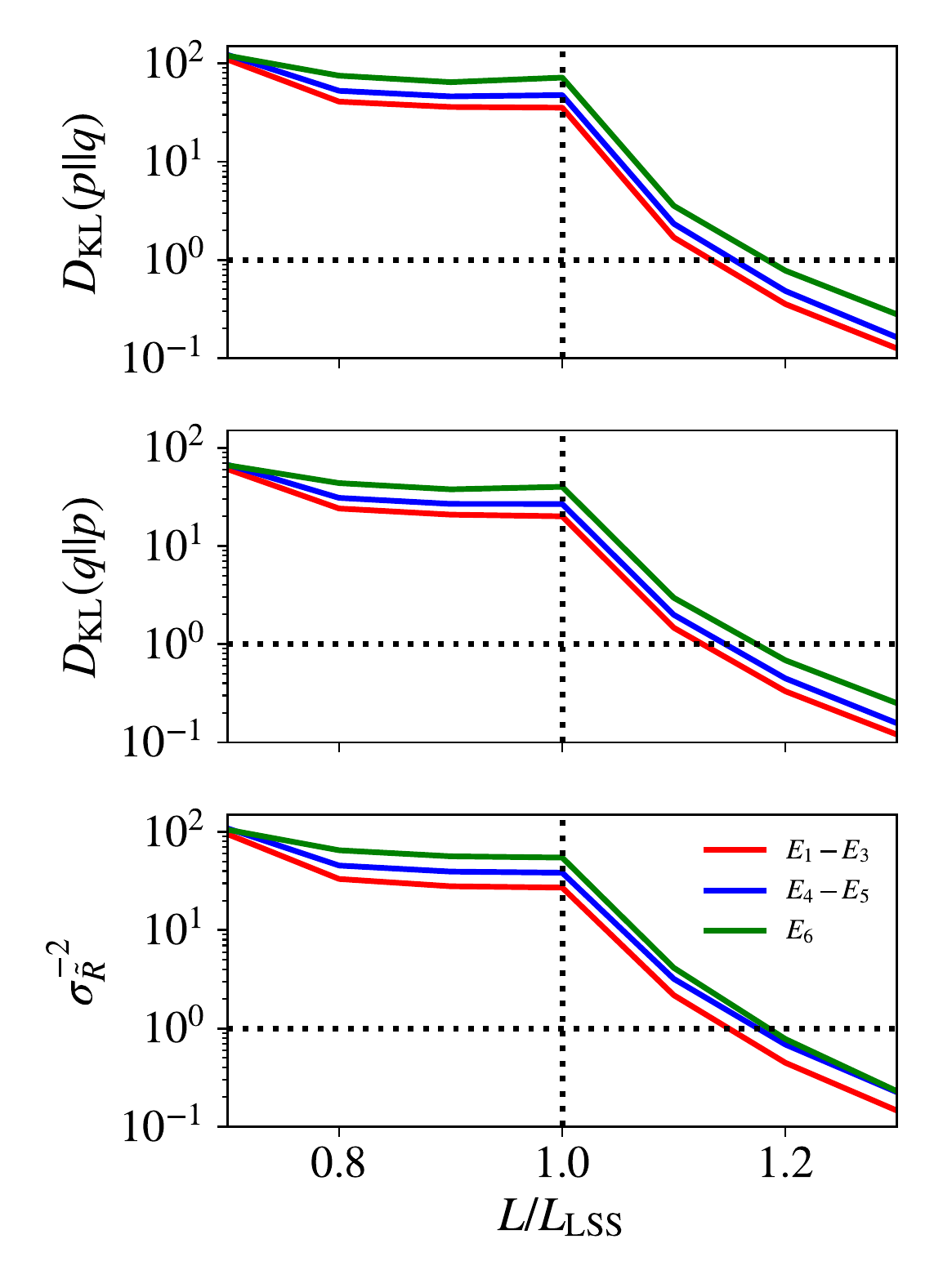}
\caption{{\bf Left panel:} Absolute values of the rescaled CMB temperature covariance matrix $\Xi^{\E{i};\, \mathrm{TT}}_{\ell m \ell' m'}$ (at low multipoles $\ell$) for the cubic, untilted \E{1}--\E{6}  with $L = L_{\mathrm{LSS}}$, where $L_{\mathrm{LSS}}$ is the diameter of the last scattering surface, and for an on-axis observer.
    {\bf Right panel:} The two KL divergences $D_{\mathrm{KL}}(p || q)$ and $D_{\mathrm{KL}}(q || p)$, and the off-diagonal signal-to-noise ratio $\sigma_{\tilde{R}}^{-2}$ for the cubic, untilted \E{1}--\E{6} with an on-axis observer.
    We find that the curves for \E{1}--\E{3} are indistinguishable, meaning that the \E{1}, \E{2}, and \E{3} cases carry the same information. The same is true for \E{4} and \E{5}.
    We have varied all lengths $L$ and have set $\ell_{\textrm{max}}=30$.}
\label{fig:cov_matrix_E1_E6}
\end{figure}

The absolute values of the rescaled covariance matrices for the untilted and unshifted (or on-axis) topologies \E{1}--\E{6} with topological length scales equal to $L_{\mathrm{LSS}}$ are shown in the left panel of \cref{fig:cov_matrix_E1_E6}.
All these topologies have non-zero correlations for the $\ell - \ell' \equiv 0\Mod{2}$ blocks in their on-axis and untilted configurations, but we also find correlations in the $\ell - \ell' =$ odd blocks of \E{3}, \E{4}, and \E{5}. This can be explained by the lack of symmetry in the $z$ direction.
For \E{2}, the $z\rightarrow z\pm L_B$ transformations both give a $180^\circ$ corkscrew rotation, preserving the $z$-symmetry when the observer is on-axis.
But \E{3}, \E{4}, and \E{5} do not have this symmetry as the transformation $z\rightarrow z + L_B$ gives the positive corkscrew rotations of $90^\circ$, $120^\circ$, and $60^\circ$, respectively, while the transformation in the opposite direction $z\rightarrow z - L_B$ gives a negative rotation, breaking the $z$-direction symmetry.
Both \E{2} and \E{6} lose this symmetry when they are off-axis, creating $\ell-\ell'=$ odd correlations.
\E{1}, on the other hand, will never have $\ell-\ell'=$ odd correlations.

We can learn something about the symmetries between $(m, m')$ elements in the covariance plots. It is straightforward to see that an on-axis observer of untilted \E{2}, \E{3}, \E{4}, and \E{5} has
\begin{equation}
    \xi^{\E{i};\unitvec{k}_{\vec{n}}}_{k_{\vec{n}} \ell m} \propto \sqrt{\lambda^{\E{i}}}\Kdelta_{m n_3 \Mod{\lambda^{\E{i}}}},
\end{equation}
where $\Mod{\lambda^{\E{i}}}$ indicates that the Kronecker delta is 1 if $m \equiv n_3\Mod{\lambda^{\E{i}}}$.
Here the integer $\lambda^{\E{i}}$ is the number of times one has to translate by $z\rightarrow z + L_B$ to get a corkscrew rotation of $360^\circ$,
namely, $\lambda^{\E{2}} = 2$, $\lambda^{\E{3}} = 4$, $\lambda^{\E{4}} = 3$, and $\lambda^{\E{6}} = 6$.
By combining the two eigenmodes that enter the sum in the covariance matrix, we find
\begin{align}
    \xi^{\E{i};\unitvec{k}_{\vec{n}}}_{k_{\vec{n}} \ell m} \xi^{\E{i};\unitvec{k}_{\vec{n}}*}_{k_{\vec{n}} \ell' m'} 
    & \propto \lambda^{\E{i}}\Kdelta_{m n_3 \Mod{\lambda^{\E{i}}}}\Kdelta_{m' n_3 \Mod{\lambda^{\E{i}}}} 
    = \lambda^{\E{i}} \Kdelta_{m m'\Mod{\lambda^{\E{i}}}}\Kdelta_{n_3  m \Mod{\lambda^{\E{i}}}}.
\end{align}
This then results in a clear pattern in the covariance matrix, 
$C^{\E{i}}_{\ell m\ell'm'} \propto \Kdelta_{m m'\Mod{\lambda^{\E{i}}}}$. 
As mentioned, this pattern generically disappears once we arbitrarily rotate the coordinate system.

There are other patterns, such as an untilted, cubic \E{1} being invariant under a $\pi/4$ rotation, yielding only $m-m'\equiv 0\Mod{4}$ correlations, but we will not find and categorize all these symmetries in this work.

\subsection{KL divergence}

Searching for repeating circles in the temperature fluctuations of the CMB has been a widely used method to look for evidence for a non-trivial cosmic topology.
However, when the Dirichlet domain does not fully contain the last scattering surface, no matched circles appear in the CMB and other methods must be applied.
We can then ask the question ``At what size of the Dirichlet domain will a non-trivial topology of the Universe be indistinguishable from the trivial topology, i.e., from the covering space?''

To answer this question we can apply the KL divergence method \cite{kullback1951, kullback1959information} to a given set of ``data'' $\{a_{\ell m}\}$ to compare the probability distribution of a non-trivial topology, $p(\{a_{\ell m}\})$, to that of the trivial one, $q(\{a_{\ell m}\})$.
We imagine an ideal experiment with no noise, no foreground emission, and no mask to assess the maximum detectability of topology from the CMB\@.
The KL divergence $D_{\mathrm{KL}}(p || q)$, which quantifies the loss of information if we assume the (``correct'') model $p$ is represented by the (``incorrect'') model $q$, is given as
\begin{equation}
    D_{\mathrm{KL}}(p || q) = \int \mathrm{d}\{a_{\ell m}\} \,\, p(\{a_{\ell m}\}) \ln \left[\frac{p(\{a_{\ell m}\})}{q(\{a_{\ell m}\})} \right]\;.
\end{equation}
This can be written in terms of the eigenvalues $\lambda_j$ of the rescaled covariance matrix, $\Xi^{\E{i}}_{\ell m\ell'm'}$ \cite{Fabre:2013wia}. The KL divergence then becomes
\begin{equation}
D_{\mathrm{KL}}(p || q) = \frac12 \sum_j \left(\ln |\lambda_j|+\lambda_j^{-1} - 1 \right).\label{eq:DKLpq}
\end{equation}

We could also turn the question around: ``How much information is lost if we assume the model $q$ is represented by the model $p$?''
As the eigenvalues of the inverse of the matrix $\Xi^{\E{i}}_{\ell m\ell'm'}$ are simply $1/\lambda_j$, we find
\begin{equation}
    D_{\mathrm{KL}}(q || p) = \frac{1}{2} \sum_i \left(-\ln |\lambda_j|+\lambda_j - 1 \right).\label{eq:DKLqp}
\end{equation}

\subsection{Cosmic variance signal-to-noise ratio}

We can also evaluate the signal-to-noise ratio for detecting the off-diagonal elements of a specific correlation matrix in the ideal case of a full-sky CMB map with a given angular resolution and zero instrumental noise.
In this case, the noise is due to the cosmic variance \cite{White1993} from a particular random sky realization, and thus is in principle the best we can do to detect topology from a CMB temperature map.

The spherical-harmonic coefficients are random complex numbers with zero mean,
random phase, and amplitude variance given by 
\begin{equation}
    \sigma^2_{\ell m} \equiv \langle |a_{\ell m}|^2\rangle = C_{\ell m \ell m},
\end{equation}
ignoring instrumental noise and beam effects as appropriate for a cosmic variance-limited observation. The $a_{\ell m}$ satisfy the reality condition $a_{\ell-m}=(-1)^m a^*_{\ell m}$. For $m\ne0$, they are described by
two independent scalar Gaussian random variables for their real and imaginary parts, each with variance $\sigma_{\ell m}^2/2$, and for $m=0$ by a single scalar Gaussian random variable of variance $\sigma_{\ell m}^2$.

Products of two different spherical-harmonic coefficients are also complex random variables, which can be written as
\begin{equation}
    a_{\ell m} a^*_{\ell'm'} = C_{\ell m\ell'm'} \pm S_{\ell m\ell'm'}\,,
\label{Eq:aa_meanpluserror}
\end{equation}
where the $C_{\ell m\ell'm'}$ is the expectation value of $a_{\ell m} a^*_{\ell'm'}$, and the standard error term $S_{\ell m\ell'm'}$ is a random complex number with zero mean for both real and imaginary components.
To obtain the standard error, the product $a_{\ell m} a^*_{\ell'm'}$ can be expanded into its real and imaginary parts; each contains two terms which are products of two random normal variables (the real and/or imaginary parts of $a_{\ell m}$).
We approximate the two as being independent; this is exact for a Gaussian random curvature field in the covering space because of statistical isotropy.
It is a good  approximation whenever the correlations between spherical-harmonic coefficients are small compared to their amplitudes, which is the case for all the \E{i} topologies, except, possibly, for low $\ell$ values.
It also appears to be the case for the observed sky when analyzed assuming statistical isotropy, again except, possibly, for low $\ell$ values.
Since the variance of the product of two independent random normal variables is the product of the variances, we have
\begin{align}
    \Var\left( \Re \, a_{\ell m}\Re \,a^*_{\ell' m'}\right) &= \frac12 \sigma^2_{\ell m}\left(1+\Kdelta_{m0}\right)\frac12 \sigma_{\ell' m'}^2\left(1+\Kdelta_{m'0} \right), \nonumber\\
    \Var\left( \Re \, a_{\ell m}\Im \,a^*_{\ell' m'}\right) &= \frac12 \sigma^2_{\ell m}\left(1+\Kdelta_{m0}\right)\frac12 \sigma_{\ell' m'}^2\left(1-\Kdelta_{m'0}\right), 
\end{align}
and the same result for the other possible combinations of real and imaginary parts on the left side.
Combining these,
\begin{align}
\Var\Re \left(a_{\ell m} a^*_{\ell' m'}\right) = \frac{1}{2} \sigma^2_{\ell m}\sigma^2_{\ell' m'}\left(1+\Kdelta_{m0}\Kdelta_{m'0}\right), \nonumber\\
\Var\Im \left(a_{\ell m} a^*_{\ell' m'}\right)
= \frac{1}{2} \sigma^2_{\ell m}\sigma^2_{\ell' m'}\left(1-\Kdelta_{m0}\Kdelta_{m'0}\right).
 \label{Eq:aa_stderror}
\end{align}
Note that the random variable $a_{\ell m} a^*_{\ell'm'}$ is not itself a Gaussian variable (its amplitude is distributed as the Bessel function $K_0$), but this is not important for our conclusions.

To estimate the possible signal-to-noise ratio with which a given topology can be detected, we write the off-diagonal signal of a particular topology as $R C^{\E{i}}_{\ell m\ell'm'}$ with $R=1$ corresponding to the actual signal. We can then estimate $R$ from the off-diagonal elements, where each individual element can serve as a noisy estimator $\tilde R$ of $R$; for example, 
$\tilde{R}^{\Re}_{\ell m \ell'm'} = \Re\left(a_{\ell m}a^*_{\ell'm'}\right) / \Re C^{\E{i}}_{\ell m\ell'm'}$ 
or the same with the imaginary component (except for $m=m'=0$ where the latter vanishes).
We can also estimate $R$ using any weighted normalized sum of the individual estimates; a textbook result is that the minimum
variance estimator uses the inverse variance weight for each term. In this case, the inverse variance of the estimate is the sum of the inverse variances of the terms, so  
\begin{align}
    \label{eqn:inversevar}
    \sigma_{\tilde R}^{-2} \equiv \frac{1}{\Var{\tilde R} }  
    &= \frac12 \sum_{(\ell m)\neq(\ell'm')} 
    \frac{1+\Kdelta_{m0}+\Kdelta_{m'0}-\Kdelta_{m0}\Kdelta_{m'0}}{2} \nonumber\\
    &\qquad\qquad\times\left[\frac{2(\Re C^{\E{i}}_{\ell m\ell'm'})^2}{\left(1+\Kdelta_{m0}\Kdelta_{m'0}\right)\sigma^2_{\ell m}\sigma^2_{\ell' m'}} + \frac{2(\Im C^{\E{i}}_{\ell m\ell'm'})^2}{\sigma^2_{\ell m}\sigma^2_{\ell' m'}} \right] \nonumber\\
    &= \frac{1}{2} \sum_{(\ell m)\neq(\ell'm')}
    \frac{1+\Kdelta_{m0}+\Kdelta_{m'0}-\Kdelta_{m0}\Kdelta_{m'0}}{1+\Kdelta_{m0}\Kdelta_{m'0}} ~
    \frac{\left|C^{\E{i}}_{\ell m\ell'm'}\right|^2}{\sigma^2_{\ell m}\sigma^2_{\ell' m'}}\nonumber\\
    &\simeq \frac{1}{2}
    \sum_{(\ell m)\neq(\ell'm')}
    \frac{\left|C^{\E{i}}_{\ell m\ell'm'}\right|^2}{\sigma^2_{\ell m}\sigma^2_{\ell' m'}}
    .
\end{align}
The sums in \eqref{eqn:inversevar} are over all pairs of indices except along the diagonal where $\ell=\ell'$ and $m=m'$, between $\ell=2$ and some $\ell_\mathrm{max}$.
The Hermitian redundancy of the correlation matrix\footnote{
    For $X\neq Y$ in $C^{XY}$, we would drop the factor of $1/2$ in \eqref{eqn:inversevar}.
    } means that the sum over-counts the off-diagonal terms by exactly a factor of 2.
This is accounted for by taking $1/2$ of the sum, canceling the factors of 2 in each term.  
The imaginary-part term vanishes when $m=m'=0$. The terms then combine to give the squared amplitude. 
We account for the symmetry ${C^{\E{i}}_{\ell -m\ell'-m'}=(-1)^{m+m'}C^{\E{i}}_{\ell' m'\ell m }}$  when neither $m=0$ nor $m'=0$, with the factor ${(1+\Kdelta_{m0}+\Kdelta_{m'0}-\Kdelta_{m0}\Kdelta_{m'0})/2}$ in the sum.
Finally, in practice, we ignore the Kronecker delta terms in the final expression, which account for only a fraction $1/\ell^2_\mathrm{max}$ of terms.

The maximum possible signal-to-noise ratio of the off-diagonal elements in the covariance matrix for measuring a given topology is simply $\sigma_{\tilde R}^{-2}$, and it 
is proportional to the mean-square correlation coefficient of the correlation matrix $C^{\E{i}}_{\ell m \ell'm'}$. \cref{eqn:inversevar} can be directly compared with the KL statistics.\footnote{
     $D_{\mathrm{KL}}=1$ can be taken as a theoretical detectability threshold for 
     non-trivial topology in the absence of noise, foregrounds, and mask.
     $D_{\mathrm{KL}}$ characterizes the relative odds of two models --- here a non-trivial topology  versus the covering space.
    According to Wilks' theorem   twice the likelihood ratio is asymptotically $\chi^2$ distributed with number of degrees of freedom equal to the difference between those of the two distributions.  
    Thus, for example, $D_{\mathrm{KL}}\gtrapprox3$ implies that topology can be strongly detected in cases where there are  three orientation degrees of freedom and three more that characterize the manifold of the particular topology.}
This maximum signal-to-noise ratio is obtained if we know the topology and know our position and orientation in the Dirichlet domain. 
In practice, we cannot know these things a priori, so we must search over a space of locations and orientations for
any particular topology, with only the correct choice giving the maximum signal-to-noise ratio computed here.

\subsection{KL divergence and signal-to-noise ratio for compact, orientable, Euclidean manifolds}

\begin{figure}[t]
    \centering
    \includegraphics[width=0.9\linewidth]{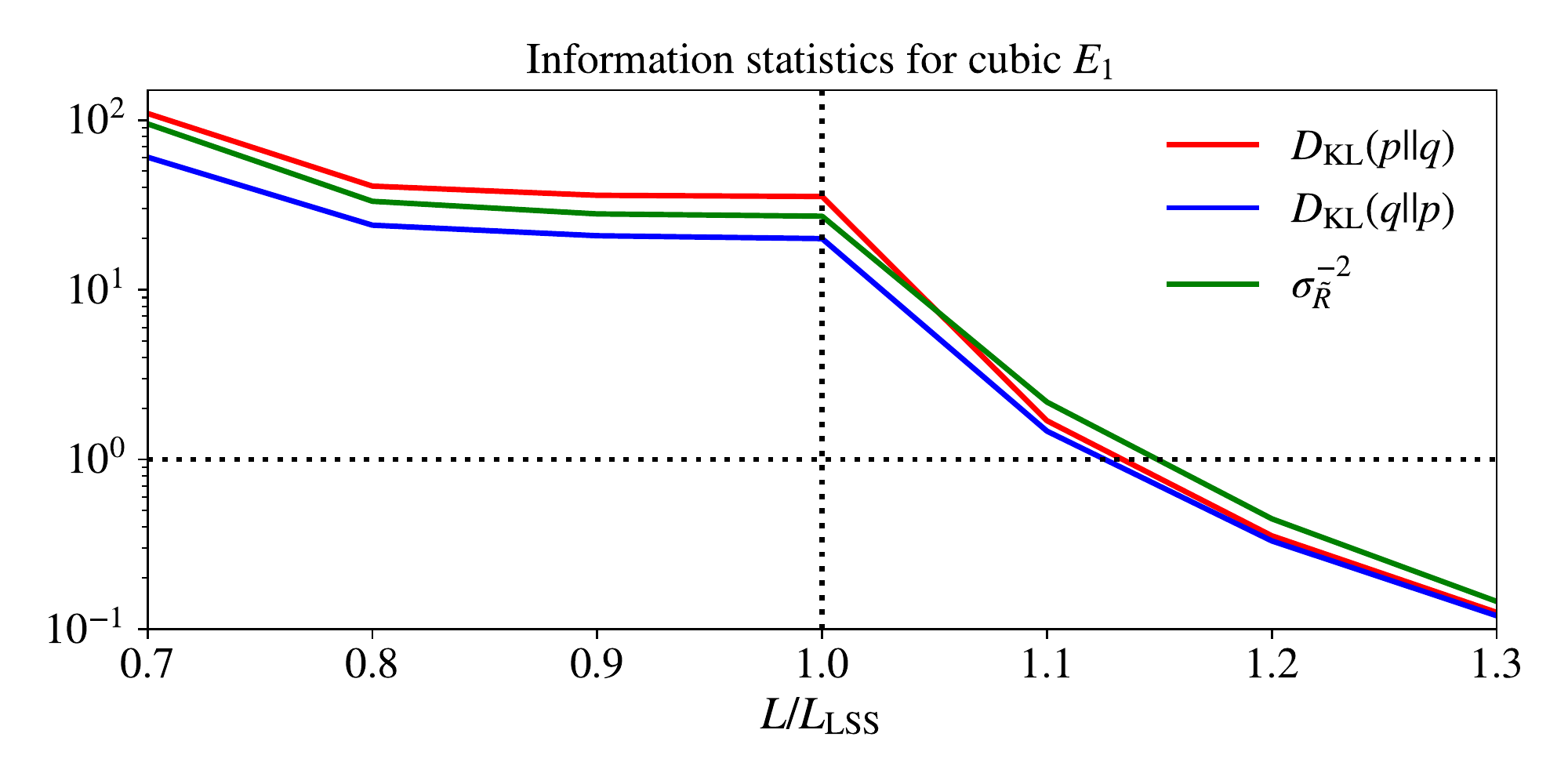}
    \caption{Comparison of the $D_{\mathrm{KL}}(p || q)$, $D_{\mathrm{KL}}(q || p)$, and $\sigma_{\tilde{R}}^{-2}$ statistics for cubic \E{1} as a function of  $L_1=L_2=L_3\equiv L$ in units of $L_{\mathrm{LSS}}$, where $L_{\mathrm{LSS}}$ is the diameter of the last scattering surface.
    }
    \label{fig:statistics-comparison}
\end{figure}

In \cref{fig:statistics-comparison} we compare the three statistics $D_{\mathrm{KL}}(p || q)$, $D_{\mathrm{KL}}(q || p)$, and $\sigma_{\tilde{R}}^{-2}$ for the cubic \E{1} as a function of  $L_1=L_2=L_3\equiv L$ in units of $L_{\mathrm{LSS}}$, and find that they convey similar information.
In the right panel of \cref{fig:cov_matrix_E1_E6}, we compare the three statistics for the untilted \E{1}--\E{6} with an on-axis observer.
We find that they qualitatively tell the same story for all cases; once the size of the Dirichlet domain is larger than the diameter of the last scattering surface, the information content of the topology of the Universe in the CMB data drops by increasing the length scale and will eventually cross the no-detection line $D_{\mathrm{KL}} = 1$.
As these three statistics qualitatively behave similarly, from now on we will only quote $D_{\mathrm{KL}}(p || q)$.
However,
the off-diagonal-signal-to-noise statistic has the advantage over the KL divergence in that calculating it does not involve inverting a large matrix. 
The sums in \eqref{eqn:inversevar} can become unwieldy, but they can be approximated by using a randomly chosen smaller number of terms and scaling, in the style of Monte Carlo integration.

\begin{figure}
  \centering
\includegraphics[width=0.48\linewidth]{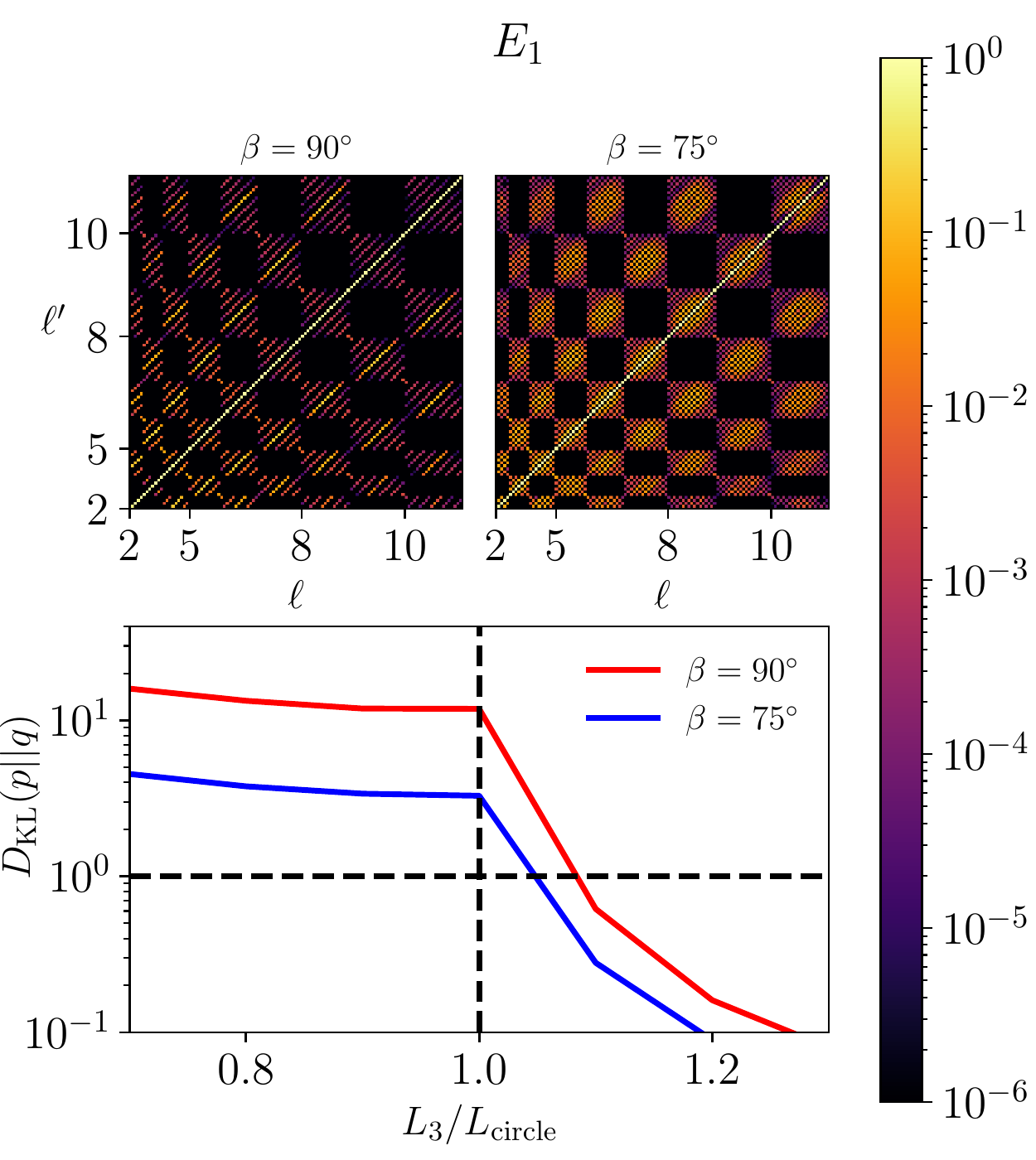}
\includegraphics[width=0.48\linewidth]{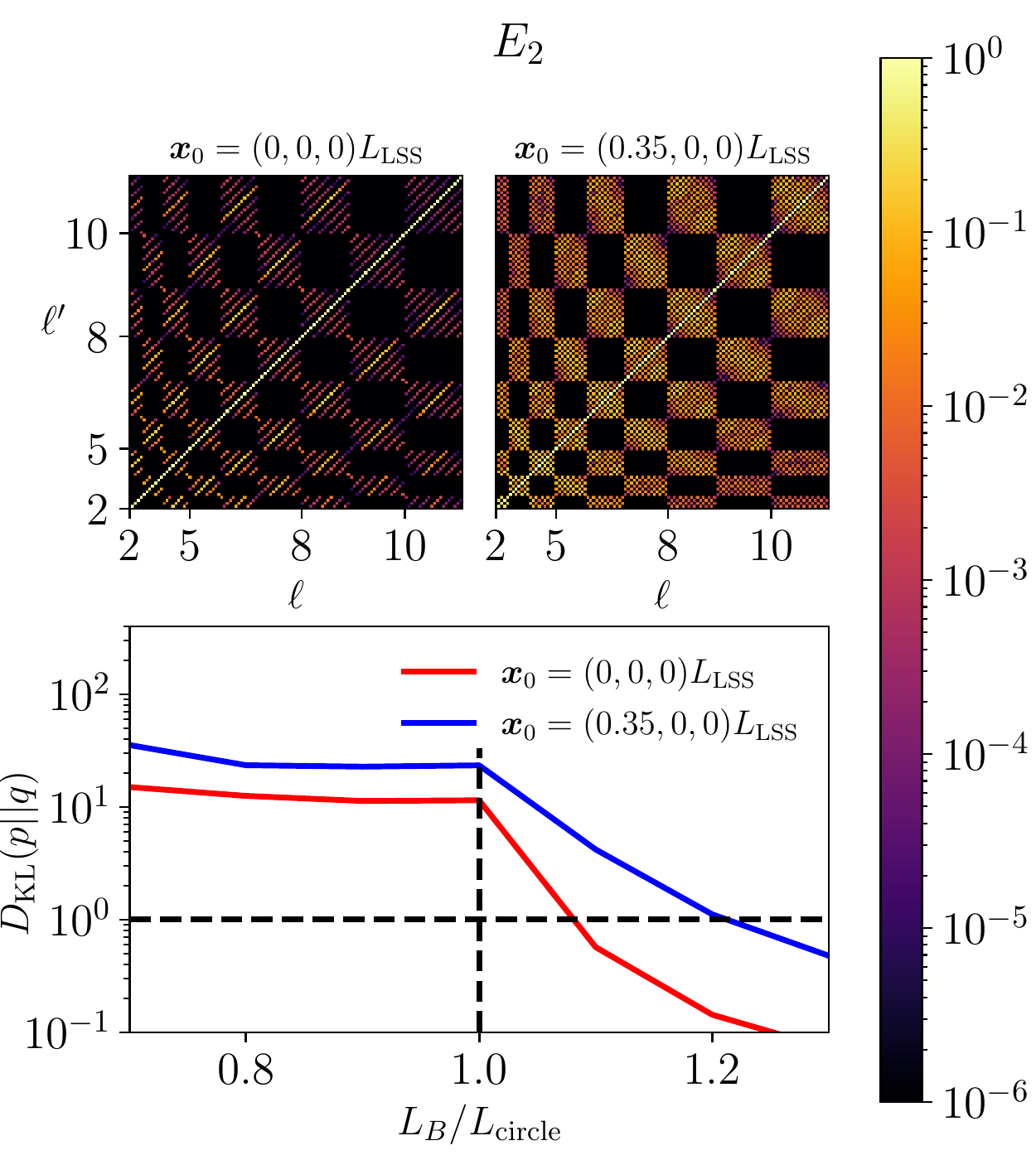}
\caption{Absolute values of the rescaled CMB temperature covariance matrix $\Xi^{\E{i};\, \mathrm{TT}}_{\ell m \ell' m'}$ (at low multipoles $\ell$) and the KL divergence for \E{1} and \E{2}, where we have set $L_1= L_2 = 1.4 L_{\mathrm{LSS}}$ for \E{1} and $L_{A_1} = L_{A_2} = 1.4 L_{\mathrm{LSS}}$ for \E{2}.
    The KL divergence is calculated with $\ell_{\mathrm{max}} = 30$.
    In \E{1}, the circle limit for both tilt angles $\beta=90^\circ$ and $\beta=75^\circ$ is $L_3=L_{\mathrm{circle}}=L_{\mathrm{LSS}}$, where $L_{\mathrm{circle}}$ is defined as the smallest $L_3$ for which no pairs of identical circles appear in the CMB\@ and $L_{\mathrm{LSS}}$ is the diameter of the last scattering surface. For \E{2}, the limit for an on-axis observer is $L_B = L_{\mathrm{circle}} = L_{\mathrm{LSS}}$, where $L_{\mathrm{circle}}$ is the smallest $L_B$ for which no pairs of identical circles appear in the CMB\@, while the observer at $\vec{x}_0=L_{\mathrm{LSS}} \transpose{(0.35, 0, 0)}$ reaches the circle limit when $L_B = L_{\mathrm{circle}} = \sqrt{1-0.7^2}L_{\mathrm{LSS}}\approx 0.71 L_{\mathrm{LSS}}$.
    For the rescaled covariance matrix plots we have used $L_3=L_{\mathrm{circle}}$ for \E{1} and $L_B = L_{\mathrm{circle}}$ for \E{2}.}
\label{fig:E1_E2}
\end{figure}

\begin{figure}
  \centering
\includegraphics[width=0.48\linewidth]{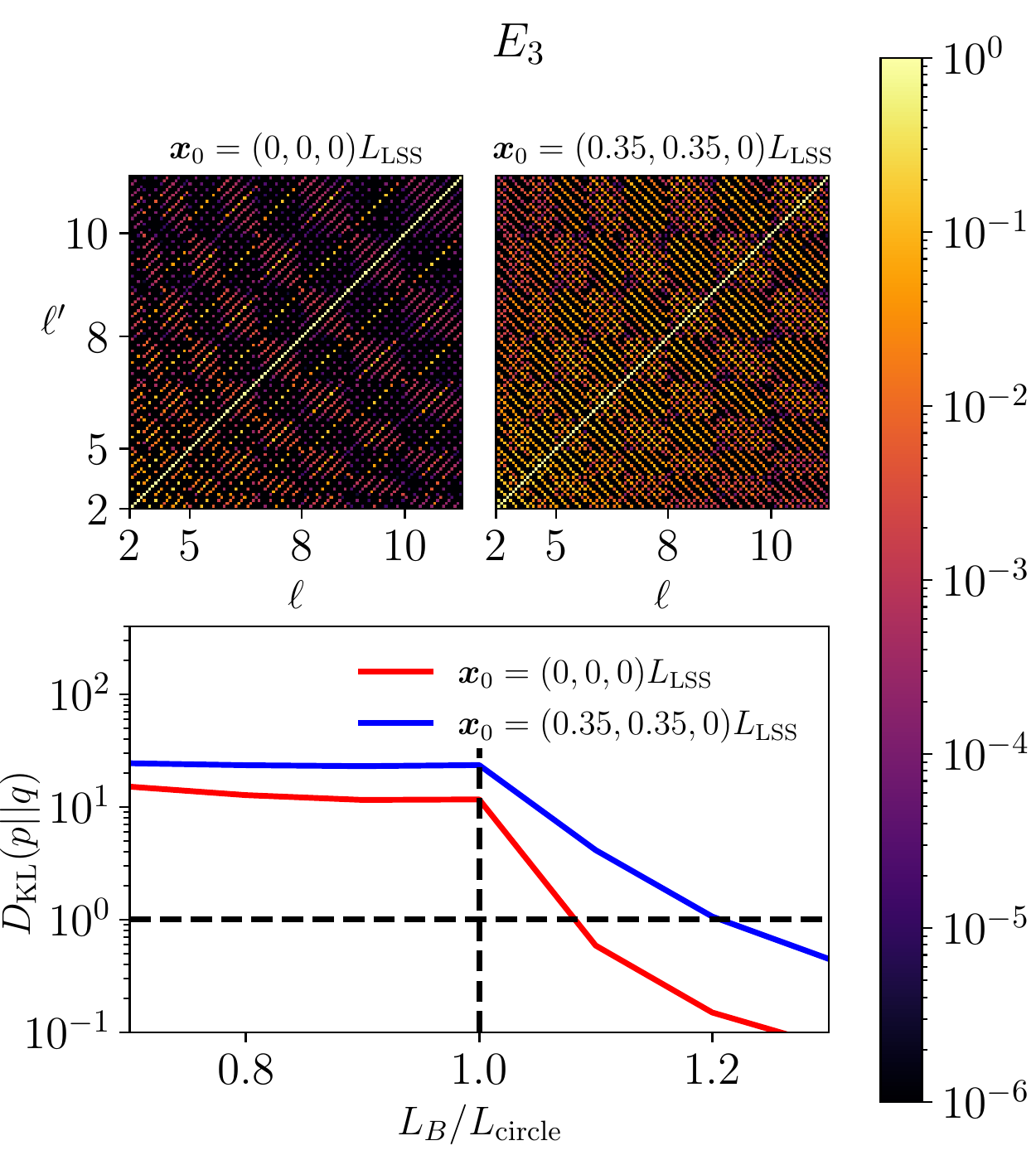}
\includegraphics[width=0.48\linewidth]{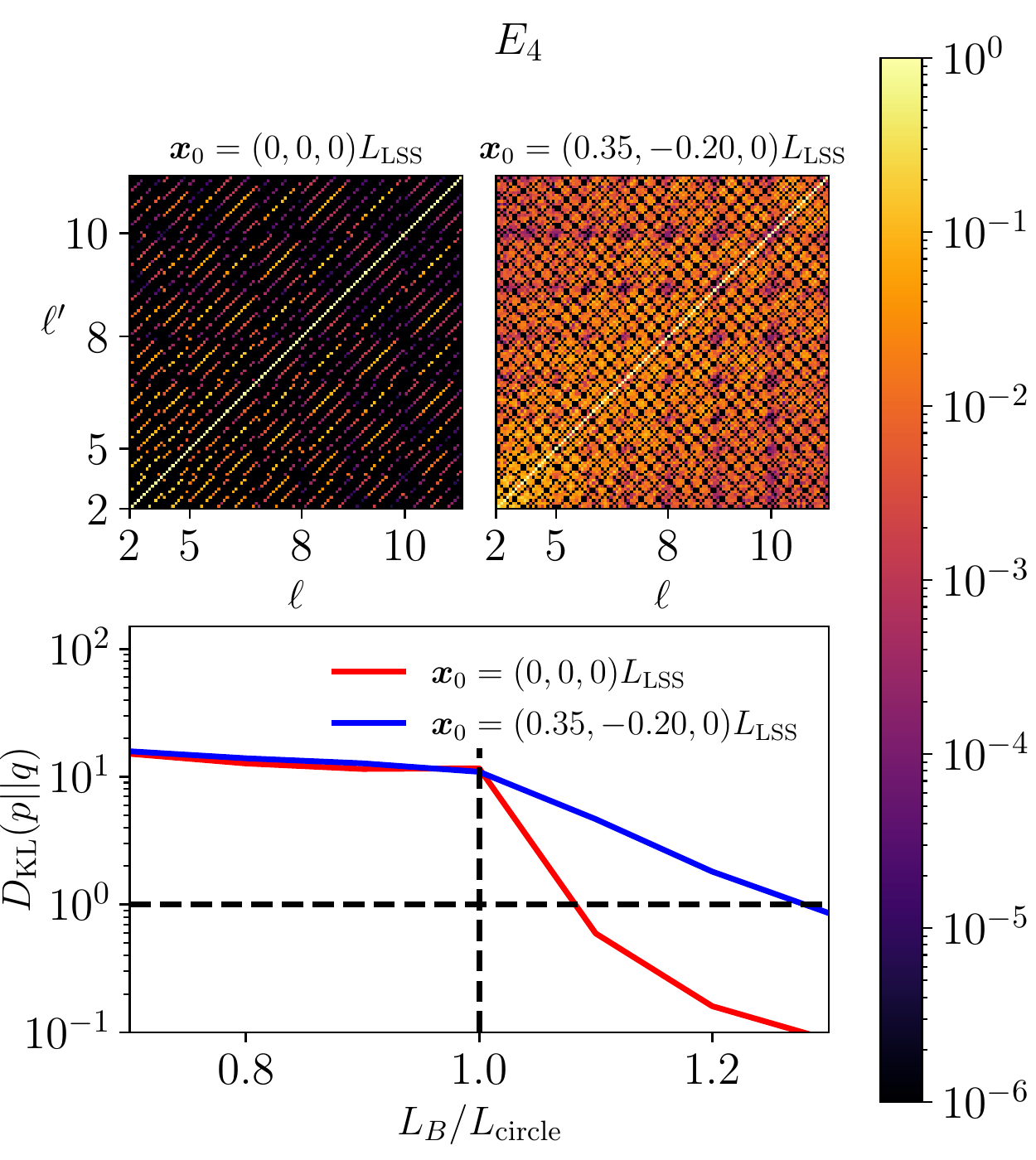}
\caption{As in \cref{fig:E1_E2}, but for \E{3} and \E{4}, where we have set $L_A = 1.4 L_{\mathrm{LSS}}$.
In both cases, the circle limit for the on-axis observers is $L_B = L_{\mathrm{circle}} = L_{\mathrm{LSS}}$, while the off-axis observers have a circle limit of $L_B = L_{\mathrm{circle}} = \sqrt{1-0.7^2}L_{\mathrm{LSS}}\approx 0.71 L_{\mathrm{LSS}}$.}
\label{fig:E3_E4}
\end{figure}

\begin{figure}
  \centering
    \includegraphics[width=0.48\linewidth]{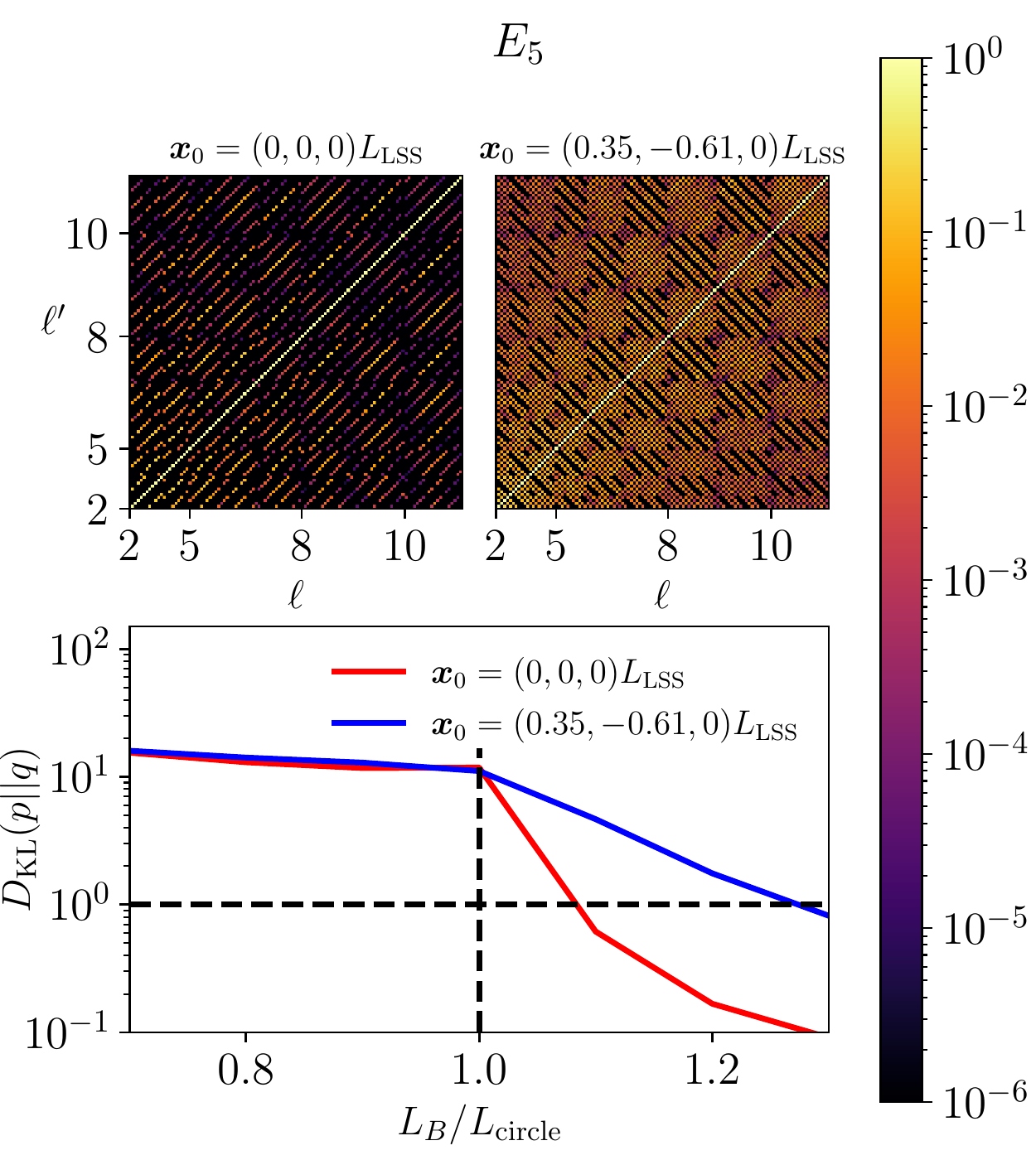}
    \includegraphics[width=0.48\linewidth]{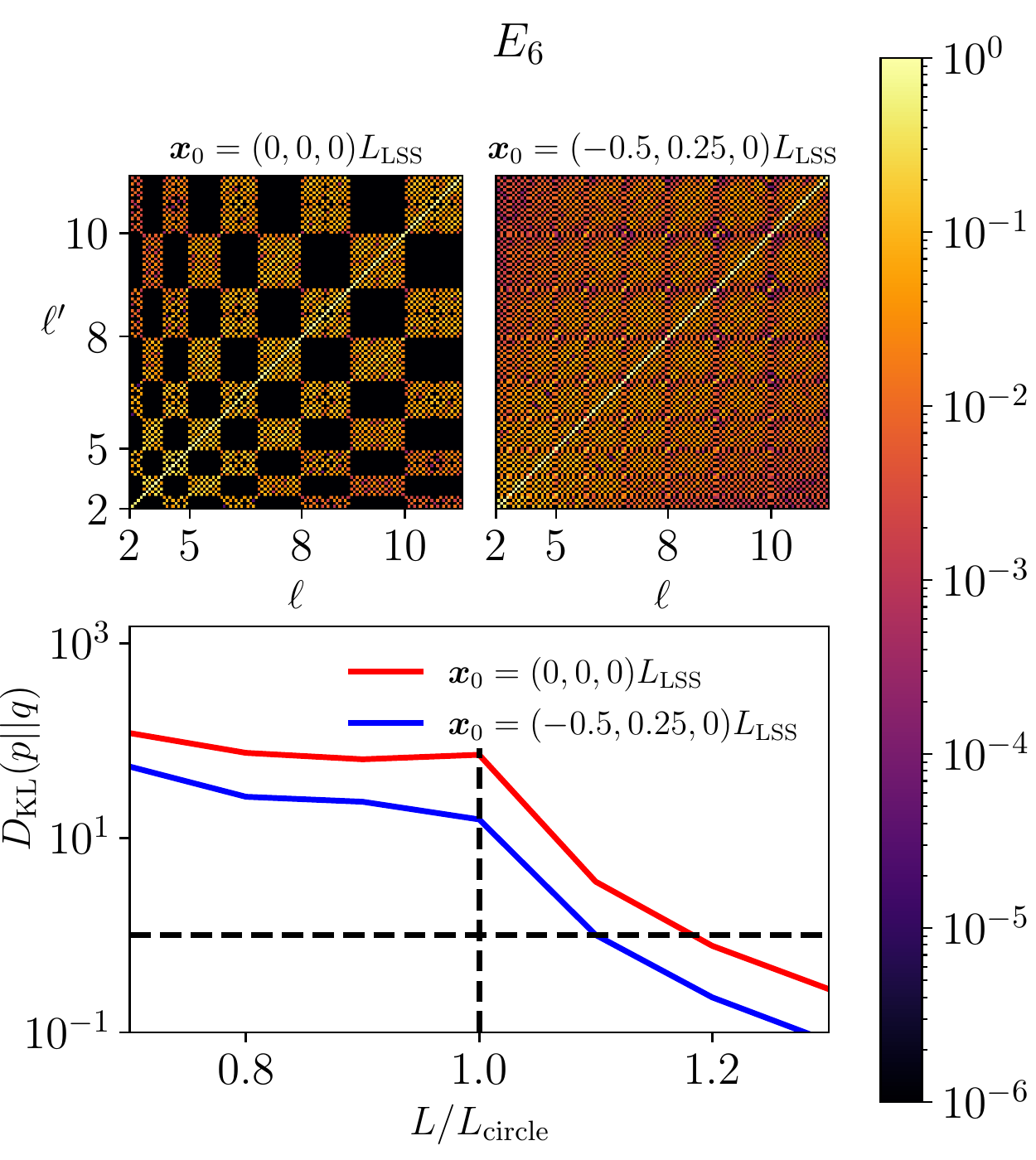}
    \caption{As in \cref{fig:E1_E2,fig:E3_E4}, but for \E{5} and \E{6}.
    For the covariance matrix plots of \E{5}, we have set $L_{A} = 1.4 L_{\mathrm{LSS}}$ and $L_B = L_{\mathrm{circle}}$. $L_{\mathrm{circle}}=L_{\mathrm{LSS}}$ and $L_{\mathrm{circle}}=\sqrt{1-0.7^2}L_{\mathrm{LSS}}\approx 0.71 L_{\mathrm{LSS}}$ for the observer locations $\vec{x}_0=L_{\mathrm{LSS}}\transpose{(0,0,0)}$ and $\vec{x}_0=L_{\mathrm{LSS}}\transpose{(0.35,0.61,0)}$, respectively. For the covariance matrix plots of \E{6}, we have set $L_{Ax}=L_{By}=L_{Cz}=L_{\mathrm{circle}}$, where $L_{\mathrm{circle}} = 1/\sqrt{2}L_{\mathrm{LSS}}\approx 0.71$ for $\vec{x}_0=L_{\mathrm{LSS}}\transpose{(0,0,0)}$ and $L_{\mathrm{circle}}\approx 0.90$ for $\vec{x}_0=L_{\mathrm{LSS}}\transpose{(-0.5,0.25,0)}$.
    In the KL divergence plot of \E{5}, we have set $L_{A} = 1.4 L_{\mathrm{LSS}}$ while varying $L_B$. For \E{6}, we have varied $L\equiv L_{Ax}=L_{By}=L_{Cz}$, and we have set $r_x=r_y=r_z=1/2$.}
    \label{fig:E5_E6}
\end{figure}

We show in \cref{fig:E1_E2,fig:E3_E4,fig:E5_E6} the absolute values of the rescaled covariance matrices $|\Xi^{\E{i}}_{\ell m \ell' m'}|$ of each topology for two different configurations: an on-axis observer $\vec{x}_0=\transpose{(0,0,0)}$ and an off-axis observer $\vec{x}_0\neq\transpose{(0,0,0)}$.
As the tilt and off-axis position of the observer are degenerate, we set the tilt parameters to zero.
The exception is for \E{1} as this topology is homogeneous.
For \E{1} we have chosen an untilted ($\beta=90^\circ$) configuration and a tilted ($\beta=75^\circ$) configuration in \cref{fig:E1_E2}.

For \E{2}--\E{5}, we set $L_A = 1.4L_{\mathrm{LSS}}$ and vary $L_B$ in the KL divergence plots.\footnote{Note that for \E{2} the length parameters are $L_{A_1}$, $L_{A_2}$, and $L_B$, instead of $L_A$ and $L_B$. Here, for simplicity, we refer to $L_{A_1}=L_{A_2}$ as $L_A$ for \E{2}.}
The $x$-axis is shown as $L_B/L_{\mathrm{circle}}$, where $L_{\mathrm{circle}}$ is defined as the smallest $L_B$ for which no pairs of identical circles appear in the CMB\@. For an on-axis observer, this is simply $L_{\mathrm{LSS}}$ as long as $L_A > L_{\mathrm{LSS}}$.\footnote{If $L_A < L_{\mathrm{LSS}}$ there would always be identical circles in the CMB for all values of $L_B$.}
Hence, when $L_B/L_{\mathrm{circle}} > 1$ no identical circles appear in the CMB and the information content of cosmic topology in the CMB drops as seen by the falling KL divergence.
For the figures of the rescaled covariance matrices, we set $L_B = L_{\mathrm{circle}}$, meaning that these covariance matrices can never be ruled out by CMB circle searches.  
For \E{1}, we do the same as above, but with $L_A\to L_1 = L_2$ and $L_B\to L_3$.
For \E{2}--\E{6}, $L_{\mathrm{circle}}$ depends on the parameters of the topology, while \E{1} always has $L_3 = L_{\mathrm{circle}} = L_{\mathrm{LSS}}$ as its circle limit even with a tilted topology ($\beta \neq 90^\circ$).

The off-axis positions in \cref{fig:E1_E2,fig:E3_E4,fig:E5_E6} are not randomly chosen.
We have chosen locations that are in the middle between two excluded regions for $L_B < L_{\mathrm{LSS}}$ of Fig.~2 of \rcite{COMPACT:2022nsu}.
What this means is that the position is chosen so that the observer has the highest number of nearest clones separated by the same distance to the observer.
We, therefore, hypothesize that these off-axis positions represent maximum possible information of non-trivial topology in the CMB\@.

For example, for \E{2} with $L_A = 1.4L_{\mathrm{LSS}}$ and $\vec{x}_0=L_{\mathrm{LSS}}\transpose{(0.35, 0, 0)}$ in \cref{fig:E1_E2} there are four nearest clones with the same distance to the observer.
Namely at positions $\vec{x} = \transpose{((1.4-0.35)L_{\mathrm{LSS}}, 0, L_B)}$, $\vec{x} = \transpose{(-0.35L_{\mathrm{LSS}}, 0, L_B)}$, $\vec{x} = \transpose{((1.4-0.35)L_{\mathrm{LSS}}, 0, -L_B)}$, and $\vec{x} = \transpose{(-0.35L_{\mathrm{LSS}}, 0, -L_B)}$.
The distance to these four clones is $\sqrt{(0.7L_{\mathrm{LSS}})^2+L_B^2}$.
When the distance is equal to $L_{\mathrm{LSS}}$, we get $L_B = L_{\mathrm{circle}} = \sqrt{1-0.7^2}L_{\mathrm{LSS}}$.
An on-axis observer $\vec{x}_0=\transpose{(0, 0, 0)}$ has only two nearest clones, $\vec{x}=\transpose{(0, 0, L_B)}$ and $\vec{x}=\transpose{(0, 0, -L_B)}$, both with a distance $L_B$ yielding $L_B = L_{\mathrm{circle}} = L_{\mathrm{LSS}}$.
Fewer nearest clones with the same distance means fewer correlations in the covariance matrix and a lower KL divergence.

\E{6} is more complex, and we found an off-axis position from numerical considerations.
From \cref{fig:E5_E6} we already find that the on-axis position contains more information than the chosen off-axis one, meaning that there could be a more optimal off-axis position yielding higher information.

We have also looked into whether including higher multipoles $\ell_{\mathrm{max}}$ increases the KL divergence for \E{1} to \E{6}. Although going to higher multipoles increases the KL divergence when the distance to the nearest clone is less than $L_{\mathrm{LSS}}$, we find that increasing $\ell_{\mathrm{max}}=30$ to $\ell_{\mathrm{max}}=50$ makes a negligible difference when there are no identical circles in the CMB\@. This agrees with the findings of \rcite{COMPACT:2022gbl} (see the lower panel of Fig.~4 in that reference).

\begin{figure}
    \centering
    \includegraphics[width=\linewidth]{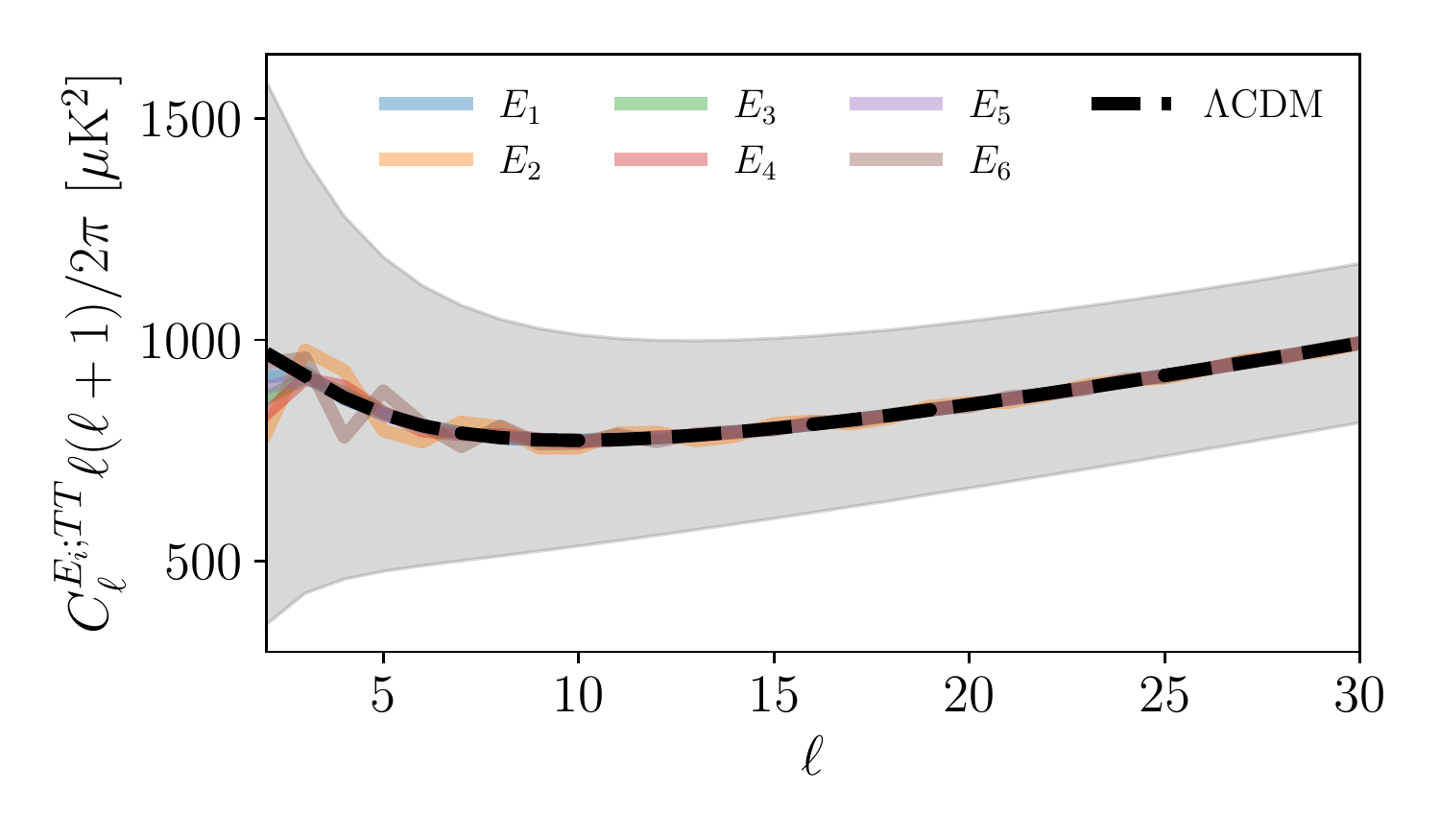}
    \caption{The quantity $C^{\E{i};TT}_\ell \equiv \frac{1}{2\ell+1}\sum_{m} C^{\E{i};TT}_{\ell m \ell m}$ for the CMB temperature sky for the configurations with off-axis observers in \cref{fig:E1_E2,fig:E3_E4,fig:E5_E6}. The gray band shows the $1\sigma$ cosmic variance for $\Lambda$CDM, i.e., the trivial topology.}
    \label{fig:C_ell}
\end{figure}

Finally, we show the diagonal $C^{\E{i};TT}_\ell$ in \cref{fig:C_ell}, defined as $C^{\E{i};TT}_\ell \equiv \frac{1}{2\ell+1}\sum_{m} C^{\E{i};TT}_{\ell m \ell m}$ for the non-trivial topologies. We have divided by $R_\ell(|\vec{k}|) = 0.99$ in \cref{eqn:cut_off_definition} to make up for the power lost by not summing over an infinite number of wavevectors. These topologies are chosen to have $L_3=L_B=L= L_{\mathrm{circle}}$ so that their nearest clones are exactly $L_{\mathrm{LSS}}$ away, meaning that the CMB contains no identical circles.

We end this section by briefly discussing how to generate $a^{\E{i}}_{\ell m}$ realizations of the CMB sky signal assuming a non-trivial topology, even though we have not used such realizations in this paper. There are three different ways for doing that. As the first possibility, one can sum over all the Fourier modes that contribute to each spherical-harmonic coefficient by generating realizations of the primordial density fluctuations $\delta^{\mathcal{R}}_{\vec{k}_{\vec{n}}}$ for each discrete wavevector $\vec{k}_{\vec{n}}$.
Namely, one would compute
\begin{equation}
    \label{eqn:a_lm_E_i}
    a^{\E{i}}_{\ell m} 
    =  \frac{4\pi}{V_{\E{i}}}
    \sum_{\vec{n}\in\setN^{\E{i}} } \delta^{\mathcal{R}}_{\vec{k}_{\vec{n}}} 
\xi^{\E{i};\unitvec{k}_{\vec{n}}}_{k_{\vec{n}} \ell m}
     \Delta_\ell(k_{\vec{n}}).
\end{equation}
As all the information is in the covariance matrix, the second possibility is to generate the $C^{\E{i}}_{\ell m \ell'm'}$ and create a realization by using Cholesky decomposition \cite{watkins2004, Mukherjee2014}.
However, calculating the full covariance matrix is computationally more expensive than performing the summation in \cref{eqn:a_lm_E_i}.
A third option is to generate three-dimensional real-space density fluctuations for a given topology and its parameters and then take a spherical slice to create the temperature fluctuations of the CMB\@.
We expect this to be computationally significantly faster than solving \cref{eqn:a_lm_E_i}, especially for topology length scales of the order of or larger than $L_\mathrm{LSS}$.

\subsection{Expectations for three-dimensional versus two-dimensional correlations}
\label{secn:3dv2d}

We have seen that the reach of the CMB for extracting topological information is likely limited to the case where the distance to our nearest clone is not much more than the diameter of the last scattering surface.
We expect that three-dimensional data from the full interior of the last scattering surface --- such as might eventually be obtained using some combination of large-scale structure surveys, intensity mapping, and other data types ---  would allow us to extend this reach, possibly substantially.

In \cref{secn:eigenmodes}, we identified two topological effects on eigenmodes that could result in measurable effects on observables --- the \emph{discretization} of the set of allowed wavevectors $\vec{k}_{\vec{n}}$, and, in the non-homogeneous manifolds, the induced \emph{correlation} between Fourier modes of different allowed wavevectors of the same magnitude.
The observability of the discretization is limited by the observer's ability to resolve different wavevectors.
The separation between allowed wavevectors is $\sim2\pi/L_{\mathrm{top}}$, where $L_{\mathrm{top}}$ is some characteristic length scale associated with the topology.
The observer's resolution is limited in principle to $\sim 2\pi/L_{\mathrm{LSS}}$, leading the detectability to fall as   $L_{\mathrm{LSS}}/L_{\mathrm{top}}\to\infty$.
However, as  the magnitude of the wavevector increases, and the number of modes with the same magnitude of wavevector  grows as $k^2$,  it may become more and more difficult to attain the target angular resolution in the wavevector space, further degrading the ability to detect topological information.

Meanwhile, if we are in an inhomogeneous space in which one (or more) of the group generators involves a rotation about some axis, then all pairs of Fourier modes with wavevectors related by that same rotation have perfectly correlated amplitudes.
This is true no matter the magnitude of the wavevector.
That sounds like a potentially very powerful signal of topology.
However, the phase angle  of the correlation grows as the magnitude of the wavevector increases.
For large enough $k$, neighboring wavevectors have essentially random phases.
This will degrade the utility of the correlation information if observations average over nearby wavevectors.

All this needs to be evaluated carefully and is the subject of an upcoming paper \cite{Anselmi:2023tbd} targeted at this and related topics.

\section{Conclusion}
\label{secn:conclusion}

The local spatial geometry of the Universe may be Euclidean (i.e., flat) when averaged over volumes that are large compared to the largest structures but small compared to our Hubble volume or to the size of the last scattering sphere; 
however, that does not mean that the   topology is that of the infinite-in-all-directions ``covering space'' of Euclidean geometry \Espace.
There are eighteen possible topologies (labeled \E{1}--\E{18}) for three-manifolds that have homogeneous local flat geometry, the covering space is just one of them --- \E{18}.
In this paper, we have considered the ten topologies that have orientable Euclidean manifolds; in an upcoming  paper, we will consider the remaining eight topologies that have non-orientable Euclidean manifolds \cite{Copi:2023tbd}.
These ten topologies of orientable manifolds fall into four classes: six with compact manifolds, i.e., all three spatial dimensions are compact, (\E{1}--\E{6}); two that have two compact dimensions (\E{11} and \E{12}); one  that has one compact dimension (\E{16}); and the covering space \E{18}, which has no compact dimensions.

For each of these topologies (cf.\ \cref{secn:topologiesmanifolds}) we have  provided a completely general parametrization of all the possible homogeneous manifolds.
This builds on previous work (see especially \rcite{Riazuelo2004:prd}), which included all these topologies, but not, or not explicitly, with their most general parametrization.
In particular, we have allowed the full range of possibilities for the translation vectors associated with the  generators of \E{2}--\E{5} (that was already the case for \E{1}), the two generators of \E{11} and \E{12}, and the generator of \E{16}.
Most new, however, is the recognition that group actions that lead to the so-called slab space \E{16} are not generically pure translations, but corkscrew motions; the standard group action based on pure translations is a very special case because it is the only \E{16} manifold that is homogeneous.
We have also identified an ``associated \E{1}'' for each manifold of topologies \E{2}--\E{6}, an associated \E{11} for each \E{12} manifold, and an associated unrotated, and therefore homogeneous, \E{16} (i.e., \slabh) for each rotated \E{16} (i.e., \slabi) where that is possible, i.e, where the rotation is a rational multiple of $2\pi$.

In \cref{secn:eigenmodes}, we have provided the general analytic expressions for the eigenmodes of the (scalar) Laplacian for each of these orientable manifolds, which are linear combinations of Fourier modes.
Once again, we realize that \slabh\  is a very special case of \E{16}.
In considering a generic slab manifold, \slabi, based on a corkscrew motion we find that the rotation associated with that corkscrew must be a rational multiple of $2\pi$ in order for the eigenmodes (of the scalar Laplacian) not to have azimuthal symmetry around the corkscrew axis, and thus to be a basis set for general functions on the manifold.
This is a situation unlike any of the other \Espace\ topologies.

We have also presented the Fourier-mode correlation matrices $C^{\E{i};XY}_{\vec{k}_{\vec{n}}\vec{k}_{\vecnp}}$ for the amplitudes of the Fourier modes of the associated \E{1}, \E{11}, and \slabh, 
under the standard assumption that the primordial amplitudes of the eigenmodes of a particular  scalar field, e.g., the gauge-invariant curvature potential $\mathcal{R}$, are Gaussian random statistically independent variables of zero mean with a variance that depends only on the eigenvalue of the Laplacian.
The $X$ and $Y$ label two scalar fields (potentially at some later epoch) that are linearly related to that primordial field through transfer functions $\Delta^{X}(k)$ and $\Delta^{Y}(k)$.
($X$ and $Y$ are not necessarily different.)
Finally, we have presented, for similar scalar fields ``projected'' onto the sky, the spherical-harmonic-spherical-harmonic correlation matrix $C^{\E{i};XY}_{\ell m\ell'm'}$  for each topology as a function of manifold parameters.
Once again, this builds on previous work, especially \rcite{Riazuelo2004:prd}, but allows for the most general parametrization of each manifold and of the observer's position therein.

We have explicitly included the effect of the choice of origin $\vec{x}_0$ on the analytic expressions for the eigenmodes, and the correlation matrices.
This is an important addition because, except for \E{1}, \E{11}, \slabh, and \E{18}, the manifolds are inhomogeneous (i.e. expectation values of observables are inhomogeneous), and the expectations for the distributions of observables  depend on the location of an observer in the manifold.

In \cref{secn:numerical_results}, we have studied the CMB temperature spherical-harmonic correlation matrices $C^{\E{i};TT}_{\ell m\ell'm'}$ for \E{1}--\E{6}.
We have shown, in \cref{fig:rot_C}, how a preferred orientation of the observer concentrates the correlations in a small fraction of the matrix elements, while a random orientation diffuses it broadly across the matrix.
We have suggested for future consideration that this existence of a preferred orientation of the observer may constitute a signature of topology that could be used to identify the presence of a non-trivial topology.
In \cref{fig:cov_matrix_E1_E6}, as well as \cref{fig:E1_E2,fig:E3_E4,fig:E5_E6}, we have presented the spherical-harmonic correlation matrices for \E{1}--\E{6} for specific simple choices of the manifold parameters and the observer location.

\cref{fig:cov_matrix_E1_E6}, as well as \cref{fig:statistics-comparison} and \cref{fig:E1_E2,fig:E3_E4,fig:E5_E6}, also displays the KL divergence  $D_{\mathrm{KL}}(\E{i}||\E{18})$ of the usual (diagonal) \Espace\  covering-space CMB temperature correlation matrix $C^{\E{18};TT}_{\ell m\ell'm'}$ given a true underlying correlation matrix $C^{\E{i};TT}_{\ell m\ell'm'}$ of $\E{i}$ ($i\neq18$)  (assuming identical cosmological parameters).
This represents the information potentially available from measured CMB temperature fluctuations to distinguish between the possibility that the Universe is an \E{i} manifold and the possibility that it is the covering space.
We have also plotted $D_{\mathrm{KL}}(\E{18}||\E{i})$ in \cref{fig:cov_matrix_E1_E6,fig:statistics-comparison}, and an alternative cosmic variance signal-to-noise statistic \eqref{eqn:inversevar} which is more straightforward to calculate for large ranges of $\ell$, and which conveys the same message. We have not displayed these two statistics in \cref{fig:E1_E2,fig:E3_E4,fig:E5_E6} as they contain no clear new information.

These measures of the information content suggest that once the clone-to-last-scattering ratio --- the ratio of the distance to an observer's nearest topological clone to the diameter of the last scattering surface --- is greater than 1, the information available to distinguish between an \E{i} manifold and the covering space declines rapidly.
In most cases, we find that  the ``standard'' manifold parameter choices and observer locations minimize the available information and that other manifold parameters or observer positions are more favorable for detecting topology.
Nevertheless, in all cases,  by the time the clone-to-last-scattering ratio reaches somewhere between 1.1 and 1.3 the KL divergences of the CMB temperature fluctuations have fallen below 1 and there is no prospect of using those fluctuations to differentiate between topologies.
On the one hand, this is disappointing given that we already know observationally (with ~95\% confidence) \cite{Cornish:2003db,Cornish:2011ys} that this ratio is greater than $0.985$ --- we might have hoped to explore much larger values of the ratio.
On the other hand, if a non-trivial cosmic topology is the heart of the explanation for large-angle anomalies in the CMB temperature fluctuations \cite{Yoho:2010pb, Copi:2010na, Planck:2013lks, Rassat2014, Planck:2015igc, Schwarz:2015cma, Gruppuso:2017nap, Planck:2019evm, Cayuso:2019hen, Abdalla:2022yfr, Jones:2023ncn}, then there \emph{is} significant information about the topology of the Universe in the CMB, teaching us that the clone-to-last-scattering ratio is likely $\lesssim1.3$, assuming that the insights gained from \Espace\ apply more broadly to all manifolds.

The CMB is fundamentally poor at  determining the  amplitudes of individual Fourier modes because the CMB signal is dominated by the Sachs-Wolfe effect on the last scattering surface, and is thus a measure of the gravitational potential on that spherical shell centered on us.
The projection from the statistically simple correlation between Fourier modes to the correlation between spherical harmonics is certainly not invertible and leads to a considerable loss of information.
We are optimistic that three-dimensional observations (large-scale structure surveys, intensity mapping, \textit{etc}) will not experience the same precipitous drop in the KL divergence for clone-to-last-scattering ratios greater than $1$.
This will be explored in an upcoming paper \cite{Anselmi:2023tbd}.

\acknowledgments
We thank Jeffrey Weeks and David Singer for valuable conversations, and Javier Carr\'on Duque
for providing extensive constructive comments on a draft of this paper.
J.R.E.\ acknowledges support from the European Research Council under the Horizon 2020 Research and Innovation Programme (Grant agreement No.~819478).
Y.A.\ acknowledges support by the Spanish Research Agency (Agencia Estatal de Investigaci\'on)'s grant RYC2020-030193-I/AEI/10.13039/501100011033, by the European Social Fund (Fondo Social Europeo) through the  Ram\'{o}n y Cajal program within the State Plan for Scientific and Technical Research and Innovation (Plan Estatal de Investigaci\'on Cient\'ifica y T\'ecnica y de Innovaci\'on) 2017-2020, by the Spanish Research Agency through the grant IFT Centro de Excelencia Severo Ochoa No CEX2020-001007-S funded by MCIN/AEI/10.13039/501100011033, and by the Richard S.\ Morrison Fellowship at CWRU.
C.J.C., A.K., D.P.M., and G.D.S.\ acknowledge partial support from NASA ATP grant RES240737; G.D.S.\ from DOE grant DESC0009946; Y.A., G.D.S., P.P., S.S., Q.T., and O.G.\ from the Simons Foundation; Y.A., A.H.J., and G.D.S.\ from the Royal Society (UK); and A.H.J.\ from STFC in the UK\@.
A.T.\ is supported by the Richard S.\ Morrison Fellowship.
Y.A. and G.D.S. thank the INFN (Sezione di Padova); J.R.E., S.A., D.P.M.,  G.D.S., and A.T. thank the IFT; and  A.K. thanks CWRU and ICL
for support and hospitality where part of this work was accomplished.

\appendix

\section{Appendix: Construction of general generators}

In this work, the most general allowed set of three generators $g_{a_j}$ for each of the orientable \E{i} has been provided.
Here we describe a constructive, algebraic approach to determining the parametrization of these generators.

The topology of each orientable manifold can be set by the choice of elements of $SO(3)$, i.e., by the matrices $\mat{M}^{\E{i}}_a$.
There is limited freedom to adjust the $\mat{M}^{\E{i}}_a$: only the freedom to rotate the coordinate system to choose the orientations of the axes of rotation.\footnote{
    The set of matrices is not always unique even up to rotations of the coordinate system.
    The Hantzsche-Wendt and non-orientable manifolds can be described by multiple different sets of matrices.
    Further, for the non-orientable manifolds the same set of $O(3)$ elements $\mat{M}^{\E{i}}_a$ can be associated with discretely different translation vectors $\vec{T}^{\E{i}}_{a_j}$ and thus generate topologically distinct manifolds.
}
The remainder and essence of our task will then be to determine the most general allowed vectors $\vec{T}^{\E{i}}_{a_j}$ associated with the $\mat{M}^{\E{i}}_a$ of each \E{i}.
Once the orientation of the coordinate system has been set, any remaining freedom will be used to simplify the three vectors $\vec{T}^{\E{i}}_{a_j}$.
There are two principal tools available to constrain the components of $\vec{T}^{\E{i}}_{a_j}$:
\begin{enumerate}
    \item Any finite sequence of generators and their inverses are group elements.
    Functionally, this is enforced by checking that any such sequence that is a pure translation is an integer linear combination of some ``basis set'' of three linearly independent translations, so that the group is a discrete group, $\Gamma^{\E{i}}$.
    In \cref{secn:topologiesmanifolds} we call this basis set (the generators of) ``the associated \E{1}''.
    Part of this process is the determination of the set of such pure translations that can be chosen as the basis set.
    
    For example, consider two generators $g^{\E{n}}_{ai}$ and $g^{\E{n}}_{bj}$ of $\Gamma^{\E{n}}$ for one of the orientable Euclidean manifolds $\E{n}$, associated with the $SO(3)$ elements $\mat{M}^{\E{n}}_a$ and $\mat{M}^{\E{n}}_b$, respectively, and let $\vec{T}^{\E{n}}_k$ ($k=1,2,3$) represent the pure translations of the associated \E{1}.
    Since $(\mat{M}^{\E{n}}_a)^{-1} (\mat{M}^{\E{n}}_b)^{-1} \mat{M}^{\E{n}}_a \mat{M}^{\E{n}}_b=\identity$ for all $a$ and $b$ in all such $\Gamma^{\E{n}}$,\footnote{
        Note that this is not true for arbitrary elements of $O(3)$ since it is a non-abelian group. However, it is true for the particular elements of $O(3)$ used in the generators for each of the \E{n}.}
    we must insist that 
    \begin{equation}
        (g^{\E{n}}_{ai})^{-1} (g^{\E{n}}_{bj})^{-1} g^{\E{n}}_{ai} g^{\E{n}}_{bj}: \vec{x} \to \vec{x} + \sum_{k=1}^3 m_k \vec{T}^{\E{n}}_k
        \label{eqn:topologycondition}
    \end{equation}
    for some triplet of integers $m_k$.
    
    This may not be sufficient.
    Consider the case where $\mat{M}^{\E{n}}_a\neq \identity$ but $(\mat{M}^{\E{n}}_a)^2 = \identity$ and where there are (at least) two generators associated with this matrix.
    Then the application of any two of these generators should result in a pure translation.
    In other words, all combinations of the form
    \begin{equation}
        g^{\E{n}}_{ai} g^{\E{n}}_{aj}, \quad g^{\E{n}}_{a{i}} (g^{\E{n}}_{a{j}})^{-1}, \quad (g^{\E{n}}_{a{i}})^{-1} g^{\E{n}}_{a{j}}, \quad (g^{\E{n}}_{a{i}})^{-1} (g^{\E{n}}_{a{j}})^{-1}
    \end{equation}
    for each of $i,j\in \{1, 2\}$ must lead to pure translations that are integer linear combinations of the basis set.
    If this is true, then \eqref{eqn:topologycondition} is trivially satisfied.
    For some topologies two of these combinations will be used to define the translation vectors and the rest will lead to constraints that must be satisfied.
    \item We must also ensure that the set of transformations (the group elements) consists only of freely acting transformations, i.e., no transformation (other than the identity transformation) has a fixed point.
\end{enumerate}
    
Once these conditions have been enforced, we may find that there appear to be distinct sets of ``solutions'', i.e., parametrizations of the $\vec{T}^{\E{i}}_{a_j}$ that cannot be transformed into one another by rotations, reorderings, or rescalings.
We must still prove that two such sets do not generate the same lattice of clones for a given starting point.
A simple way that this can happen is if the $\vec{T}^{\E{i}}_{a_j}$ of one set are just integer linear combinations of the vectors of the other set.

We will show individually that we are able to bring this program to a successful conclusion for each of the orientable manifolds.

\subsection{\E{1}: 3-torus}
\label{app:E1}

We have established in the main body of the paper that the three translations for \E{1} can be parametrized most generally as
\begin{equation} 
    \vec{T}^{\E{1}}_{1} = \begin{pmatrix} L_{1x} \\ 0 \\ 0 \end{pmatrix}, \quad
    \vec{T}^{\E{1}}_{2} = \begin{pmatrix} L_{2x} \\ L_{2y} \\ 0 \end{pmatrix}, \quad
    \vec{T}^{\E{1}}_{3} = \begin{pmatrix} L_{3x} \\ L_{3y} \\ L_{3z} \end{pmatrix} ,
\end{equation}
where the three $0$ entries have been fixed through a rotation.
Specifics on the ranges of the parameters are given in the main body.

\subsection{\E{2}: Half-turn space}
\label{app:E2}

The half-turn space has one generator, $g_B$, that is a corkscrew motion, while the other two, $g_{\A{j}}$ ($j=1,2$), are pure translations.
Here we develop the techniques employed in the rest of the orientable manifolds to algebraically determine an action for the most general generators.

To begin we first use two of the rotational degrees of freedom to align the corkscrew axis with the $z$-axis, i.e., $\mat{M}_B = \diag(-1, -1, 0)$, and use the remaining rotational degree of freedom to set $L_{1y}=0$.
Next, we can choose a special origin that allows us to set $L_{Bx} = L_{By} = 0$ (cf.\ \cref{eqn:actionofgeneratoraxisshift}).
We will restore an arbitrary origin at the end of the calculation.
With all of this, we can choose our general starting case to be\footnote{Here and throughout this appendix we will refrain from adding topology-specific superscripts to quantities until the final results.
All intermediate quantities should be recognized as being specific to the topology under consideration.}
\begin{equation}
    \vec{T}_{\A{1}} = \begin{pmatrix} L_{1x} \\ 0 \\ L_{1z} \end{pmatrix}, \quad
    \vec{T}_{\A{2}} = \begin{pmatrix} L_{2x} \\ L_{2y} \\ L_{2z} \end{pmatrix}, \quad
    \vec{T}_{B} = \begin{pmatrix} 0 \\ 0 \\ L_{Bz} \end{pmatrix}.
\end{equation}
From these a set of pure translations, an associated \E{1}, can be constructed.
The actions of both the $g_{\A{i}}$ are already pure translations, so two of the needed translation vectors can be chosen as $\vec{T}_1 \equiv \vec{T}_{\A{1}}$ and $\vec{T}_2 \equiv \vec{T}_{\A{2}}$.
A third translation vector can be defined based on the fact that $(\mat{M}^{\E{2}}_B)^2 = \identity$.
Note that
\begin{equation}
    g^{\E{2}}_3 \equiv (g^{\E{2}}_B)^2: \vec{x} \to \vec{x} + (\identity + \mat{M}^{\E{2}}_B) \vec{T}^{\E{2}}_B.
\end{equation}
From this the third translation vector is
\begin{equation}
    \vec{T}^{\E{2}}_3 \equiv (\identity + \mat{M}^{\E{2}}_B) \vec{T}^{\E{2}}_B \equiv \mat{M}^{\E{2}}_{02} \vec{T}^{\E{2}}_B,
\end{equation}
for $\mat{M}^{\E{2}}_{02} = \diag(0, 0, 2)$ from \eqref{eqn:M0jdef}.
Thus the third translation vector of the associated \E{1} for \E{2} is as quoted in the main text \eqref{eqn:E2assocE1},
\begin{equation}
    \vec{T}^{\E{2}}_3 = \begin{pmatrix} 0 \\ 0 \\ 2L_{Bz} \end{pmatrix}.
\end{equation}
For our subsequent development here we choose $\vec{T}_3 \equiv \vec{T}^{\E{2}}_3$.

The remaining condition to impose (all other combinations of generators will either already be pure translations or reduce to pure translations of these) is that for $i$ and $j\in\{1, 2\}$ 
\begin{equation}
    \inverse{g}_B \inverse{g}_{\A{i}} g_B g_{\A{j}}: \vec{x}  \to \vec{x} + \vec{T}_{\A{j}} - \mat{M}_B \vec{T}_{\A{i}}
    = \vec{x} + \sum_{k} m^{(i)}_k \vec{T}_k,
\end{equation}
for some set of $m_k^{(i)}\in \integers$.
Since $\vec{T}_{\A{j}}$ is already one of the pure translations, $\vec{T}_j$, this is equivalent to requiring that
\begin{equation}
    \label{eqn:E2groupcondition}
    \mat{M}_B \vec{T}_{\A{i}} = \sum_{k} m^{(i)}_k \vec{T}_k, \quad \mbox{for some } m_k^{(i)}\in \integers ,
\end{equation}
leading to the set of conditions
\begin{align}
    \begin{pmatrix}
    -L_{1x}\\
    \hphantom{-} 0\\
    \hphantom{-} L_{1z}
    \end{pmatrix}
    &= 
    \begin{pmatrix}
    m^{(1)}_1 L_{1x} + m^{(1)}_2 L_{2x}  \\
    m^{(1)}_2 L_{2y}  \\
    m^{(1)}_1 L_{1z} + m^{(1)}_2 L_{2z} 
        + m^{(1)}_3 2 L_{Bz}
    \end{pmatrix},
    \\
    \begin{pmatrix}
    -L_{2x}\\
    -L_{2y}\\
    \hphantom{-} L_{2z}
    \end{pmatrix}
    &= 
    \begin{pmatrix}
    m^{(2)}_1 L_{1x} + m^{(2)}_2 L_{2x}  \\
    m^{(2)}_2 L_{2y}  \\
    m^{(2)}_1 L_{1z} + m^{(2)}_2 L_{2z} 
        + m^{(2)}_3 2 L_{Bz}
    \end{pmatrix}.
\end{align}

From the $x$-components of these equations and the fact that $L_{2y}\neq 0$ (since at least one of the translation vectors needs to have a nonzero $y$-component) we immediately see that $m_2^{(1)}=0$ and $m_2^{(2)}=-1$, which in turn means that $m_1^{(1)}=-1$ and $m_1^{(2)}=0$, so that we are finally left with
\begin{equation}
    L_{1z} = m^{(1)}_3  L_{Bz} \mbox{ and } L_{2z} = m^{(2)}_3  L_{Bz} .
\end{equation}
Since $T_{3z} = 2 L_{Bz}$, we can always subtract integer multiples of $\vec{T}_3$, which is equivalent to subtracting $2 n L_{Bz}$ for $n\in\integers$ from $L_{1z}$ and $L_{2z}$, meaning that we only need to consider $m^{(1)}_3\in\{0,1\}$ and $m^{(2)}_3\in\{0,1\}$.
However, if either of $m^{(i)}_3$ are 1, then $\inverse{g}_{\A{i}} g_B$ has a fixed point.
Explicitly, consider the case $m^{(1)}_3=1$ so that $L_{1z}=L_{Bz}$.
Notice that
\begin{equation}
    \inverse{g}_{\A{1}} g_B: \vec{x} \to \begin{pmatrix} -x - L_{1x} \\ -y \\ \hphantom{-} z \end{pmatrix} \,,
\end{equation}
and so $\inverse{g}_{\A{1}} g_B:\vec{x} \to\vec{x}$,
for the special location $\vec{x} = \transpose{(-L_{1x}/2, 0, z)}$.
Hence $\inverse{g}_{\A{1}} g_B$ has a fixed point and this case is invalid.
A similar argument for $m^{(2)}_3 = 1$ leads to the requirement $L_{1z} = L_{2z} = 0$.

Restoring an arbitrary origin (which reintroduces $L_{Bx}$ and $L_{By}$), we are left with the generic result quoted in the main text (\ref{eqn:E2generalT}),
\begin{equation}
    \vec{T}^{\E{2}}_{\A{1}} = \begin{pmatrix} L_{1x} \\ 0 \\ 0 \end{pmatrix}, \quad
    \vec{T}^{\E{2}}_{\A{2}} = \begin{pmatrix} L_{2x} \\ L_{2y} \\ 0 \end{pmatrix}, \quad
    \vec{T}^{\E{2}}_{B} = \begin{pmatrix} L_{Bx} \\ L_{By} \\ L_{Bz} \end{pmatrix}\,,
\end{equation}
or equivalently (\ref{eqn:E2generalTalt}),
\begin{equation}
    \vec{T}^{\E{2}}_{\A{1}} = L_1 \begin{pmatrix} 1 \\ 0 \\ 0 \end{pmatrix}, \quad
    \vec{T}^{\E{2}}_{\A{2}} = L_2 \begin{pmatrix} \cos\alpha \\ \sin\alpha \\ 0 \end{pmatrix}, \quad
    \vec{T}^{\E{2}}_{B} = L_B \begin{pmatrix} \cos\beta\cos\gamma \\ \cos\beta\sin\gamma \\ \sin\beta \end{pmatrix}.
\end{equation}

\subsection{\E{3}: Quarter-turn space}
\label{app:E3}

The quarter-turn space is similar to \E{2}.
Since there is a quarter turn, instead of a half turn, the computations are a bit more involved, but follow similar logic.
Here we will explicitly impose the ordering conditions discussed in the main text.

To begin we first use the freedom to rotate the coordinate system such that the axis of rotation of $\mat{M}_B$ is parallel to $\unitvec{e}_z$, and then to choose $\vec{T}_{\A{1}}$ to be in the $xz$-plane.
For \E{3}
\begin{equation}
    \mat{M}_B = \mat{R}_{\unitvec{z}}(\pi/2) =
        \begin{pmatrix}
        0 & -1 & 0\\
        1 & \hphantom{-}0 & 0\\
        0 & \hphantom{-}0 & 1
        \end{pmatrix} ,
\end{equation}
which has the properties
\begin{equation}
    (\mat{M}_B)^2 = \diag(-1, -1, 1), \quad
    (\mat{M}_B)^3 = \transpose{\mat{M}}_B, \quad
    (\mat{M}_B)^4 = \identity .
\end{equation}
Once again $\mat{M}_A=\identity$, so $g_{\A{1}}$ and $g_{\A{2}}$ are pure translations by $\vec{T}_{\A{1}}$ and $\vec{T}_{\A{2}}$, respectively.
We therefore assign them the alternative labels $\vec{T}_{1}\equiv\vec{T}_{\A{1}}$ and $\vec{T}_{2}\equiv\vec{T}_{\A{1}}$.
Finally, as in \E{2} we can shift to the origin where $L_{Bx}=L_{By}=0$ and again have as our starting case
\begin{equation}
    \vec{T}_{\A{1}} = \begin{pmatrix} L_{1x} \\ 0 \\ L_{1z} \end{pmatrix}, \quad
    \vec{T}_{\A{2}} = \begin{pmatrix} L_{2x} \\ L_{2y} \\ L_{2z} \end{pmatrix}, \quad
    \vec{T}_{B} = \begin{pmatrix} 0 \\ 0 \\ L_{Bz} \end{pmatrix}.
\end{equation}
Similar to \E{2}, an associated \E{1} can be constructed from three pure linearly independent translations.
We again have the obvious translation vectors $\vec{T}_1 \equiv \vec{T}_{\A{1}}$ and $\vec{T}_2 \equiv \vec{T}_{\A{2}}$.
A third translation vector can be defined based on the fact that $(\mat{M}^{\E{3}}_B)^4 = \identity$.
Noting that
\begin{equation}
    g^{\E{3}}_3 \equiv (g^{\E{3}}_B)^4: \vec{x} \to \vec{x} + \mat{M}^{\E{3}}_{04} \vec{T}^{\E{3}}_B,
\end{equation}
where $\mat{M}^{\E{3}}_{04} = \diag(0, 0, 4)$ from \eqref{eqn:M0jdef},
the third translation vector of the associated \E{1} for \E{3} is, as quoted in the main text \eqref{eqn:E3assocE1},
\begin{equation}
    \vec{T}^{\E{3}}_3 = \begin{pmatrix} 0 \\ 0 \\ 4 L_{Bz} \end{pmatrix}.
\end{equation}
For our subsequent development here we choose $\vec{T}_3 \equiv \vec{T}^{\E{3}}_3$.

To proceed we can always choose $\vec{T}_{\A{1}}$ such that 
$\vert\vec{T}_{\A{1}\perp}\vert \leq \vert\vec{T}_{\A{2}\perp}\vert$, where $\perp$ means the projection into the $xy$-plane.
Also, by adding or subtracting integer multiples of $\vec{T}_1$ from $\vec{T}_{\A{2}}$, we can ensure that $\vert L_{2x}\vert < \vert L_{1x}\vert$; thus we generically have
\begin{equation}
     (L_{2x})^2 < (L_{1x})^2 \leq (L_{2x})^2 + (L_{2y})^2 .
\end{equation}
Similarly, shifting by integer multiples of $\vec{T}_3$ allows us to insist that $\vert L_{1z}\vert \leq 2\vert L_{Bz}\vert$ and $\vert L_{2z}\vert \leq 2\vert L_{Bz}\vert$.

Given the basis set for the associated \E{1} of $\vec{T}_1$, $\vec{T}_2$, and $\vec{T}_3$ we can proceed to find the most general action of the generators.
Whereas for \E{2} we had one set of conditions \eqref{eqn:E2groupcondition} on the products of generators and their inverses being integer linear combination of $\vec{T}_{i}^{\E{2}}$, for \E{3} we have two independent sets, which we can take for $i$ and $j\in\{1, 2\}$ to be
\begin{align}
    \label{eqn:E3groupcondition1}
    {g}_B \inverse{g}_{\A{i}} \inverse{g}_B g_{\A{j}}: \vec{x} &\to \vec{x} + \vec{T}_{\A{j}} - {\mat{M}_B} \vec{T}_{\A{i}} 
    = \vec{x} + \vec{T}_{\A{j}}+ \sum_{k} m^{(i)}_k \vec{T}_k,\\
    \label{eqn:E3groupcondition2}
    (\inverse{g}_B)^2 \inverse{g}_{\A{i}} (g_B)^2 g_{\A{j}}: \vec{x} &\to \vec{x} +  \vec{T}_{\A{j}} - (\mat{M}_B)^2 \vec{T}_{\A{i}} 
    = \vec{x} + \vec{T}_{\A{j}}+ \sum_{k} n^{(i)}_k \vec{T}_k.
\end{align}
Just like in \E{2}, these reduce to
\begin{align}
    \label{eqn:E3groupconditionSimple1}
    \mat{M}_B \vec{T}_{\A{i}} 
    &=  \sum_{k} m^{(i)}_k \vec{T}_k, \quad \mbox{for some } m^{(i)}_k \in \integers,\\
    \label{eqn:E3groupconditionSimple2}
    (\mat{M}_B)^2 \vec{T}_{\A{i}} 
    &=  \sum_{k} n^{(i)}_k \vec{T}_k,  \quad \mbox{for some } n^{(i)}_k \in \integers.
\end{align}
Since $\mat{M}_B\vec{T}_{\A{1}}=\vec{T}_{\A{2}}$, and since we could have used $ \inverse{g}_B \inverse{g}_{\A{i}} {g_B} g_{\A{j}}$ instead of  $ {g}_B \inverse{g}_{\A{i}} \inverse{g_B} g_{\A{j}}$ for the left-hand side of \eqref{eqn:E3groupcondition1}, which would have replaced $\mat{M}_B$ with $\transpose{\mat{M}}_B=\mat{M}_B^3$ on the right-hand side of \eqref{eqn:E3groupcondition1},
we need to only consider \eqref{eqn:E3groupconditionSimple1}.
Thus, despite all the apparent freedom, this once again leads to only two sets of conditions:
\begin{eqnarray}
    \label{eqn:E3conditions1}
    \begin{pmatrix}
    0\\
    L_{1x}\\
    L_{1z}
    \end{pmatrix}
    &= 
    \begin{pmatrix}
    m^{(1)}_1 L_{1x} + m^{(1)}_2 L_{2x}  \\
    m^{(1)}_2 L_{2y}  \\
    m^{(1)}_1 L_{1z} + m^{(1)}_2 L_{2z} 
        + m^{(1)}_3 4 L_{Bz}
    \end{pmatrix},
    \\
    \label{eqn:E3conditions2}
    \begin{pmatrix}
    -L_{2y}\\
    \hphantom{-}L_{2x}\\
    \hphantom{-}L_{2z}
    \end{pmatrix}
    &= 
    \begin{pmatrix}
    m^{(2)}_1 L_{1x} + m^{(2)}_2 L_{2x}  \\
    m^{(2)}_2 L_{2y}  \\
    m^{(2)}_1 L_{1z} + m^{(2)}_2 L_{2z} 
        + m^{(2)}_3 4 L_{Bz}
    \end{pmatrix}\,.
\end{eqnarray}

As in \E{2} we must have $L_{2y}\neq 0$.
From the $y$-component of \eqref{eqn:E3conditions1} we see immediately that
\begin{equation}
    \label{eqn:E3L1xcondition}
    L_{1x} = m_2^{(1)} L_{2y} \neq 0 .
\end{equation} 
To show this, suppose that $L_{1x}=0$.
From \eqref{eqn:E3L1xcondition} this means that $m_2^{(1)} = 0$.
Applying these relations to the $x$- and $y$-components of \eqref{eqn:E3conditions2} leads to $L_{2y} = - (m^{(2)}_2)^2 L_{2y}$.
Since $L_{2y}\neq 0$ this would then require $m^{(2)}_2=0$, but from the $x$-component of \eqref{eqn:E3conditions2} this would force $L_{2y}=0$, which is inconsistent.
Thus, $L_{1x}\neq 0$ and we must have $m_2^{(1)}\neq 0$.

Proceeding, we note that $\vert\vec{T}_{\A{1}\perp}\vert = \vert L_{1x} \vert = \vert m^{(1)}_2 \vert \vert L_{2y} \vert$ and $\vert\vec{T}_{\A{2}\perp}\vert = \sqrt{L_{2x}^2 + L_{2y}^2} \leq \vert L_{2y} \vert$.
Thus $\vert\vec{T}_{\A{1}\perp}\vert \leq \vert\vec{T}_{\A{2}\perp}\vert$ requires $\vert m_2^{(1)}\vert\leq1$, which means that $m_2^{(1)}=\pm 1$.
Next, from the $x$-component of \eqref{eqn:E3conditions1}, we require $L_{2x} = -m_1^{(1)} L_{2y}$.
However, we can always add or subtract integer multiples of $\vec{T}_1$ from $\vec{T}_{\A{2}}$, so $\vert L_{2x} \vert$ cannot be larger than $\frac{1}{2}L_{1x}$, but again, since $\vert\vec{T}_{\A{1}\perp}\vert \leq \vert\vec{T}_{\A{2}\perp}\vert$, we must have 
\begin{equation}
    0<L_{1x}^2\leq L_{2x}^2 + L_{2y}^2 =  \frac{1+ (m_1^{(1)})^2}{(m_1^{(1)})^2} L_{2x}^2 \leq 
      \frac{1+ (m_1^{(1)})^2}{(m_1^{(1)})^2} \frac{L_{1x}^2}{4} \,.
\end{equation}
This is satisfied if and only if $m_1^{(1)} = 0$.
Since $m_2^{(1)} \neq 0$, the $x$-component of \eqref{eqn:E3conditions1} informs us that 
\begin{equation}
    L_{2x} = 0 .
\end{equation}
The $y$-component of \eqref{eqn:E3conditions2} then requires that $m_2^{(2)} = 0$.

Next, the $x$-component of \eqref{eqn:E3conditions2} together with \eqref{eqn:E3L1xcondition} and $m_2^{(2)}=0$
gives us that $m_1^{(2)} m_2^{(1)} = -1$, thus (since $m_2^{(1)}=\pm1$) we have $m_1^{(2)} = -m_2^{(1)} = \mp 1$.
We thus conclude that 
\begin{equation}
    L_{2y} = \pm L_{1x}.
\end{equation}

The $z$-components of the two conditions now give us
\begin{align}
    \label{eqn:E3conditionagain}
    0 &= - L_{1z} \pm L_{2z} + m_3^{(1)} 2 L_{Bz}, \nonumber \\
    0 &= \mp  L_{1z} -  L_{2z}  + m_3^{(2)} 4 L_{Bz} .
\end{align}
Thus $L_{1z}$ and $L_{2z}$ are both integer multiples of $L_{Bz}$.
But, as in \E{2}, unless these integers are both $0$, at least one of $\inverse{g}_{\A{1}} (g_B)^n$ for $n\in \{1, 2, 3\}$ will have a fixed point.
From this, we can show that we must have 
\begin{equation}
    L_{1z} = L_{2z} = 0.
\end{equation}

Putting all of this together, shifting to an arbitrary origin, and renaming $L_{1x}$ to $L_A$ we arrive at the general form from the text \eqref{eqn:E3generalT}
\begin{equation}
    \vec{T}^{\E{3}}_{\A{1}} = \begin{pmatrix} L_{A} \\ 0 \\ 0 \end{pmatrix}, \quad
    \vec{T}^{\E{3}}_{\A{2}} = \begin{pmatrix} 0 \\ L_{A} \\ 0 \end{pmatrix}, \quad
    \vec{T}^{\E{3}}_{B} = \begin{pmatrix} L_{Bx} \\ L_{By} \\ L_{Bz} \end{pmatrix} .
\end{equation}
Note that we might have thought we should put $\pm L_A$ in $\vec{T}^{\E{3}}_{\A{2}}$, however this can be removed by rotating about the $z$-axis by $\pi/2$ or by replacing $\vec{T}^{\E{3}}_{\A{2}}$ with $-\vec{T}^{\E{3}}_{\A{2}}$.

\subsection{\E{4}: Third-turn space}
\label{app:E4}

The third-turn space is similar to the previous two cases, particularly \E{3}, from which the development follows.
We again begin by using the rotational freedom to set the axis of $\mat{M}_B$ parallel to $\unitvec{e}_z$ and then to choose $\vec{T}_{\A{1}}$ to be in the $xz$-plane.
For \E{4}
\begin{equation}
    \mat{M}_B = \mat{R}_{\unitvec{z}}(2\pi/3) =
        \begin{pmatrix}
        -1/2 & -\sqrt{3}/2 & 0\\
        \sqrt{3}/2 & -1/2 & 0\\
        0 & \hphantom{-}0 & 1
        \end{pmatrix} ,
\end{equation}
which has the properties
\begin{equation}
    (\mat{M}_B)^2 = \transpose{\mat{M}}_B \mbox{ and } (\mat{M}_B)^3 = \identity.
\end{equation}
As in \E{2} and \E{3} and shifting the origin so that $L_{Bx}=L_{By}=0$, we again start with
\begin{equation}
    \vec{T}_{\A{1}} = \begin{pmatrix} L_{1x} \\ 0 \\ L_{1z} \end{pmatrix} , \quad
    \vec{T}_{\A{2}} = \begin{pmatrix} L_{2x} \\ L_{2y} \\ L_{2z} \end{pmatrix} , \quad
    \vec{T}_{B} = \begin{pmatrix} 0 \\ 0 \\ L_{Bz} \end{pmatrix}.
\end{equation}
Similar to \E{2}, an associated \E{1} can be constructed from three pure translations.
We again have the obvious translation vectors $\vec{T}_1 \equiv \vec{T}_{\A{1}}$ and $\vec{T}_2 \equiv \vec{T}_{\A{2}}$.
A third translation vector can be defined based on the fact that $(\mat{M}^{\E{4}}_B)^3 = \identity$.
Noting that
\begin{equation}
    g^{\E{4}}_3 \equiv (g^{\E{4}}_B)^3: \vec{x} \to \vec{x} + \mat{M}^{\E{4}}_{03} \vec{T}^{\E{4}}_B,
\end{equation}
where $\mat{M}^{\E{4}}_{03} = \diag(0, 0, 3)$ from \eqref{eqn:M0jdef},
the third translation vector of the associated \E{1} for \E{4} is as quoted in the main text \eqref{eqn:E4assocE1},
\begin{equation}
    \vec{T}^{\E{4}}_3 = \begin{pmatrix} 0 \\ 0 \\ 3 L_{Bz} \end{pmatrix}.
\end{equation}
For our subsequent development here, we choose $\vec{T}_3 \equiv \vec{T}^{\E{4}}_3$.

To proceed, we can always choose $\vec{T}_{\A{1}}$ such that $\vert\vec{T}_{\A{1}\perp}\vert \leq \vert\vec{T}_{\A{2}\perp}\vert$, where $\perp$ means the projection into the $xy$-plane.
Also, by adding or subtracting integer multiples of $\vec{T}_1$ from $\vec{T}_{\A{2}}$, we can ensure that $\vert L_{2x}\vert < \vert L_{1x}\vert$; thus we generically have
\begin{equation}
    \label{eqn:E4A1A2ordering}
     (L_{2x})^2 < (L_{1x})^2 \leq (L_{2x})^2 + (L_{2y})^2  .
\end{equation}

Since $\mat{M}_B^2 = \transpose{\mat{M}}_B$, we only need to consider the condition with one factor of $g_B$,
\begin{equation}
    \label{eqn:E4groupconditions}
    \inverse{g}_B \inverse{g}_{\A{i}} g_B g_{\A{j}}: \vec{x} \to \vec{x} + \vec{T}_{\A{j}} - \transpose{\mat{M}}_B \vec{T}_{\A{i}} 
    = \vec{x} + \sum_{k} m^{(i)}_k \vec{T}_k,
\end{equation}
and not the condition analogous to \eqref{eqn:E3groupcondition2}.
Again this reduces to
\begin{align}
    \mat{M}_B \vec{T}_{\A{i}} = \sum_{k} m^{(i)}_k \vec{T}_k, \quad \mbox{for some } m_k^{(i)}\in \integers, 
\end{align}
and again leads to the two sets of conditions
\begin{align}
    \label{eqn:E4puretranslations1}
    \begin{pmatrix}
    -\frac{1}{2}L_{1x}\\
   -\frac{\sqrt{3}}{2}L_{1x}\\
    L_{1z}
    \end{pmatrix}
    &= 
    \begin{pmatrix}
    m^{(1)}_1 L_{1x} + m^{(1)}_2 L_{2x}  \\
    m^{(1)}_2 L_{2y}  \\
    m^{(1)}_1 L_{1z} + m^{(1)}_2 L_{2z} 
        + m^{(1)}_3 3 L_{Bz}
    \end{pmatrix},
    \\
    \label{eqn:E4puretranslations2}
    \begin{pmatrix}
    -\frac{1}{2}L_{2x}-\frac{\sqrt{3}}{2}L_{2y}\\
    \frac{\sqrt{3}}{2}L_{2x}-\frac{1}{2}L_{2y}\\
    L_{2z}
    \end{pmatrix}
    &= 
    \begin{pmatrix}
    m^{(2)}_1 L_{1x} + m^{(2)}_2 L_{2x}  \\
    m^{(2)}_2 L_{2y}  \\
    m^{(2)}_1 L_{1z} + m^{(2)}_2 L_{2z} 
        + m^{(2)}_3 3 L_{Bz}
    \end{pmatrix} \,.
\end{align}

Similar to \E{3}, the $y$-component of \eqref{eqn:E4puretranslations1} requires that $m^{(1)}_2\neq 0$.
To see this, we note that $m^{(1)}_2 = 0$ means that $L_{1x}=0$.
Applying these two relations to the $x$- and $y$-components of \eqref{eqn:E4puretranslations2} leads to contradictions.
Explicitly, if $m^{(2)}_2 = 0$ then $L_{2y}=0$, whereas if $m^{(2)}_2 \neq 0$ then $(L_{2x}/L_{2y})^2 = -1$; neither of these are valid solutions.
Thus we require $m^{(1)}_2\neq 0$.

Returning to the $x$- and $y$-components of \eqref{eqn:E4puretranslations1} we have
\begin{equation}
    L_{2x} = - \frac{2 m_1^{(1)}+1}{2 m_2^{(1)}}  L_{1x} \mbox{ and }
    L_{2y} = -\frac{\sqrt{3}}{2 m_2^{(1)}} L_{1x}\,,
\end{equation}
for $m_1^{(1)}\in \integers$ and $m_2^{(1)}\in \integers^{\neq 0}$.
For \eqref{eqn:E4A1A2ordering} to be satisfied with these relations, we must have
\begin{equation}
    (m_1^{(1)})^2 + m_1^{(1)} + \frac{1}{4} \leq (m_2^{(1)})^2 \leq (m_1^{(1)})^2 + m_1^{(1)} + 1 .
\end{equation}
The only (integer) solutions to these inequalities are 
\begin{equation}
    m_1^{(1)} \in \{-1, 0\} \mbox{ and } m_2^{(1)} = \pm 1.
\end{equation}
These translate to 
\begin{equation}
    \label{eqn:E4L2xysolutions}
    L_{2x} = \pm \frac{1}{2} L_{1x} \mbox{ and }
    L_{2y} = \pm \frac{\sqrt{3}}{2} L_{1x},
\end{equation}
with independent sign choices.
Plugging the expressions for $L_{2x}$ and $L_{2y}$ into the $x$- and $y$-components of \eqref{eqn:E4puretranslations2} leads to 
\begin{equation}
    m_1^{(2)} \in \{-1, 0\} \mbox{ and } m_2^{(2)} = \pm 1.
\end{equation}

The many choices for the sets of integers $m^{(i)}_j$ with $i, j\in \{1, 2\}$, when plugged into the $z$-components of \eqref{eqn:E4puretranslations1} and \eqref{eqn:E4puretranslations2}, lead to the requirements that $L_{1z}$ and $L_{2z}$ must be integer multiples of $L_{Bz}$.
We can always make these integer multiples $0$, $1$, or $2$ by adding or subtracting integer multiples of $\vec{T}_3$.
But, as in the previous cases, choosing either $1$ or $2$ leads to fix points.
Thus $L_{1z}=L_{2z}=0$.

The independent choices of signs for $L_{2x}$ and $L_{2y}$ are equivalent as it is immaterial whether we take $\vec{T}_{\A{2}}$ or $-\vec{T}_{\A{2}}$, since it is a pure translation.
Using this, we can always choose $L_{2y} > 0$.
Finally, we can always switch between $L_{2x}=-\frac{1}{2} L_{1x}$ and $L_{2x}=+\frac{1}{2} L_{1x}$ by adding (subtracting) $\vec{T}_{\A{1}}$ to (from) $\vec{T}_{\A{2}}$.
We choose $L_{2x}=-\frac{1}{2}L_{1x}$.

Putting all of this together, shifting to an arbitrary origin, and renaming $L_{1x}$ to $L_A$ we arrive at the general form from the text \eqref{eqn:E4generalT},
\begin{equation}
    \vec{T}^{\E{4}}_{\A{1}} = \begin{pmatrix} L_{A} \\ 0 \\ 0 \end{pmatrix} , \quad
    \vec{T}^{\E{4}}_{\A{2}} = \begin{pmatrix} -L_{A}/2 \\ \sqrt{3} L_{A} / 2 \\ 0 \end{pmatrix} , \quad
    \vec{T}^{\E{4}}_{B} = \begin{pmatrix} L_{Bx} \\ L_{By} \\ L_{Bz} \end{pmatrix} .
\end{equation}

\subsection{\E{5}: Sixth-turn space}
\label{app:E5}

The sixth turn space is very much like \E{4}, though with a richer matrix structure.
We begin by using the rotational freedom to set the axis of $\mat{M}_B$ parallel to $\unitvec{e}_z$ and then to choose $\vec{T}_{\A{1}}$ to be in the $xz$-plane.
For \E{5}
\begin{equation}
    \mat{M}_B = \mat{R}_{\unitvec{z}}(\pi/3) =
        \begin{pmatrix}
        1/2 & -\sqrt{3}/2 & 0\\
        \sqrt{3}/2 & 1/2 & 0\\
        0 & \hphantom{-}0 & 1
        \end{pmatrix} ,
\end{equation}
 which has the properties
 \begin{align}
    &(\mat{M}_B)^2 = \mat{R}_{\unitvec{z}}(2\pi/3), \quad
    (\mat{M}_B)^3 = \mat{R}_{\unitvec{z}}(\pi) = \diag(-1, -1, 1), \nonumber \\
    &(\mat{M}_B)^4 = \mat{R}_{\unitvec{z}}(4\pi/3)
    = (\transpose{\mat{M}}_B)^2, \quad
    (\mat{M}_B)^5 = \mat{R}_{\unitvec{z}}(5\pi/3)
    = \transpose{\mat{M}}_B, \quad
    (\mat{M}_B)^6 = \identity .
\end{align}
As in the previous cases, we shift the origin so that $L_{Bx} = L_{By} = 0$, and we again start with
\begin{equation}
    \vec{T}_{\A{1}} = \begin{pmatrix} L_{1x} \\ 0 \\ L_{1z} \end{pmatrix} , \quad
    \vec{T}_{\A{2}} = \begin{pmatrix} L_{2x} \\ L_{2y} \\ L_{2z} \end{pmatrix} , \quad
    \vec{T}_{B} = \begin{pmatrix} 0 \\ 0 \\ L_{Bz} \end{pmatrix}.
\end{equation}
An associated \E{1} can again be constructed from three pure translations,
starting with the obvious translation vectors $\vec{T}_1 \equiv \vec{T}_{\A{1}}$ and $\vec{T}_2 \equiv \vec{T}_{\A{2}}$.
A third translation vector can be defined based on the fact that $(\mat{M}^{\E{5}}_B)^6 = \identity$.
Noting that
\begin{equation}
    g^{\E{5}}_3 \equiv (g^{\E{5}}_B)^6: \vec{x} \to \vec{x} + \mat{M}^{\E{5}}_{06} \vec{T}^{\E{5}}_B,
\end{equation}
where $\mat{M}^{\E{5}}_{06} = \diag(0, 0, 6)$ from \eqref{eqn:M0jdef},
the third translation vector of the associated \E{1} for \E{5} is as quoted in the main text \eqref{eqn:E5assocE1},
\begin{equation}
    \vec{T}^{\E{5}}_3 = \begin{pmatrix} 0 \\ 0 \\ 6 L_{Bz} \end{pmatrix}.
\end{equation}
For our subsequent development here we choose $\vec{T}_3 \equiv \vec{T}^{\E{5}}_3$.

To proceed, we can again choose $\vec{T}_{\A{1}}$ such that $\vert\vec{T}_{\A{1}\perp}\vert \leq \vert\vec{T}_{\A{2}\perp}\vert$, where $\perp$ means the projection into the $xy$-plane.
Also, by adding or subtracting integer multiples of $\vec{T}_{\A{1}}$ from $\vec{T}_{\A{2}}$, we can ensure that $\vert L_{2x}\vert < \vert L_{1x}\vert$; thus we generically have
\begin{equation}
    \label{eqn:E5A1A2ordering}
     (L_{2x})^2 < (L_{1x})^2 \leq (L_{2x})^2 + (L_{2y})^2 .
\end{equation}

Whereas for \E{4} we had two sets of conditions, \eqref{eqn:E4groupconditions}, on the products of generators and their inverses being integer linear combination of $\vec{T}_{i}^{\E{4}}$, which we were able to limit to one, for \E{5} we have 5 sets, which we can limit to 3 
\begin{align}
    \label{eqn:E5groupcondition1}
    \mat{M}_B \vec{T}_{\A{i}} 
    &= \sum_{k} m^{(i)}_k \vec{T}_k, \quad \mbox{for some } m_k^{(i)}\in \integers , \\
    \label{eqn:E5groupcondition2}
    (\mat{M}_B)^2 \vec{T}_{\A{i}} 
    &= \sum_{k} n^{(i)}_k \vec{T}_k, \quad \mbox{for some } n_k^{(i)}\in \integers , \\
    \label{eqn:E5groupcondition3}
    (\mat{M}_B)^3 \vec{T}_{\A{i}} 
    &= \sum_{k} p^{(i)}_k \vec{T}_k, \quad \mbox{for some } p_k^{(i)}\in \integers .
\end{align}
This follows since the condition for $(\mat{M}_B)^4$ is equivalent to that for $(\mat{M}_B)^2$ and the condition for $(\mat{M}_B)^5$ is equivalent to that for $\mat{M}_B$.

This leads to a daunting 6 sets of conditions.
However, if we focus on the $(\mat{M}_B)^2$ equations \eqref{eqn:E5groupcondition2} we see they are identical to the conditions from \E{4} \eqref{eqn:E4groupconditions}, thus the solution is the same and we have an immediate candidate for the general solution.
By direct computation, we can verify that this solution also satisfies \eqref{eqn:E5groupcondition1} and \eqref{eqn:E5groupcondition3}, so there are no additional constraints.

Finally, shifting to an arbitrary origin, as in \E{4} we have the general form from the text \eqref{eqn:E5generalT},
\begin{equation}
    \vec{T}^{\E{5}}_{\A{1}} = \begin{pmatrix} L_{A} \\ 0 \\ 0 \end{pmatrix} , \quad
    \vec{T}^{\E{5}}_{\A{2}} = \begin{pmatrix} -L_{A}/2 \\ \sqrt{3} L_{A} / 2 \\ 0 \end{pmatrix} , \quad
    \vec{T}^{\E{5}}_{B} = \begin{pmatrix} L_{Bx} \\ L_{By} \\ L_{Bz} \end{pmatrix} .
\end{equation}

\subsection{\E{6}: Hantzsche-Wendt space}
\label{app:E6}

In contrast to the other orientable, compact spaces, the Hantzsche-Wendt space has rotations around multiple axes.
These are conventionally chosen as half turns about three axes which must be orthogonal (see, e.g., \rcite{Thurston1982ThreeDM}).
Here, we use the freedom to rotate the coordinate system such that these three axes are aligned with the coordinate axes choosing $\mat{M}_A = \mat{R}_{\unitvec{x}}(\pi)$, $\mat{M}_B = \mat{R}_{\unitvec{y}}(\pi)$, and $\mat{M}_C = \mat{R}_{\unitvec{z}}(\pi)$. For \E{6} we then have
\begin{equation}
    \mat{M}_{A} =
    \begin{pmatrix}
        1 & \hphantom{-}0 & \hphantom{-}0\\
        0 & -1 & \hphantom{-}0\\
        0 & \hphantom{-}0 & -1
    \end{pmatrix}, \quad
        \mat{M}_{B} = 
    \begin{pmatrix}
        -1 & \hphantom{-}0 & \hphantom{-}0\\
        \hphantom{-}0 & \hphantom{-}1 & \hphantom{-}0\\
        \hphantom{-}0 & \hphantom{-}0 & -1
    \end{pmatrix} , \quad
        \mat{M}_{C} = 
    \begin{pmatrix}
        -1 & \hphantom{-}0 & \hphantom{-}0\\
        \hphantom{-}0 & -1 & \hphantom{-}0\\
        \hphantom{-}0 & \hphantom{-}0 & \hphantom{-}1
    \end{pmatrix},
\end{equation}
so that
\begin{equation}
    (\mat{M}_{A})^2 = (\mat{M}_{B})^2 = (\mat{M}_{C})^2 = \identity.
\end{equation}
This has used up all the rotational freedom.
There is still the freedom to choose a special origin so as to simplify the translation vectors (as we have done in the previous cases), however, we will begin without using this freedom and thus start with the most general vectors
\begin{equation}
    \vec{T}_{a} = \begin{pmatrix} L_{a x} \\ L_{a y} \\ L_{a z} \end{pmatrix} \quad \mbox{for } a \in \{ A,B,C \} .
\end{equation}
An associated \E{1} can again be constructed from three pure translations.
In contrast to \E{2}--\E{5}, the action of each generator in \E{6} includes a rotation.
However, since the rotations are all half turns, we can choose the three translations as was done for $g^{\E{2}}_3$.
From the properties of the matrices $\mat{M}_a$, we construct the pure translations from $(g_a)^2:\vec{x} \to \vec{x} + (\identity + \mat{M}_a) \vec{T}_a$.
This leads to the pure translation vectors quoted in the main text \eqref{eqn:E6assocE1},
\begin{equation}
    \vec{T}_1 \equiv \vec{T}^{\E{6}}_1 = \begin{pmatrix} 2 L_{Ax} \\ 0 \\ 0 \end{pmatrix}, \quad
    \vec{T}_2 \equiv \vec{T}^{\E{6}}_2 = \begin{pmatrix} 0 \\ 2 L_{By} \\ 0 \end{pmatrix}, \quad
    \vec{T}_3 \equiv \vec{T}^{\E{6}}_3 = \begin{pmatrix} 0 \\ 0 \\ 2 L_{Cz} \end{pmatrix}.
\end{equation}

To proceed, we note that $\mat{M}_a \mat{M}_b = \mat{M}_c$ for any $\{ a, b, c \}$ permutation of $\{ A, B, C \}$, which means that we must have $g_a g_b = g_c + \vec{t}$ for $\vec{t}$ a pure translation constructed from integer linear combinations of the $\vec{T}_i$.
This leads to the set of conditions
\begin{align}
    \mat{M}_A \vec{T}_B + \vec{T}_A &= \vec{T}_C + \sum_i m_i^C \vec{T}_i , \\
    \mat{M}_B \vec{T}_C + \vec{T}_B &= \vec{T}_A + \sum_i m_i^A \vec{T}_i , \\
    \mat{M}_C \vec{T}_A + \vec{T}_C &= \vec{T}_B + \sum_i m_i^B \vec{T}_i ,
\end{align}
for some $m^a_i \in \integers$.
Written out explicitly, these conditions are
\begin{align}
    \label{eqn:E6condition1}
    \begin{pmatrix}
        \hphantom{-}L_{Bx} + L_{Ax} \\
        -L_{By} + L_{Ay} \\
        -L_{Bz} + L_{Az} 
    \end{pmatrix}
    &= 
    \begin{pmatrix}
        2 m^C_1  L_{Ax} + L_{Cx} \\
        2 m^C_2  L_{By} + L_{Cy} \\
        2 m^C_3  L_{Cz} + L_{Cz}
    \end{pmatrix} , \\
    \label{eqn:E6condition2}
    \begin{pmatrix}
        -L_{Cx} + L_{Bx} \\
        \hphantom{-}L_{Cy} + L_{By} \\
        -L_{Cz} + L_{Bz}
    \end{pmatrix}
    &= 
    \begin{pmatrix}
        2 m^A_1  L_{Ax} + L_{Ax} \\
        2 m^A_2  L_{By} + L_{Ay} \\
        2 m^A_3  L_{Cz} + L_{Az}
    \end{pmatrix} , \\
    \label{eqn:E6condition3}
    \begin{pmatrix}
        -L_{Ax} + L_{Cx} \\
        -L_{Ay} + L_{Cy} \\
        \hphantom{-}L_{Az} + L_{Cz}
    \end{pmatrix}
    &= 
    \begin{pmatrix}
        2 m^B_1  L_{Ax} + L_{Bx} \\
        2 m^B_2  L_{By} + L_{By} \\
        2 m^B_3  L_{Cz} + L_{Bz}
    \end{pmatrix} .
\end{align}
From these, we first focus on the $x$-component of \eqref{eqn:E6condition2}, the $y$-component of \eqref{eqn:E6condition3}, and the $z$-component of \eqref{eqn:E6condition1}.  This gives
\begin{align}
    \label{eqn:E6lengthdifferences}
    L_{Cx}-L_{Bx} &= -(2m^A_1+1) L_{Ax}, \nonumber \\
    L_{Ay}-L_{Cy} &= -(2m^B_2+1) L_{By}, \\
    L_{Bz}-L_{Az} &= -(2m^C_3+1) L_{Cz}, \nonumber
\end{align}
reducing the original set to 6 parameters.
Shifting by integer multiples of the translation vectors $\vec{T}_i$ allows us to remove all the even multiples on the right-hand side of these expressions.
In particular, the choice $m^A_1=m^B_2=m^C_3=-1$ reproduces the general form from the text \eqref{eqn:E6generalT},
\begin{equation}
    \vec{T}^{\E{6}}_{A} = \begin{pmatrix} L_{Ax} \\ L_{By} + L_{Cy} \\ L_{Az} \end{pmatrix} , \quad
    \vec{T}^{\E{6}}_{B} = \begin{pmatrix} L_{Bx} \\ L_{By} \\ L_{Cz} + L_{Az} \end{pmatrix} , \quad
    \vec{T}^{\E{6}}_{C} = \begin{pmatrix} L_{Ax} + L_{Bx} \\ L_{Cy} \\ L_{Cz} \end{pmatrix} .
\end{equation}

The behavior of the generators under a shift of origin suggests an alternative form for the translation vectors.
A shift in the origin preserves the difference between the lengths on the left-hand side of \eqref{eqn:E6lengthdifferences} but not their sums.
(In fact, this is another way to understand why the three lengths on the right-hand sides appear in the translation vectors of the associated \E{1} \eqref{eqn:E6assocE1}.)
The sums of the lengths can be used to parametrize the generators in another form.
Notice that the sum of the $x$-components of \eqref{eqn:E6condition1} and \eqref{eqn:E6condition3} gives
\begin{equation}
    L_{Bx} + L_{Cx} = 2( m^C_1 + m^B_1 ) L_{Ax} + L_{Bx} + L_{Cx},
\end{equation}
in other words that
\begin{equation}
    2( m^C_1 + m^B_1 ) L_{Ax} = 0.
\end{equation}
But this is precisely the freedom used above to shift by integer multiples of $\vec{T}_1$.
In other words, the sum of lengths $L_{Bx} + L_{Cx}$ is unconstrained by the set of conditions.
Similar arguments hold for the sums $L_{Cy} + L_{Ay}$ and $L_{Az} + L_{Bz}$.
This motivates the definition of new continuous, dimensionless parameters given by the average of these length pairs as
\begin{align}
   r_x &\equiv \frac{L_{Bx}+L_{Cx}}{2L_{Ax}} = \frac{2L_{Bx}+L_{Ax}}{2L_{Ax}} , \nonumber\\
   r_y &\equiv \frac{L_{Cy}+L_{Ay}}{2L_{By}} = \frac{2L_{Cy}+L_{By}}{2L_{By}} , \\
   r_z &\equiv \frac{L_{Az}+L_{Bz}}{2L_{Cz}} = \frac{2L_{Az}+L_{Cz}}{2L_{Cz}} , \nonumber
\end{align}
where the second equality in each line follows from the same choices that went into the general form \eqref{eqn:E6generalT}.
Solving for one of the lengths in each equation, we can write these as
\begin{equation}
    L_{Bx} = \left( r_x - \frac{1}{2} \right) L_{Ax}, \quad
    L_{Cy} = \left( r_y - \frac{1}{2} \right) L_{By}, \quad
    L_{Az} = \left( r_z - \frac{1}{2} \right) L_{Cz}.
\end{equation}
Finally, these give the alternative form quoted in the main text \eqref{eqn:E6generalTalt},
\begin{equation}
    \vec{T}_{A}^{\E{6}} = \begin{pmatrix} L_{Ax} \\ (r_y + \frac{1}{2})L_{By} \\ (r_z- \frac{1}{2})L_{Cz} \end{pmatrix} , \quad
    \vec{T}_{B}^{\E{6}} = \begin{pmatrix} (r_x- \frac{1}{2})L_{Ax} \\ L_{By} \\ (r_z + \frac{1}{2})L_{Cz} \end{pmatrix} , \quad
    \vec{T}_{C}^{\E{6}} = \begin{pmatrix} (r_x + \frac{1}{2})L_{Ax} \\ (r_y- \frac{1}{2})L_{By} \\ L_{Cz} \end{pmatrix} .
\end{equation}

\subsection{\E{11}: Chimney space}
\label{app:E11}

The general form of the generators for the chimney space can be determined in a manner similar to \E{1}, though it requires two generators, not three.
Since $\mat{M}^{\E{11}}=\identity$, \E{11} is homogeneous and we can use the rotational freedom to define one of the translation vectors as the $x$-axis of the coordinate system and the remaining freedom to define the $xy$-plane.
In other words, two vectors define a plane so the two translation vectors can be used to define the $xy$-plane.
Thus the two translations $\vec{T}^{\E{1}}_{i}$ can be parametrized most generally as quoted in the text \eqref{eqn:E11generalT},
\begin{equation} 
    \label{eqn:E11translations}
    \vec{T}^{\E{11}}_{1} = \begin{pmatrix} L_{1x} \\ 0 \\ 0 \end{pmatrix} , \quad
    \vec{T}^{\E{11}}_{2} = \begin{pmatrix} L_{2x} \\ L_{2y} \\ 0 \end{pmatrix} ,
\end{equation}
or equivalently
\begin{equation} 
        \vec{T}^{\E{11}}_{1} = L_1 \begin{pmatrix} 1 \\ 0 \\ 0 \end{pmatrix} , \quad
        \vec{T}^{\E{11}}_{2} = L_2 \begin{pmatrix} \cos\alpha \\ \sin\alpha \\ 0 \end{pmatrix} .
\end{equation}

\subsection{\E{12}: Chimney space with half turn}
\label{app:E12}

The chimney space with half turn is a root of \E{11} and is a limit of the half-turn space \E{2} with one non-compact dimension.
It has one generator, $g_A$, that is a translation and one generator, $g_B$, that is a corkscrew motion.
The rotational freedom to align the coordinate system allows the axis of rotation to be chosen as the $z$-axis.
Freedom to rotate around the $z$-axis allows the choice $T_{Ay}=0$.
Finally, we can choose a special origin that allows us to set $L_{Bx}=L_{By}=0$; we will restore an arbitrary origin at the end of the calculation.
With all of this, we can choose our general starting case to be
\begin{equation}
    \vec{T}_A = \begin{pmatrix} L_{Ax} \\ 0 \\ L_{Az} \end{pmatrix}, \quad
    \vec{T}_B = \begin{pmatrix} 0 \\ 0 \\ L_{Bz} \end{pmatrix}.
\end{equation}
From these a set of pure translations, an associated \E{11}, can be constructed.
The action of $g_A$ is already a pure translation, so one of the needed translation vectors can be chosen as $\vec{T}_1 \equiv \vec{T}_A$.
A second translation vector can be defined based on the fact that $(\mat{M}^{\E{12}}_B)^2 = \identity$.
Note that 
\begin{equation}
    g^{\E{12}}_2 \equiv (g^{\E{12}}_B)^2: \vec{x} \to \vec{x} + (\identity + \mat{M}^{\E{12}}_B) \vec{T}^{\E{12}}_B \equiv \vec{x} + \mat{M}^{\E{12}}_{02} \vec{T}^{\E{12}}_B,
\end{equation}
for $\mat{M}^{\E{12}}_{02} = \diag(0, 0, 2)$ from \eqref{eqn:M0jdef}.
Thus the second translation vector of the associated \E{11} for \E{12} is as quoted in the main text \eqref{eqn:E12assocE11},
\begin{equation}
   \vec{T}^{\E{12}}_2 \equiv \begin{pmatrix} 0 \\ 0 \\ 2 L_{Bz} \end{pmatrix}.
\end{equation}
For our subsequent development here we choose $\vec{T}_2 \equiv \vec{T}^{\E{12}}_2$.

The remaining condition to impose is that
\begin{equation}
    \inverse{g}_B \inverse{g}_{A} g_B g_A: \vec{x} \to \vec{x} + \vec{T}_A - \mat{M}_B \vec{T}_A
    = \vec{x} + \sum_{k} m_k \vec{T}_k,
\end{equation}
for $m_k \in \integers$.
Since $\vec{T}_A$ is already the pure translation $\vec{T}_1$, this is equivalent to requiring
\begin{equation}
    \mat{M}_B \vec{T}_A = \sum_{k} m_k \vec{T}_k, \quad \mbox{for some } m_k \in \integers,
\end{equation}
leading to the set of conditions
\begin{equation}
    \begin{pmatrix} -L_{Ax} \\ \hphantom{-} 0 \\ \hphantom{-} L_{Az}  \end{pmatrix}
    = \begin{pmatrix} m_1 L_{Ax} \\ 0 \\ m_1 L_{Az} + 2 m_2 L_{Bz} \end{pmatrix} .
\end{equation}

From the $x$-component of these conditions, we see that $m_1 = -1$.
Applying this to the $z$-component we immediately see that $L_{Az}=0$ (and $m_2=0$) is the general solution.
Shifting back to an arbitrary origin we have the generic result quoted in the main text \eqref{eqn:E12generalT},
\begin{equation}
    \vec{T}^{\E{12}}_A = \begin{pmatrix} L_{Ax} \\ 0 \\ 0 \end{pmatrix}, \quad
    \vec{T}^{\E{12}}_B = \begin{pmatrix} L_{Bx} \\ L_{By} \\ L_{Bz} \end{pmatrix}.
\end{equation}

As stated, we notice that \E{12} is a limit of \E{2} with $\vert\vec{T}^{\E{2}}_{\A{2}}\vert\to\infty$.
As shown above, we were able to use the same arguments as for \E{2} to argue that $L_{Ay}=0$, or equivalently that $\alpha=0$.

\subsection{\E{16}: Slab space including rotations}
\label{app:E16}

The slab space is distinct from the other topologies as it allows a corkscrew motion.
The standard definition of the slab space does not contain this rotation since the rotated slab is topologically the same as the unrotated slab.
However, a rotation does lead to physically observable effects, and thus unrotated and rotated slab manifolds must be treated separately.
For this reason, we split the description of \E{16} into two cases labeled \slabh\ and \slabi.

\subsubsection{\slabh: Conventional unrotated slab space}
\label{app:E16h}

The general form of the generators for the unrotated slab space follows directly from the fact that it contains only a single generator that is a pure translation.
Since $\mat{M}^{\slabh}_A=\identity$, \slabh\ is homogeneous and we can use the rotational freedom to define the translation vector as the $z$-axis of the coordinate system.
Thus we immediately arrive at the general form quoted in the text \eqref{eqn:E16hgeneralT},
\begin{equation} 
    \vec{T}^{\slabh}_A = L\begin{pmatrix} 0 \\ 0 \\ 1 \end{pmatrix}.
\end{equation}

\subsubsection{\slabi: General rotated slab space}
\label{app:E16i}

An arbitrary rotation by an angle $\zeta$ is allowed in the slab space.
We can use part of the rotational freedom to choose the $z$-axis of the coordinate system as the rotation axis.
We can then use the freedom of rotation about this axis to set the $y$-component of the translation vector to zero.
Thus we can start with
\begin{equation}
    \mat{M} = \mat{R}_{\unitvec{z}} (\zeta), \quad \mbox{with} \quad \vec{T} = \begin{pmatrix} L_x \\ 0 \\ L_z \end{pmatrix}.
\end{equation}

In principle $0 < \zeta < 2\pi$ (where the endpoints are excluded so that \slabi\ does not include \slabh) and this would be the general action of a group element.
However, in this work, we focus on the observable consequences of our universe having a non-trivial topology.
To this end, in \cref{secn:eigenmodes} the eigenmodes of each topology are determined.
In order to use a Fourier basis to expand all functions on the manifold, a required condition is (effectively) that $\mat{M}^q = \identity$ for some finite $q \in \integers^{>0}$.
This in turn requires that $q \zeta = 2\pi p$ for $p\in \integers^{\neq 0}$.
In other words, $\zeta$ must be a rational multiple of $2\pi$.
If $\zeta$ is not a rational multiple of $2\pi$ then only cylindrically symmetric functions can be expanded in a Fourier basis.
The translation vector given above is general, it is the rotation matrix that has further constraints.
With this, we arrive at the general form given in the text \eqref{eqn:E16igeneralT},
\begin{align} 
    &\mat{M}^{\slabi}_B = \mat{R}_{\unitvec{z}}(2\pi p / q)
     = \begin{pmatrix} \cos(2\pi p/q) & -\sin(2\pi p/q) & 0 \\
       \sin(2\pi p/q) & \hphantom{-}\cos(2\pi p/q) & 0 \\
       0 & \hphantom{-}0 & 1 \end{pmatrix} ,
       \quad \mbox{with} \nonumber \\
     &\vec{T}^{\slabi}_B = \begin{pmatrix} L_x \\ 0 \\ L_z \end{pmatrix} = L\begin{pmatrix} \cos\beta \\ 0 \\ \sin\beta \end{pmatrix},
\end{align}
and $p\in \integers^{\neq 0}$, $q\in \integers^{>0}$, and $|p|$ and $q$ relatively prime.
From this, a pure translation, an associated \slabh, can be constructed.
Based on the fact that $(\mat{M}^{\slabi}_B)^q = \identity$ we note that
\begin{equation}
    (g^{\slabi}_B)^q: \vec{x} \to \vec{x} + \mat{M}^{\slabi}_{0q} \vec{T}^{\slabi}_B,
\end{equation}
for $\mat{M}^{\slabi}_{0q} = \diag(0, 0, q)$ from \eqref{eqn:M0jdef}.
Thus the translation vector of the associated \slabh\ for \slabi\ is as quoted in the main text \eqref{eqn:E16iassocE16h},
\begin{equation}
    \vec{T}^{\slabi}_1 = \begin{pmatrix} 0 \\ 0 \\ q L_{z} \end{pmatrix}.
\end{equation}

\bibliographystyle{utphys}
\bibliography{topology,additional}
\end{document}